\documentclass[fleqn]{LMCS}

\usepackage{url}
\usepackage{latexsym}
\usepackage{amsfonts}
\usepackage{amssymb,amsmath}
\usepackage{enumerate}
\usepackage[breaklinks=true]{hyperref}

\theoremstyle{plain}
\theoremstyle{plain}
\theoremstyle{plain}\newtheorem{hyps}[thm]{Hypotheses}
\theoremstyle{definition}\newtheorem{note}[thm]{Note}
\def\eg{{\em e.g.}}
\def\cf{{\em cf.}}

\newcommand{\To}{T$_0$}
\newcommand{\fo}{f$_0$}

\newcommand{\EE}{\mathcal{E}}

\newcommand{\restr}{\upharpoonright}

\newcommand{\lqq}{\mbox{``}}
\newcommand{\rqq}{\mbox{''}}

\newcommand{\defis}{\mbox{-}}

\newcommand{\setof}[1]{\{#1\}}
\newcommand{\App}{{\rm App}}

\newcommand{\bYdef}{\mathrel{\BYDEF}}
\newcommand{\BYDEF}%
{\mbox{$\hspace{.2em}%
\rightleftharpoons\hspace{.2em}$}}

\newcommand{\pr}{^{\prime}}
\newcommand{\pointwiselub}{\biguplus}

\newcommand{\arr}{\rightarrow}
\newcommand{\arrr}{\longrightarrow}
\newcommand{\Arr}{\Rightarrow}
\newcommand{\Arrr}{\Longrightarrow}
\newcommand{\Larr}{\Leftarrow}
\newcommand{\Arra}{\Leftrightarrow}
\newcommand{\embed}{\hookrightarrow}

\newcommand{\la}{\langle}
\newcommand{\ra}{\rangle}
\newcommand{\lla}{\langle\!\langle}
\newcommand{\rra}{\rangle\!\rangle}
\newcommand{\Osem}[1]{\lla #1 \rra}
\newcommand{\Dsem}[1]{[\![ #1 ]\!]}
\newcommand{\tuple}[1]{\la #1 \ra}
\newcommand{\AND}{\mathrel{\&}}

\newcommand{\G}{G}
\newcommand{\NN}{\mathbf{N}}
\newcommand{\bNN}{\bar{\NN}}
\newcommand{\bMM}{\bar{\MM}}

\newcommand{\BB}{\mathbf{B}}
\newcommand{\bQQ}{\bar{\QQ}}
\newcommand{\bbQ}{\mathbb{Q}}
\newcommand{\tbbQ}{\mathbb{Q}}%
\newcommand{\bbW}{\mathbb{W}}
\newcommand{\tbbW}{\mathbb{W}}%
\newcommand{\bbD}{\mathbb{D}}

\newcommand{\Basictype}{\iota}
\newcommand{\PCF}{\mbox{\bf PCF}}
\newcommand{\cPCF}{\mbox{$\mathcal PCF$}}

\newcommand{\LCF}{\mbox{\bf LCF}}

\newcommand{\I}{\mbox{\bf I}}
\newcommand{\K}{\mbox{\bf K}}
\newcommand{\Ss}{\mbox{\bf S}}
\newcommand{\Y}{\mbox{\bf Y}}
\newcommand{\IF}{\mbox{\bf if}}
\newcommand{\PIF}{\mbox{\bf pif}}
\newcommand{\THEN}{\mbox{\bf then}}
\newcommand{\ELSE}{\mbox{\bf else}}
\newcommand{\mPIF}{m_{\rm\bf pif}}

\newcommand{\CC}{{\mathcal C}}
\newcommand{\KA}{{\mathcal K}^A}
\newcommand{\KbarA}{{\mathcal K}^{\bar{A}}}
\newcommand{\KAeff}{{\mathcal K}^A_{\rm eff}}

\newcommand{\jj}{\underline{j}}

\newcommand{\bottom}{{\perp}}
\newcommand{\Undef}{{\perp}}

\newcommand{\fla}[1]{$#1$}   %

\newcommand{\Lub}{\bigsqcup}
\newcommand{\MM}{{\mathcal M}}
\newcommand{\sM}{M^{\bigstar}}
\newcommand{\sm}{m^{\bigstar}}
\newcommand{\sMM}{\MM^{\bigstar}}
\newcommand{\HH}{{\mathcal H}}
\newcommand{\bx}{\bar{x}}
\newcommand{\bY}{\bar{y}}
\newcommand{\bc}{\bar{c}}
\newcommand{\bC}{\bar{C}}
\newcommand{\bq}{\bar{q}}
\newcommand{\bj}{\bar{j}}
\newcommand{\bi}{\bar{i}}
\newcommand{\bpsi}{\bar{\psi}}

\newcommand{\dx}{\dot{x}}
\newcommand{\dy}{\dot{y}}
\newcommand{\dz}{\dot{z}}
\newcommand{\df}{\dot{f}}
\newcommand{\dg}{\dot{g}}

\newcommand{\dQ}{\dot{\bbQ}}
\newcommand{\dW}{\dot{\bbW}}

\newcommand{\dE}{\dot{E}}
\newcommand{\dEE}{\dot{\EE}}

\newcommand{\Id}[1]{I(#1)}

\newcommand{\QQ}{{\mathcal Q}}

\newcommand{\tQ}{\tilde{Q}}
\newcommand{\tW}{\tilde{W}}
\newcommand{\WW}{{\mathcal W}}

\newcommand{\lee}{\preceq}
\newcommand{\sqle}{\sqsubseteq}
\newcommand{\?}{\mbox{?}}

\newcommand{\false}{\mbox{\bf false}}
\newcommand{\true}{\mbox{\bf true}}

\def\doi{3 (3:7) 2007}
\lmcsheading%
{\doi}
{1--50}
{}
{}
{Feb.~17, 2006}
{Sep.~\phantom{0}6, 2007}
{}

\begin{document}

\title[Fully Abstract Models for \PCF\ and $\PCF^+$]{Inductive Definition and 
Domain Theoretic Properties \\
of 
Fully Abstract Models for \PCF\ and $\PCF^+$}

\author[V.~Sazonov]{Vladimir Sazonov}	%
\address{Department of Computer Science, 
the University of Liverpool, 
Liverpool, L69 3BX, U.K.
}	%
\email{vyszuk@yahoo.co.uk, \url{v.sazonov@csc.liv.ac.uk}}  %


\begin{abstract}
A construction of fully abstract typed models 
for PCF and PCF+ 
(i.e., PCF + ``parallel conditional function''), respectively, is presented. 
It is based on general notions of 
sequential computational strategies and wittingly 
consistent 
non-deterministic strategies
introduced by the author 
in the seventies. 
Although these notions of strategies are old, the definition of the 
fully abstract models is new, in that it 
is given level-by-level in the finite type hierarchy. 
To prove full abstraction and non-dcpo 
domain theoretic properties of these models, 
a theory of computational strategies is developed. 
This is also an alternative and, in a sense, an analogue 
to the later game strategy semantics approaches of 
Abramsky, Jagadeesan, and Malacaria; 
Hyland and Ong; and  Nickau. 
In both cases of PCF and PCF+ 
there are definable universal (surjective) 
functionals from numerical functions to any given type, 
respectively, 
which also makes each of these models unique up to isomorphism. 
Although such models are non-omega-complete and therefore 
not continuous in the traditional terminology, 
they are also proved to be 
sequentially complete 
(a weakened form of omega-completeness), 
``naturally'' continuous 
(with respect to existing directed ``pointwise'', or ``natural'' lubs) 
and also 
``naturally'' omega-algebraic and 
``naturally'' bounded complete---appropriate generalisation 
of the ordinary notions 
of domain theory to the case of non-dcpos. 

\end{abstract}
\maketitle

\section{Introduction}\label{sec-intro}

\noindent
\LCF, a \emph{Logic for Computable Functions}, 
was introduced in 1969 by Scott 
in a seminal paper \cite{Scott93} (published only in 1993). 
Its term language \PCF---a typed version of the lambda calculus 
over integers and booleans with the 
least fixed point operator \fla{\Y}---was further considered in the middle 
of the seventies by Plotkin \cite{Plotkin77}, 
Milner \cite{Milner77}, and the author 
\cite{Saz76d,Saz76AL,Saz76SMZH,Saz76t}. 
In particular, the expressive power of \fla{\PCF} in the framework 
of a standard continuous model $\setof{\bbD_{\alpha}}$ 
for \PCF\ was described 
in terms of 
\emph{(sequential) computational strategies} 
as the Theorem:  
``\emph{definable in \fla{\PCF}} = 
\emph{sequentially computable}'' \cite{Saz76AL}. 
Also, a 
\emph{precise correspondence between operational 
and denotational semantics}  
in various formulations (and even an untyped version) 
was 
obtained in \cite{Saz76SMZH} (and independently, 
in somewhat different terms, by Hyland~1976, 
Plotkin~1977 and Wadsworth~1976). 
The full abstraction property of the standard continuous model for 
$\PCF^+=\textrm{``}\PCF+\mbox{parallel OR}$ (or parallel \IF)''
(by definability of all finite continuous functionals) 
was stated, as well as the fact that 
\fla{\PCF^{++}=\textrm{``}\PCF+\mbox{ parallel OR}+\exists}'' 
defines  
all computable 
continuous functionals (Plotkin \cite{Plotkin77} and, 
without publishing proofs, 
the author \cite{Saz76d,Saz76AL}).
Degrees of parallelism for continuous 
finite type functionals with various examples 
were introduced in \cite{Saz76d} 
(see also \eg\ \cite{Bucciarelli97Degrees,MTrakhte76TCS}). 

A first, essentially syntactic, construction of 
a continuous \emph{fully abstract model} for \PCF\ 
was given in 1977 by Milner 
\cite{Milner77}. The characteristic property of fully abstract models 
is as follows: %
\[
\forall\; {\rm ground\ type\ program\ contexts\ } 
\CC\;(\Dsem{\CC[M]}\sqle\Dsem{\CC[N]})\Arrr\Dsem{M}\sqle\Dsem{N}
\]
which says (for `$=$' in place of `$\sqle$') that, if two program fragments 
behave equivalently in all computational contexts, then they 
should have the same denotational semantics. The main reason for focusing 
particular attention on this definition and on Milner's model 
is that for the standard continuous 
model  $\setof{\bbD_{\alpha}}$ and \PCF\ this natural property 
of the denotational 
semantics does \emph{not} hold. As mentioned above, \PCF\ defines 
(exactly) all sequentially computable functionals, whereas the 
standard model contains some `extra' elements, such as 
`parallel' disjunction OR $\in \bbD_{o,o\arr o}$ and 
`parallel' existential quantification 
$\exists\in \bbD_{(\iota\arr o)\arr o}$. This is essentially the reason 
for the violation of full abstraction. 
But although Milner's fully abstract model 
satisfies desirable properties of continuity, 
it is not a satisfactory domain theoretic characterization of sequentiality 
due to the existence in it of non-sequential 
limit functionals (Normann \cite{Normann2004}).

Also, having a syntactic nature, 
the definition of this model was considered as 
not very satisfactory 
in comparison with the standard model of \emph{all} continuous functionals. 
Non-syntactic
\emph{game semantic} approaches 
to defining fully abstract models 
were developed by  
Abramsky, Jagadeesan, Malacaria~\cite{Abramsky-Jagadeesan2000}; 
Hyland, Ong~\cite{Hyland-Ong2000} and Nickau~\cite{Nickau96}.  
Various approaches to sequentiality and full abstraction were considered also 
by 
Kahn and Plotkin~\cite{KP93},  
Berry and Curien~\cite{BC82}, 
Bucciarelli and 
Ehrhard~\cite{BE91,BE93}; Curien~\cite{Cur92}; 
Jung and Stoughton~\cite{Jung-Stoughton93}; 
O'Hearn and Riecke~\cite{OR95}; 
Marz, Rohr and Streicher~\cite{Marz-Thesis,Marz-Rohr-Streicher}; 
Sieber~\cite{Sieber92}, 
Cartwright and Felleisen~\cite{Cartw-Fell92}
and others. Unlike this paper some of them consider more general sequentiality concepts 
going outside \PCF\ and even outside the class 
of monotonic functionals 
such as sequentially realizable functionals (equivalent to some other 
approaches implicitly mentioned above); a unifying approach is presented by 
Longley~\cite{Longley-seq-realizable}.

Hyland and Ong~\cite{Hyland-Ong2000} 
identified a very close analogy between the old approach 
to sequentiality of functionals via computational strategies in \cite{Saz76AL} 
and their game theoretic framework. 
One of the goals of this paper is to demonstrate 
\emph{how computational 
sequential strategies could define a fully abstract model $\setof{\bbQ_\alpha}$ 
for \fla{\PCF} inductively, level-by-level in the finite type hierarchy 
in a direct computational way} corresponding to 
the original definition and characterization of higher type sequentiality 
in \cite{Saz76SMZH,Saz76AL}. (The latter was applied only to the standard, 
non-fully-abstract continuous 
model $\setof{\bbD_\alpha}$ containing not only sequential functionals.) 
It is important to stress the straightforward, inductive 
character of the definition of $\setof{\bbQ_\alpha}$ which may be compared, 
at least partly, with the inductive definition of the continuous model 
$\setof{\bbD_{\alpha}}$. 
Assuming, by induction, that we have 
the class of sequential functionals 
of types up to level $l$, we define what are sequential functionals 
of the level $l+1$ as those computable by sequential strategies. 
In this respect our approach differs from the game-semantic one 
based on a quotient construction for all types simultaneously. 
However, 
proving the essential properties of the inductively defined model $\setof{\bbQ_\alpha}$ 
of hereditarily sequential functionals is not so direct and 
requires the quite involved theory of computational strategies 
and a quotient construction $\tQ\cong\bbQ$ giving an alternative, non-inductive 
definition of the same model. Reference in the inductive step of the definition 
of $\bbQ$ to all 
sequential functionals of the previous levels also reflects the complicated 
character of this inductive definition. A finitary version referring only to 
the immediate subtypes of the given type should not be possible due to 
the undecidability result 
of Loader \cite{LoaderTCS2001}.

However, the definition of $\bbQ$ is sufficiently straightforward, although involving 
some technical complications to make it mathematically correct and, additionally, 
to crucially simplify the correctness proof of the induction step. 

As in \cite{Abramsky-Jagadeesan2000}, ``we want to capture just those 
sequential computations in which the different parts 
or modules interact with each other in purely functional 
fashion'' and, as in \cite{Hyland-Ong2000}, ``without 
recourse to the syntax or operational semantics 
of the language'' (\PCF). More precisely, 
we will use computability by sequential strategies to define (hereditarily) 
sequential functionals. 
Although  
\PCF\ is a partial case of the general concept of a system 
of strategies, our definitions will not be reduced simply to doing things in \PCF. 
We will work in the quite general 
terms of abstract computability in higher types 
in a ``functional fashion'', by using ``interpreted computations'' 
(involving applicative terms) in the style of \emph{denotational semantics},
to define a fully abstract 
model for \PCF. 
Also note that the very term ``sequential'' primarily assumes 
``sequentially computable''. That is why involving some kind of 
computability approach 
at the level of denotational semantics is quite natural. In fact, we 
will also provide an alternative, generalized \emph{operational semantics} 
of strategies---not only for ground types---and 
demonstrate that it is coherent with the denotational one (the approach 
originally presented in \cite{Saz76SMZH} 
but not in the ``fully abstract framework'' 
as in the present paper.)
This distinction together with the interplay 
between operational ($\Osem{\defis}$) and denotational ($\Dsem{\defis}$) semantics 
($\Dsem{\Osem{A}}=\Dsem{A}$ for arbitrary finite type combinations of strategies) 
is one of the crucial points 
of this paper.

On the other hand, we read in 
\cite{Hyland-Ong2000} that: 
``we do not have a proper definition of higher-type 
sequentiality from first principles''. 
There could probably be various philosophical 
views concerning 
what are these ``first principles''. 
However, for the simpler case of non-higher-type sequentiality, 
we see that its definition (say, for the conditional 
function {\bf if-then-else}), 
reduces to the existence of a sequential strategy of 
computation of a function by asking of an Oracle the 
values of the arguments---here of a basic type. 
For higher types, 
we just extend this idea by allowing more general queries 
to the Oracle---applicative 
combinations 
(of a basic type) of the arguments and strategies. 
This approach recalls and generalizes that 
of Kleene \cite{Kleene60,Kleene62} for Turing computability of finite 
type functionals and is essentially an extensional one, 
despite its somewhat intensional-computational features, 
and can be also considered as a natural generalisation both of combinators 
and the conditional operation {\bf if-then-else} having an evidently 
functional/extensional character. 
Moreover, this allows us to characterise, in abstract computational terms, 
the expressive power of \PCF\ both in the standard model $\setof{\bbD_\alpha}$ 
of all continuous finite type functionals \cite{Saz76AL} and 
in the fully abstract model $\setof{\bbQ_\alpha}$ considered in this paper where 
all functionals prove to be definable in \PCF\ + ``all (one place 
numeric) functions of the type $\iota\arr\iota$''. 
By the way, the ordinary 
concept of continuous functions over dcpo domains, 
usually considered as non-intensional, 
is nothing more than a very abstract version of the idea of 
computability: $fx = \bigsqcup_n fx_n$ for $x=\bigsqcup_n x_n$ with $fx$ of a basic type 
means that the value of $fx$ can be ``computed'' by extracting ``finite'' information 
$x_n$ from the argument $x$; we abstract all other details of a computation process. 
That is, it has some hidden 
intensional features. We should have just a natural balance, or interplay, 
between ``intensional'' and ``extensional''. 
For computational strategies the former aspect corresponds to the operational 
semantics of strategies, and the latter is represented by the concept of 
interpreted computations leading to denotational semantics of strategies and 
to the extensional inductive definition of the fully abstract model of 
sequentially computable functionals. 

Let us stress again, as this is an important point: 
denotational semantics of strategies, and thus 
the corresponding inductive definition of the fully abstract model, is based on 
\emph{interpreted computations} in terms of ``real'' (applications of) 
finite type functionals. Therefore it has, despite computations involved, 
rather an extensional character, whereas the operational semantics of 
(combinations of) strategies is based on purely ``syntactical'', 
\emph{non-interpreted computations} 
in terms of strategies only (like in terms of the language \PCF\ only) 
and without invoking ``real'' finite type 
functionals.

The main drawback of our approach, in comparison with 
game-theoretical ones, 
is the lack of a construction 
for a general category (not referring to finite types) 
like that of games with arrows representing suitable game strategies. However, 
this more concrete view allows 
us to construct, inductively, 
a monotonic order extensional fully abstract model for \PCF,  
in a straightforward and natural way. 
Unfortunately, this inductive definition contrasts with the 
proof of the main domain-theoretic properties of $\bbQ$ 
which involves a significant amount of machinery 
of computational strategies, 
including an isomorphic quotient construction 
$\setof{\tQ_{\alpha}}\cong\setof{\bbQ_{\alpha}}$ 
(reflected by the tilde symbol). 
In comparison, the game theoretic approach is based on a quotient 
construction in the very definition of the fully abstract model. 
In this respect, it looks 
more intensional.

It turns out that this model consists only of continuous 
functionals with respect to existing ``pointwise'', or 
``natural'' lubs. We need to consider this generalized and novel version 
of continuity, called \emph{natural continuity}, because the poset of 
sequential functionals of a given type (starting with the level 3) is not 
$\omega$-complete, 
as was recently shown by Normann \cite{Normann2004}, 
and therefore this model is not isomorphic 
to the `limit-term' model in \cite{Milner77}. 
Note that the model $\setof{\bbQ_\alpha}$ 
satisfies the corresponding uniqueness 
property 
(the property formally different from, but similar to, that of 
the continuous fully abstract model of Milner) 
and is therefore isomorphic to the game models 
defined in~\cite{Abramsky-Jagadeesan2000,Hyland-Ong2000}. 
This leads to a generalized concept of \emph{natural non-dcpo domains} 
most appropriate for describing the properties of the models 
of finite type functionals considered in this paper 
which will be shown to be  
sequentially complete 
(a weakened form of $\omega$-completeness), 
naturally continuous 
and also 
naturally $\omega$-algebraic and 
naturally bounded complete. This domain theoretic framework 
plays a crucial role in this paper and can serve as a kind of substitute for 
the categories of games mentioned above.

The more general concept of \emph{wittingly consistent} 
non-deterministic computational 
stra\-tegies defined in 
\cite{%
Saz76t} (Part II, \S 4) is also successfully used in the current 
paper to 
construct the fully abstract model $\setof{\bbW_\alpha}\cong\setof{\tW_\alpha}$ 
for $\PCF^+$ satisfying  
definability properties such as  
the fully abstract model $\setof{\bbQ_\alpha}\cong\setof{\tQ_\alpha}$ 
for \PCF\ discussed above. 
This gives a positive answer to the question 
stated in \cite{Longley-Plotkin} 
(before Proposition~6): 
\begin{quote}
\emph{``It is worth remarking that 
there is no corresponding definability result for $\PCF^+$. 
It may well be that there can be none; it is not at all clear, however, 
how to even formulate a precise statement to that effect''.}
\end{quote}
Although this question was seemingly related to the possibility 
of extending the game semantics results for \PCF\ to $\PCF^+$, 
our approach via computational 
strategies is a natural and quite general alternative 
with some analogy to the game approach and might probably 
lead also to a corresponding extended game semantics solution. 
Note also that the fully abstract model $\setof{\bbW_\alpha}$ 
for $\PCF^+$ is also 
\emph{not} $\omega$-complete (even at the level 2)---this is clear from the known 
result that $\exists$ is not definable in $\PCF^+$. 
But it is wittingly-$\omega$-complete 
and satisfies all the above mentioned generalized, ``natural'' 
versions of (non-dcpo) domain theoretic properties.

\paragraph*{\bf Organization.} 
We start with the generalized, ``natural'' version of non-dcpo 
(finite type) domain theory 
in Section~\ref{sec:prelim}. 
We define computational sequential strategies 
in Section~\ref{sec-seq-stra} and their denotational semantics 
on the base of interpreted computations in Section~\ref{sec-seq-stra-den-sem}. 
Then hereditarily 
sequential functionals are defined inductively (level-by-level) 
in Section~\ref{sec-seq-func}.  
Sections~\ref{sec-quotient} and \ref{sec:full-abs} are devoted to 
demonstrating 
the full abstraction property of the resulting model $\bbQ\cong\tQ$ for \PCF. 
The definability of a universal functional 
$U_{\alpha}:(\iota\arr\iota)\arr\alpha$
for each type $\alpha$ is also stated, but not proved (see the details in \cite{Saz76AL}).
Finitary ranked and other finite versions of 
strategies are introduced computing exactly all 
``naturally'' finite sequential functionals to demonstrate the ``natural'' continuity 
of $\tQ$ (implying other ``natural'' domain theoretic 
properties of $\tQ$) 
which is actually used in the proof of the full abstraction property of this model. 
The class of finitary strategies is also shown to be effectively closed under 
application (on the base of a kind of normalizability property).
Section~\ref{sec:full-abstr-PCF+} is devoted 
to a sketchy definition (by a very close analogy to the case of $\bbQ\cong\tQ$ and \PCF) 
of a fully abstract model $\bbW\cong\tW$ for $\PCF^+$ based on the concept of 
wittingly consistent non-deterministic strategies. 
Unlike the case of \PCF, some details are given 
(but still with a reference to the old approach for \PCF\ \cite{Saz76AL}) 
of a construction in $\PCF^+$ of a universal (surjective) 
functional $U_{\alpha}^{+}:(\iota\arr\iota)\arr\alpha$ 
for each type $\bbW_\alpha$. 
It is also demonstrated in 
Section~\ref{seq:non-complete} 
that the model $\bbW$ is not $\omega$-complete at level~2. 
Section~\ref{sec:conclusion} contains some concluding remarks 
and directions for further research. 
Finally, Appendix~\ref{appendix:univ-sys-strategies} 
presents an explicit construction of the typed version of a universal system 
of sequential strategies $\tuple{Q,\QQ}$ from \cite{Saz76t} which is used 
in previous sections for constructing~$\tQ$.

\section{Domains and Types---a Generalisation}
\label{sec:prelim}

\subsection{Basic Definitions}\label{sec:basic-def}

\noindent
Let us recall and generalize several well-known notions from domain theory 
(see, for example, \cite{Abramsky-Jung1994,Plo81}), 
emphasizing some more subtle points related with 
their usage in this paper.  
Importantly,  
some of the known terms here have a meaning different from the traditional one. 
The goal is to find a version of domain theory most appropriate for the case 
of sequential (and other kinds of) functionals.

The term {\em poset} means a set $D$ partially ordered 
by an {\em approximation} relation $\sqle_D$. 
Any poset $D$ with the least ({\em bottom}, or {\em undefined}) 
element $\Undef$ will be called a \emph{domain}. 
If $A$ is any set, then $A_{\Undef}\bYdef A\cup\{\Undef\}$ is the
corresponding {\em flat} domain where $x\sqle y\Arra (x=\Undef) \vee (x=y)$. 
A nonempty set $X\subseteq D$ is called \emph{directed} 
if, for any $x,y\in X$, we have $x\sqle z$ and $y\sqle z$ 
for some $z\in X$. 
The least upper bound (lub) of a set $X$ is denoted by $\Lub X$. 
If all directed sets have a lub in $D$ 
then it is called a {\em directed complete} 
poset, or briefly, \emph{dcpo}. 
However, the domains we will consider are typically not assumed to be dcpos. 
An element $a$ of a domain (not necessarily a dcpo) is called 
\emph{finite} (or \emph{compact}) 
if $a\sqle\Lub X$ implies $\exists x\in X.a\sqle x$ for any
directed set $X$ for which $\Lub X$ exists. 
All elements of a flat domain are evidently finite. 
A 
domain $D$ in which 
there are only countably many finite elements 
and each element $x\in D$ is a directed lub of all its finite approximations 
is called 
$\omega$-\emph{algebraic}. 
A monotonic mapping $f$ between domains is called 
\emph{continuous} 
if it preserves 
existing lubs $\Lub X$ of directed sets: 
$f(\bigsqcup X)=\bigsqcup f(X)$ 
(that is, if $\bigsqcup X$ exists then $\bigsqcup f(X)$ 
is required to exist and satisfy this equality). 
Let $(D\arr E)$ or $D\stackrel{\rm mon}{\arrr} E$ 
denote the set of all monotonic mappings ordered pointwise: 
$
f\sqsubseteq g\iff\forall x\in D(fx\sqsubseteq gx).
$
For dcpos, let $[D\arr E]$ denote the set of continuous mappings also 
ordered pointwise. 
(We can suitably extend this denotation also for some special kinds 
of non-dcpo domains, called natural domains, by taking $[D\arr E]$ to be 
the set of all naturally continuous mappings; see Section~\ref{sec:natural}.) 
If any two upper bounded elements $c,d$ have least upper bound $c\sqcup d$ 
in $D$ then $D$ is called \emph{bounded complete}. 
A domain is called \emph{finitely bounded complete} if, in the above, 
only finite $c,d$, and therefore $c\sqcup d$, are considered. 
If $D$ is an algebraic dcpo then 
it is bounded complete if, and only if, it is finitely bounded complete. 
In fact, for dcpos bounded completeness is equivalent to existence of a lub for 
any bounded set, not necessarily finite. 
Algebraic and bounded complete dcpos are also known as \emph{Scott domains} 
or as \emph{complete f$_0$-spaces} of Ershov~\cite{Ershov72}. 

The above definitions are well-known and quite natural in the context of dcpos. 
We extended them to non-dcpos rather 
as a formal intermediate step 
before introducing in Section~\ref{sec:natural} so called ``natural'' versions 
of these notions. The general idea is that nonexistence of lubs of some 
directed sets is an indication that even existing lubs might be non-natural 
(existing ``by a wrong reason''), and 
therefore the definitions of continuity, finite elements, etc.\ should 
be relativized to ``natural'' lubs only.

{\em Types} (or functional types) are defined as formal expressions
built inductively from some {\em basic\/} types, in our case $\iota$
and $o$ 
(with the generic name \mbox{Basic-type}), 
by the arrow construct: if $\alpha$ and $\beta$ are types
then $(\alpha\arr\beta)$ is a type. We usually write
$\alpha_1\arr\alpha_2\arr\cdots\arr\alpha_n\arr\beta$ or 
$\alpha_1,\alpha_2,\ldots,\alpha_n\arr\beta$ instead of
$(\alpha_1\arr(\alpha_2\arr(\cdots\arr(\alpha_n\arr\beta)\cdots)))$.
The \emph{level} of any type
$\alpha = (\alpha_1,\ldots,\alpha_n \arr \mbox{Basic-type})$
is defined as
\[
\mbox{level}(\alpha) = \mbox{max}\{1+\mbox{level}(\alpha_i)\mid 1\le i\le n\}
\]
and, in particular, 
$
\mbox{level}(\mbox{Basic-type}) = 0.
$
The \emph{arity} (or the number of arguments) 
of $\alpha$ is the number $n$ above. 

For any type $\alpha$ we define inductively, as usual, 
the corresponding (\emph{standard}) domain $\bbD_{\alpha}$ of
all {continuous} functionals of type $\alpha$ 
with $\bbD_o=\BB_{\Undef}$, $\bbD_\iota=\NN_{\Undef}$, 
and $\bbD_{\alpha\arr\beta}=[\bbD_{\alpha}\arr \bbD_{\beta}]$,
where $\BB=\setof{\true,\false}$ and 
$\NN=\setof{0,1,2,\ldots}$. 
All these $\bbD_\alpha$ are $\omega$-algebraic, bounded complete dcpos. 
More general, 
\begin{defi}%
A \emph{(typed monotonic order extensional 
applicative) structure} $\setof{E_{\alpha}}$ is a system of domains 
(with the least element $\Undef_{\alpha}$ in each) such that 
for any types $\alpha$ and $\beta$ there is a monotonic 
mapping 
$\App_{\alpha\beta}:E_{\alpha\arr\beta}\times E_{\alpha}\arr 
E_{\beta}$ (with 
$\App(f,x)$ abbreviated as $fx$ and 
$((\cdots(fx_1)x_2)\cdots x_n)$ abbreviated as $fx_1\cdots x_n$) 
satisfying 
\begin{itemize}
\item[(i)] $\Undef_{\alpha\arr\beta}\,x=\Undef_{\beta}$ for all 
$x\in E_\alpha$, and 
\item[(ii)]the extensionality condition: for all $\alpha,\beta$ and 
$f,f'\in E_{\alpha\arr\beta}$,
\[
f\sqle f'\iff \forall x\in E_{\alpha}(fx\sqle f'x).
\]
\end{itemize}
Elements of $E_{\alpha}$ 
are called \emph{functionals} of type $\alpha$.
An extensional structure $\setof{E_{\alpha}}$ is called a $\lambda$-\emph{model} 
if it is sufficiently rich to contain all 
$\lambda$-definable functionals. 
\end{defi}
For the closure under $\lambda$-definability we can equivalently require that 
$\setof{E_{\alpha}}$ contains combinators 
$\Ss_{\alpha\beta\gamma}:
(\alpha\arr(\beta\arr\gamma))\arr((\alpha\arr\beta)\arr(\alpha\arr\gamma))
$ and 
$\K_{\alpha\beta}:\alpha\arr(\beta\arr\alpha)$
for all types $\alpha,\beta,\gamma$ satisfying identities 
$\Ss xyz=xz(yz)$ and $\K uv=u$ for all $x,y,z,u,v$ 
with $\Ss,\K,x,y,z,u,v$ of appropriate types, omitted for brevity. 
We will also always assume that 
$E_{\iota}=\bbD_{\iota}=\NN_\Undef$ and $E_{o}=\bbD_{o}=\BB_\Undef$. 
To simplify the exposition, let us take that 
$\BB_\Undef\subseteq \NN_\Undef$ 
with $\Undef_o=\Undef_\iota$, $\true=1$ and $\false=0$ 
and, hence, avoid using the Boolean type $o$ at all in the ``official'' exposition. 
(However, we will use $o$ in some examples for the convenience.)
Then $\mbox{Basic-type}$ will mean just~$\iota$. 
Although in general the sets $E_{\sigma\arr\tau}$ and $(E_\sigma\arr E_\tau)$
may even not intersect, 
there is the natural embedding $E_{\sigma\arr\tau}\embed (E_\sigma\arr E_\tau)$ 
induced by the application operation. 
Moreover, without restricting generality we may also consider that the set
\begin{equation}\label{eq:EE}
E_{\alpha} = E_{(\alpha_1,\ldots,\alpha_n \arr \Basictype)}
\subseteq
(E_{\alpha_1}\times\cdots\times E_{\alpha_n}\arr E_{\Basictype}) 
\end{equation}
consists of some monotonic mappings of the type shown, ordered pointwise, 
\[
f\sqsubseteq f'\iff\forall\bx(f\bx\sqsubseteq_\iota f'\bx), 
\]
and for all $f\in E_{\alpha}$ and $x_1\in E_{\alpha_1}$ 
\begin{equation}\label{eq:App}
fx_1=
\lambda x_2,\ldots,x_n.f(x_1,x_2,\ldots,x_n) \in
E_{\alpha_2}\times\cdots\times E_{\alpha_n}\arr E_{\Basictype} 
\end{equation}
is the ``residual'' map. 
Indeed, any $\setof{E_{\alpha}}$ satisfying (\ref{eq:EE}) 
and (\ref{eq:App}) and containing constant undefined functions 
$\Undef_{\bar{\alpha}\arr\iota}=\lambda \bx^{\bar{\alpha}}.\Undef_{\iota}$ 
is a monotonic, order extensional applicative structure.  
It is clear that such an $\setof{E_{\alpha}}$ is a restricted 
class of monotonic finite-type functionals. 
\begin{defi}\label{def:continuous-structure}%
A structure $\setof{E_\alpha}$ (with $E_\alpha$ not necessarily a dcpo) 
is called \emph{continuous} 
if for each type $\alpha=(\alpha_1,\alpha_2,\ldots,\alpha_k\arr\iota)$ 
and variables $f:\alpha$ and $\bx:\bar{\alpha}$, the \emph{full application map} 
$
\lambda f\bx.f\bx:
E_{\alpha}\times E_{\alpha_1}\times E_{\alpha_2}\times\cdots\times E_{\alpha_k}\arr E_{\iota}
$
is \emph{{continuous}}. 
Equivalently, we can require the continuity 
of the application maps of two arguments 
$
\lambda fx_1.fx_1:
E_{\alpha}\times E_{\alpha_1}\arr E_{\alpha_2,\ldots,\alpha_k\arr\iota}
$. 
\end{defi}
\label{page:continuity}

\subsection{Natural Non-dcpo Domains}\label{sec:natural}

\noindent
More generally, 
\begin{defi}%
In any monotonic, order extensional applicative structure 
a \emph{pointwise lub} 
$\pointwiselub_i f_i$ of an arbitrary 
(not necessarily directed) family of functionals (of the same type) is the ordinary 
lub $\bigsqcup_i f_i$, in the case of the basic type, and, for higher types, it is  
the ordinary lub
which is also required to satisfy, inductively, the pointwise identity 
$(\pointwiselub_i f_i)x=\pointwiselub_i (f_i x)$ 
(with $\pointwiselub_i (f_i x)$ also pointwise)
for all $x$ of appropriate type.
\end{defi}

\noindent
Thus, $f=\pointwiselub_i f_i$ implies $f=\bigsqcup_i f_i$, but, in general, 
not vice versa. 
That is, $\pointwiselub$ is a restricted version of~$\bigsqcup$. 
(See an example below.) 
Equivalently, we may require from $\bigsqcup_i f_i$ the identity 
$(\bigsqcup_i f_i)\bx=\bigsqcup_i (f_i \bx)$ 
in the basic type.%
\label{page:pointwise-lub}
In fact, 
\begin{equation}\label{eq:pointwise-lub}
f=\pointwiselub_i %
f_i\textrm{ iff } f\bx=\bigsqcup_i (f_i\bx)
\textrm{ for all }\bx,
\end{equation}
assuming $f\bx$ is of the basic type. 
The concept of pointwise lub is quite natural and could also be called 
just \emph{union}, or \emph{natural lub}. 
This is even the ordinary set theoretic union 
if to identify monotonic functionals of the type 
$\alpha=(\alpha_1,\alpha_2,\ldots,\alpha_k\arr\iota)$ 
with corresponding graph subsets of 
$E_{\alpha_1}\times E_{\alpha_2}\times\cdots\times E_{\alpha_k}\times\NN$. 
In this case $\sqsubseteq$ also coincides with the set theoretic notion of inclusion 
$\subseteq$. 
Respectively, non-pointwise lubs are considered as non-natural in this sense. 
(However note that neither finite nor also ``naturally'' finite functionals 
considered below are necessarily represented as finite graph sets in the above sense.)

\begin{exa}\label{ex:O_i}%
To illustrate the above definition, 
consider a simple example in $\setof{\bbQ_\alpha}$ (the monotonic, 
order extensional $\lambda$-model of sequential functionals 
to be defined later) of a finite non-natural lub of two 
elements. 
Define two first order sequential functions 
$O_i(x_1,x_2)$, $i=1,2$, 
as $0$ if the corresponding $x_i=0$, and $\Undef$ otherwise. 
Then  
$O_1\sqcup O_2=\lambda x_1,x_2.0$ is the constant zero function 
in $\bbQ$, and this is not a natural  lub. 
The natural  lub, if it would exist in $\bbQ$, should satisfy 
$(O_1\pointwiselub O_2)(x_1,x_2)=0$ if $x_1=0$ or $x_2=0$, 
and ${}=\Undef$ otherwise. But this is not a sequential function, 
that is, it lies outside of $\bbQ$. 
\end{exa}

\begin{defi}%
A structure $\setof{E_{\alpha}}$ is called \emph{naturally  {continuous}} 
if for all types $\alpha=\tau\arr\sigma$ and $f\in E_\alpha$ the map
$
\lambda x.fx:
E_{\sigma}\arr E_{\tau}
$ 
preserves directed natural lubs of the arguments whenever they exist:
\label{page:pointwise-continuity}
$
f(\pointwiselub_i x_i)=\pointwiselub_i fx_i
$.
That is, if the directed natural lub to the left exists then 
the natural lub to the right exists too, and the equality holds. 
\end{defi}

\noindent
We can require, equivalently, for each type 
$\alpha=(\alpha_1,\alpha_2,\ldots,\alpha_k\arr\iota)$ and 
$f\in E_\alpha$,  
that the map 
$
\lambda \bx:\bar{\alpha}.f\bx:
E_{\alpha_1}\times E_{\alpha_2}\times\cdots\times E_{\alpha_k}\arr E_{\iota}
$
is naturally  {continuous} (preserves natural  lubs) in each argument. 
Evidently, natural  continuity of $fx$ or $f\bx$ in $f$ 
is automatically satisfied by the definition of natural lub as the pointwise one. 
Also, \emph{in a continuous structure} 
(that is, with continuous full application maps)
\emph{all existing directed lubs are natural (pointwise), and therefore 
any continuous structure is naturally continuous}. 
Further, 
\begin{defi}%
\emph{Naturally finite} functionals 
are defined like the ordinary finite ones, but by using the natural  lubs. 
\end{defi}

\noindent
Each finite functional is also naturally finite 
(but probably not vice versa; see the discussion below). 
\begin{defi}\label{def:alg-bound-compl-struct}%
\hfill  %
\begin{enumerate}[(a)]
\item
A structure $\setof{E_{\alpha}}$ 
is called \emph{naturally $\omega$-algebraic} if each of its elements 
is a directed natural lub of naturally finite elements, 
and there are only countably many naturally finite elements 
in the structure. 

\item
It is called \emph{naturally bounded complete} if any two 
upper bounded 
naturally finite elements 
have a lub (not necessarily a natural lub, but evidently also naturally 
finite element). 
\end{enumerate}
\end{defi}

\noindent
For any naturally algebraic and naturally bounded complete structure $\setof{E_\alpha}$ 
the sets of the form $\check{a}\bYdef\setof{x\in E_\alpha\mid a\sqsubseteq x}$, 
for $a$ naturally finite, constitute a base of a \mbox{(\To-)}~topology in each $E_\alpha$ 
which makes $E_\alpha$ satisfying this definition 
a (non-necessarily complete) \fo-space of Ershov \cite{Ershov72}. 
Note that open sets in this topology are exactly those \emph{naturally Scott open} 
(defined as usual, but with respect to the natural directed lubs).

By using  
Lemma~\ref{lemma:algebraicity} presented below, 
we will prove in Theorem~\ref{th:continuity} 
the natural  continuity and the last two properties (a) and (b) defined above 
for the special case of the model of sequential 
functionals $\setof{\bbQ_\alpha}$. 
That $\bbQ$ 
is not a dcpo was actually shown by Normann \cite{Normann2004}.
\begin{hyps}
It seems quite plausible
that in $\setof{\bbQ_\alpha}$ 
there exist
\begin{enumerate}[\em(1)]
\item
a directed non-natural  lub, 
\item 
a naturally finite, but not a finite functional 
(being a proper directed lub), 
\item
a non-continuous (but naturally continuous) functional, and 
\item 
a naturally finite (and naturally continuous), but not a continuous functional.
\end{enumerate}
We could also expect that 
\begin{enumerate}[\em(1)]
\setcounter{enumi}{4}
\item
a continuous (and therefore naturally continuous) lambda model exists 
whose higher type domains are not dcpos. 
\end{enumerate}
\end{hyps}

\begin{note}\hfill
\begin{enumerate}[(a)]
\item We see that these hypotheses reveal a terminological problem 
(``naturally finite, but not finite'', etc.). Properly speaking, 
these are naturally finite functionals which are most naturally 
considered as full-fledged finite objects in the framework of~$\bbQ$. 
Moreover, together with naturally continuous functionals, these concepts 
give rise to an appropriate non-dcpo generalisation 
of continuous, $\omega$-algebraic and bounded complete $\lambda$-models 
(originally considered over dcpos). 
This will be seen from the following considerations and 
Lemma~\ref{lemma:algebraicity}. 
The more traditional definitions of continuous and finite functionals in terms 
of the ordinary directed lubs prove to be not very adequate 
in the framework of non-dcpos. 

\item
Another important point is that, being based on types, the natural lub 
\fla{\pointwiselub} as well as other related ``natural'' 
concepts are not purely order-theoretic ones. 
However, 
one can give an abstract definition of \emph{natural (non-dcpo) domains} 
with a primitive partially defined operator 
$\pointwiselub:2^D\mathrel{\dot{\arr}}D$ in each domain which is 
a restricted version of $\bigsqcup$ and has appropriate postulated properties. 
Then the special case of these natural domains satisfying 
the conditions (a) and (b) of Definition~\ref{def:alg-bound-compl-struct} 
corresponds exactly to the \fo-spaces of Ershov \cite{Ershov72}. 
More detailed and general discussion on this generalized theory 
of non-dcpo domains and the mentioned correspondence requires 
a separate consideration to be presented elsewhere. 
It is also worth noticing that these domains appear in our presentation as 
natural non-dcpo domains rather than \fo-spaces. They prove to be 
\fo-spaces only a posteriori by using quite involved technical theory 
of computational strategies and applying Lemma~\ref{lemma:algebraicity} below.
\end{enumerate}
\end{note}

\subsection{Finitely Restricted Functionals} \ \

\paragraph*{\bf Conditions on $\setof{E_{\alpha}}$} 
For the rest of Section~\ref{sec:prelim} 
let $\setof{E_{\alpha}}$ be any monotonic, order extensional 
$\lambda$-model which contains the first order 
equality predicate $x=y$ 
(monotonic and strict in $x$ and~$y$) 
and the ordinary (monotonic and sequential) 
conditional function $\IF\ x\ \THEN\ y\ \ELSE\ z$ 
for the basic type (and hence for all types by $\lambda$-definablity). 

\bigskip

Recall that a monotonic function $\Psi:E\arr E$ 
is called a \emph{projection} 
if $\Psi x\sqsubseteq x$ for all $x\in E$, and  $\Psi\circ\Psi=\Psi$. 
We say also that $\Psi$ is a projection from $E$ \emph{onto} its range ${}\subseteq E$ 
which is also the set of all fixed points of $\Psi$. 
For any two projections, $\Psi\sqsubseteq\Psi'$ iff 
range$(\Psi)$ $\subseteq$ range$(\Psi')$. 
Note that 
$\Psi x$ is the largest $\sqsubseteq$-approximation to $x$ 
from the range of $\Psi$.

\bigskip

\noindent
Now, we will follow Milner \cite{Milner77}, 
slightly simplifying and generalizing to the ``natural'' non-dcpo case. 

\begin{defi}\label{def:proj-finitely-restricted}%
Define projections 
$\Psi_\alpha ^{[k]} : E_\alpha \arr E_\alpha$ for all types and any $k\ge 0$ 
by letting
\begin{align*}\label{eq:proj}
&\Psi^{[k]}_{\iota}:E_\iota\stackrel{\rm onto}{\arrr} E_\iota^{[k]},\;
E_\iota^{[k]}\bYdef\setof{\Undef,0,1,\ldots,k}, 
\\
&\Psi^{[k]}_{\sigma\arr\tau}f\bYdef\Psi^{[k]}_{\tau}\circ f\circ\Psi^{[k]}_{\sigma}, 
\\
&E_\alpha^{[k]}\bYdef\textrm{Range}(\Psi^{[k]}_{\alpha}). 
\end{align*}
Denote $x^{[k]}\bYdef\Psi^{[k]}_{\alpha}x$. 
Elements $x^{[k]}$ in 
$E_\alpha^{[\omega]}\bYdef\bigcup_{k} E_\alpha^{[k]}\subseteq E_\alpha$ 
are called \emph{finitely restricted}. 
\end{defi}
\noindent
These all are monotonic sequences on $k$. 
That $\Psi^{[k]}_{\iota}$ and hence all other $\Psi^{[k]}_{\sigma\arr\tau}$ are 
(representable by) elements of the $\lambda$-model 
(we write $\Psi_\alpha ^{[k]} \in E_{\alpha \arr \alpha}$)
follows from existence in it of both $=$ and $\IF$. 
By induction on types, each $E_\alpha^{[k]}$ is a finite set 
since $x^{[k]} y = (x y^{[k]})^{[k]}$ at all types. 
Also, the application of $k$-restricted functionals to any argument is $k$-restricted. 
In particular, each finitely restricted functional  
$\varphi$ 
has a tabular representation
\begin{equation}
\label{eq:fin-main}
\varphi x =
		 \underbrace{\left[\,^{b_0,\ldots,b_{n-1}}
		 _{a_0,\ldots,a_{n-1}}\right]}_{\varphi}
		x
		=\bigsqcup_{a_i \sqsubseteq x} b_i
		\textrm{ where }\varphi a_i=b_i \textrm{ and } 
		a_i,b_i\textrm{ are finitely restricted}.  		
\end{equation}
In each model $\setof{E_\alpha}$ (over $E_\iota=\NN_\Undef$) 
satisfying the above conditions 
there are only countably many finitely restricted elements. 
This is another approach to the finiteness 
of higher type functionals.
Without assuming any further conditions on $\setof{E_\alpha}$, 
each $\Psi^{[k]}_\alpha$ considered as a map 
\mbox{$\Psi^{[k]}_\alpha:E_\alpha\arr E_\alpha$} 
\emph{is naturally continuous and, moreover, 
preserves all existing natural  lubs} (not necessarily directed). 
This follows by induction on the types: 
\begin{align*}
&(\Psi^{[k]}_\tau{}\circ(\pointwiselub_i f_i)\circ{\Psi^{[k]}_\sigma})x
=\Psi^{[k]}_\tau((\pointwiselub_i f_i)(\Psi^{[k]}_\sigma x))
=\Psi^{[k]}_\tau(\pointwiselub_i (f_i(\Psi^{[k]}_\sigma x)))=
\\
&\pointwiselub_i\Psi^{[k]}_\tau(f_i(\Psi^{[k]}_\sigma x))
=\pointwiselub_i((\Psi^{[k]}_\tau\circ f_i\circ\Psi^{[k]}_\sigma) x)
=(\pointwiselub_i(\Psi^{[k]}_\tau\circ f_i\circ\Psi^{[k]}_\sigma)) x.  
\end{align*}
It also follows that \emph{each finitely restricted element 
$x^{[k]}$ is naturally finite}: 
$x^{[k]}\sqsubseteq\pointwiselub Z$ for a directed set $Z$ 
implies $x^{[k]}\sqsubseteq\pointwiselub\setof{z^{[k]}\mid z\in Z}=z^{[k]}\sqsubseteq z$ 
for some $z$ by natural  continuity of $\Psi^{[k]}$ 
and because $E^{[k]}_\alpha$ is finite. 

Moreover, \emph{if the model is naturally continuous then
$x=\pointwiselub_k x^{[k]}$ holds for all $x$}. 
Indeed, assuming by induction on types that $\bY=\pointwiselub_k\bY^{[k]}$, 
we have
$x\bY=\bigsqcup_k (x\bY^{[k]})=\bigsqcup_k (x\bY^{[k]})^{[k]}=\bigsqcup_k (x^{[k]}\bY)$. 
Thus $\bx=\pointwiselub_k\bx^{[k]}$ by (\ref{eq:pointwise-lub}), as required. 

Finally we note that, without any further assumptions on the model, 
\emph{any two upper bounded finitely restricted elements $d,e$ 
have a (not necessarily natural) lub $d\sqcup e$ which is also finitely restricted}. 
Indeed, it can be obtained as the greatest lower bound 
$\sqcap\setof{x^{[k]}\mid x\sqsupseteq d,e}$ 
for any fixed $k$ such that $d,e\in E_\alpha^{[k]}$ 
because the glb of any finite nonempty set is definable from $\IF$ and $=$.

The following Lemma is a generalisation of the Algebraicity Lemma of Milner 
in \cite{Milner77} to the case of non-dcpos and to the ``natural'' case, 
but formulated for simplicity only for the models with the numerical basic values 
$E_\iota=\NN_\Undef$. 
It clearly demonstrates that the generalisations introduced are 
quite adequate and natural.

\begin{lem}\label{lemma:algebraicity} 
Let $\setof{E_{\alpha}}$ be any monotonic, order extensional 
$\lambda$-model, with $E_\iota=\NN_\Undef$, 
which contains first order equality and the conditional. Then
\begin{enumerate}[\em(a)]
\item
this model is naturally continuous 
if, and only if, 
   \begin{itemize}
   \item[(*)] for any type 
$\alpha=\alpha_1,\ldots,\alpha_n\arr\iota$ 
and elements 
$f\in E_\alpha$ and $\bx\in E_{\bar{\alpha}}$, 
$f\bx=f\bar{d}$ holds 
for some finitely restricted 
$\bar{d}\sqsubseteq\bx$; 
   \end{itemize}

\item
if the model is naturally continuous 
then 
(i) the naturally finite elements of each $E_\alpha$ 
are exactly the finitely restricted ones, 
(ii) $\setof{E_{\alpha}}$ 
is naturally $\omega$-algebraic, and 
(iii) it is 
naturally bounded complete;  

\item
repeats {\em(b)}, but with ``naturally'' omitted. 
\end{enumerate}
\end{lem}
\proof\hfill
\begin{enumerate}[(a)]
\setcounter{enumi}{1}
\item
\label{page:algebraicity}
follows easily from the above considerations on projections~$\Psi^{[k]}$. 

\item
It suffices to recall that continuous structures 
are also naturally continuous, and 
the concepts of directed lubs, and hence of finite functionals 
in these models, are equivalent to their 
``natural'' versions. 
Note that we do not assume here that the $E_{\alpha}$ are dcpos. 

\setcounter{enumi}{0}
\item
 ``If'' follows from natural  finiteness of all $x^{[k]}$. 
``Only if'' follows from~(b).\qed
\end{enumerate}

\noindent
The clause (a) of this Lemma (not considered in \cite{Milner77}) 
is used in Section~\ref{sec:ranked-finitary} below to show that 
the model of sequential functionals $\setof{\bbQ_\alpha}$ 
is naturally continuous and satisfies the conditions (i)--(iii) from (b). 
In the application of this Lemma to $\setof{\bbQ_\alpha}$ the crucial point 
is that (*) in (a) implies all the essential domain theoretic properties 
holding for this model.

Moreover, we will also show in Theorem~\ref{th:least-correct-sem} (b) 
that the model $\setof{\bbQ_\alpha}$ 
is also \emph{sequentially complete} in the sense that it is closed 
under taking natural (pointwise) lubs of a special class of increasing 
sequences (determined by sequential strategies). 
For example, in $\setof{\bbQ_\alpha}$ we have the natural lub 
$\pointwiselub_n f^n \Undef$ giving the least fixed point of $f:\alpha\arr\alpha$ 
for all types $\alpha$. 

\subsection{On Efficiency of Naturally Finite Functionals}
\label{sec:efficiency}

For the case of the standard continuous model $\setof{\bbD_\alpha}$,
the tabular representation (\ref{eq:fin-main}) of naturally finite 
(finitely restricted) functionals proves to be quite effective and 
gives rise to an effective numbering 
of these functionals \cite{Ershov72}. The main reason for that is that 
(by induction on types)
any monotonic table as in (\ref{eq:fin-main}) represents a finitely restricted 
functional in this model. This also holds 
for $\setof{\bbW_\alpha}$ (the non-dcpo fully abstract model for $\PCF^+$) 
where naturally finite functionals are the same as in $\setof{\bbD_\alpha}$. 
The latter essentially follows from 
their definability in $\PCF^+$~\cite{Plotkin77}. In fact, the predicates 
``$\varphi\sqsubseteq\psi$'', ``$\varphi,\psi$ are upper bounded (consistent)'' 
and the application operation ``$\varphi a$'', for naturally finite $\varphi,\psi,a$, 
are effectively computable in the cases of $\setof{\bbD_\alpha}$ and $\setof{\bbW_\alpha}$.

Unfortunately, in the model of hereditarily-sequential functionals 
$\setof{\bbQ_\alpha}$ no such effective numbering is possible as can be shown 
by appropriate adaptation of the undecidability result of Loader \cite{LoaderTCS2001}. 
In fact, we cannot generally, and effectively, decide which monotonic 
tables (\ref{eq:fin-main}) represent sequential functionals in $\bbQ_\alpha$, 
let even for finitely many of $k$-restricted ones. 
But we can enumerate them 
by means of the \emph{finitary strategies} introduced in Section~\ref{sec:ranked-finitary} 
instead of using non-effective (in this case) tabular representation. 
In this sense the set of $k$-restricted 
functionals of a fixed type $\alpha$ is finite and recursively enumerable but, 
in general, ``undecidable''. 
However, it will be demonstrated in Theorem~\ref{th:applications-of-finitary} that, 
under the above mentioned ``finitary'' representation of naturally finite functionals 
$\varphi,\psi, a$, 
the application ``$\varphi a$'' is computable, and it easily follows that 
``$\varphi\not\sqsubseteq\psi$'', 
unlike ``$\varphi\sqsubseteq\psi$'', is semidecidable 
in $\setof{\bbQ_\alpha}$ 
(and similarly for $\setof{\bbW_\alpha}$ in addition to the above tabular 
effective in this case and decidable representation).

This seemingly diminishes the role of naturally finite 
(= finitely restricted) functionals and their use 
(like in $\setof{\bbD_\alpha}$)
to define effective functionals as those 
which are (natural) lubs of a  recursively enumerable directed set of 
(naturally) finite approximations. 
Such a definition seems not very appropriate, 
not only for $\setof{\bbQ_\alpha}$, but even for the case of $\setof{\bbW_\alpha}$. 
At least, further research is required.
For efficiency of functionals we should, in these cases, rather use the concept 
of an effective (sequential and, respectively, wittingly consistent) computational strategy 
\eg\ as in Definition~\ref{def:seq-func}. 

Finally, let us mention one more related question on $\bbQ$ and $\bbW$: 
for naturally finite $\varphi$ and any $x$ the application $\varphi x$ is 
evidently naturally finite, but 
\emph{is its finitary representation computable from that of $\varphi$ and a strategy 
representing $x$ in general?} (However, for $\varphi x:\iota$ it is computable.)

\subsection{Ideal Completion and Uniqueness of Fully Abstract Models}
\label{sec:completion}

\noindent
Although our goal is the fully abstract non-dcpo (in fact, naturally continuous) 
models for \PCF\ and $\PCF^+$, it make sense 
to relate them with the continuous dcpo model construction of Milner \cite{Milner77} 
via the ideal completion procedure. 

\medskip

Now, let $\EE=\setof{E_\alpha}$ be any naturally continuous $\lambda$-model 
satisfying the assumption and the conclusions (i)--(iii) 
of Lemma~\ref{lemma:algebraicity}. 
Consider its \emph{ideal completion} $\dEE=\setof{\dE_\alpha}$ which is 
a continuous dcpo model defined as follows. 
A~nonempty directed set $\dx\subseteq E_\alpha$ 
of naturally finite elements is called an \emph{ideal} if 
$a\sqsubseteq b\in \dx\Arr a\in \dx$ 
for $a,b$ naturally finite. 
Let $\dE_\alpha$, 
be the set of all ideals in $E_\alpha$. 
This is evidently a dcpo ordered by set inclusion $\subseteq$ with 
$\setof{\Undef}$ the least ideal and with directed 
lubs coinciding with set unions $\bigcup_i \dx_i$. 
Let 
$\Id{X}=\setof{a\mid
\exists x\in X(a\sqsubseteq x\AND a\textrm{ naturally finite})}$ be the 
ideal generated by a directed set $X$, and $\Id{x}=\Id{\setof{x}}$. 
As $\Id{x}\subseteq \Id{y}\!\!\!\iff\!\!\! x\sqsubseteq y$, 
we have an order isomorphic embedding of posets $I:E_\alpha\embed\dE_\alpha$ which is 
onto for the basic type. Note that always 
$\dE_{\iota\arr\iota}\cong[\NN_\Undef\arr\NN_\Undef]$. 
If $a$ is naturally finite in $E_\alpha$ then $\Id{a}$ is finite
element in the dcpo~$\dE_\alpha$. 
For any $\dx\in\dE_\alpha$,
$\Id{a}\subseteq\dx\!\!\!\iff\!\!\! a\in\dx$. In fact, $\dx$ is a directed union of such $\Id{a}$, 
and $\dE_\alpha$ is an $\omega$-algebraic dcpo domain with finite elements 
$\Id{a}$ for $a$ naturally finite. It is also bounded complete because 
$E_\alpha$ is naturally bounded complete. 
Further, we may define the application operation in $\dEE$ by 
$\df\dx=\Id{\setof{\varphi a\mid \varphi\in\df, a\in\dx}}$ 
for any $\df$ and $\dx$ of appropriate types, 
which makes it a monotonic order extensional structure. 
For the latter use the fact that 
\[
\forall a\textrm{ naturally finite }
\exists\psi\in\dg(\varphi a\sqsubseteq\psi a)\Arrr
\exists\psi\in\dg(\varphi\sqsubseteq\psi). 
\]
It is easy to show that $\dEE$ is continuous, that is having the continuous 
application operation  
($(\bigcup_i\df_i )(\bigcup_j\dx_j )=\bigcup_{ij}\df_i\dx_j$ holds 
for directed families). 
The application also agrees with the embedding $I:\EE\embed\dEE$: 
\[
\Id{fx}=(\Id{f}) (\Id{x}). 
\]
Moreover, $\dEE$ is a $\lambda$-model because for the combinators 
$\Ss,\K\in\EE$ we have 
\[
\Id{\Ss} \dx\dy\dz=\dx\dz(\dy\dz),\texttt{ and }\Id{\K} \dx\dy=\dx
\] 
in~$\dEE$ (where all directed lubs are natural/pointwise and 
``naturally finite'' = ``finite''). 
Assuming additionally the existence of 
a fixed point combinator in $\EE$ satisfying the \emph{$\Y$-property}
\begin{equation}\label{eq:Y-property}
\Y f=f(\Y f)=\pointwiselub_n f^n(\Undef) 
\end{equation}
for all $f$ of appropriate type, its 
image in $\dEE$ behaves accordingly: 
\[
\Id{\Y}\df=\bigcup_n \df^n(\setof{\Undef}). 
\]

The languages $\PCF$ and $\PCF^+$ \cite{Plotkin77,Longley-Plotkin} 
considered in this paper are based on $\Ss,\K,\Y,0$, 
plus constants for some level one functions 
(successor, predecessor, first order equality and one of two versions 
of the conditional---sequential and parallel, respectively).
For any $\EE$ satisfying the $\Y$-property the meaning of all these constants 
is also not changed by the embedding $I:\EE\embed\dEE$. 
Hence, 
\begin{prop}\label{prop:meanings}
The meaning of\/ $\PCF^{(+)}$ terms in $\dEE$ 
agrees with that in~$\EE$.
\qed 
\end{prop}
\begin{defi}\label{def:full-abstract}%
Let $\CC[\;]$ denote an arbitrary ground type program context in $\PCF^{(+)}$. 
A~model $\EE$ satisfying the $\Y$-property (\ref{eq:Y-property}) 
is called \emph{fully abstract} relative to $\PCF^{(+)}$ if  
\[
\forall
\CC\;(\Dsem{\CC[M]}\sqle\Dsem{\CC[N]})\Arrr\Dsem{M}\sqle\Dsem{N}. 
\]
\end{defi}
\noindent
Evidently, $\EE$ is fully abstract iff $\dEE$ is such 
(relative to $\PCF^{(+)}$ or, equivalently, relative to $\PCF^{(+)}$ minus $\Y$; 
use the Fixed-point Lemma in \cite{Milner77} for the dcpo case of $\dEE$). 

\begin{prop}
Let $\EE$ be any fully abstract and 
naturally continuous $\lambda$-model of\/ $\PCF^{(+)}$ 
satisfying the $\Y$-property (\ref{eq:Y-property}). 
Then the model $\dEE$ is also fully abstract and 
all finite elements in $\dEE$, and therefore all naturally finite elements in $\EE$, 
are definable in\/ $\PCF^{(+)}$ without using\/ $\Y$. 
The same holds for $\EE$ fully abstract relative to 
the language $\PCF^{(+)}$ minus\/ $\Y$ 
(although still satisfying $\Y$-property). 
\end{prop}
\proof  
The definability statement for the case of fully abstract continuous dcpo models 
(here $\dEE$) was actually shown in the proof of Theorem~3 in  \cite{Milner77}. 
This implies the case of naturally continuous model $\EE$ by using 
Proposition~\ref{prop:meanings}.\qed

\noindent
It follows as in \cite{Milner77}, by taking $\CC[\ \ ]=[\ \ ]C_1\cdots C_n:\iota$ 
with $C_i$ defining finite elements, that on definable elements, 
and therefore on all finite elements such fully abstract $\dEE$, if exists at all, 
is determined uniquely, up to isomorphism. A~general construction of such a 
continuous \emph{dcpo} model from some given level one functions is presented 
in \cite{Milner77}. 

Alternatively and extending to the case of non-dcpos, we will define 
two models $\bbQ$ and $\bbW$ 
for \PCF\ and $\PCF^{+}$, respectively, 
such that it will follow from 
Theorems~\ref{th:least-correct-sem}~(b), 
\ref{th:full-abstr}, \ref{th:universal} and \ref{th:continuity} below 
(on a generalization of the $\Y$-property, 
full abstraction property, universality and natural continuity of $\bbQ$, 
and corresponding versions for~$\bbW$) that 

\begin{thm}\label{th:UNIQUE}\hfill
\begin{enumerate}[\em(a)]
\item
$\dQ$ and $\dW$, 
are the only possible  
fully abstract 
continuous dcpo models for \PCF\ and $\PCF^{+}$, respectively 
(with $\dQ$ also isomorphic to Milner's model in \cite{Milner77} 
and $\dW$ isomorphic to $\bbD$). 

\item
Therefore also $\bbQ$ and $\bbW$ are the only possible 
fully abstract naturally continuous%
\footnote{Note that the natural continuity requirement on $\EE$ here 
can be omitted and the proof of (b) can be done straightforwardly 
by showing first that (i) the denotational semantics of $\PCF^{(+)}$ terms 
(possibly involving arbitrary type $\iota\arr\iota$ functions)
of the type $\iota$ corresponds exactly to the natural operational semantics, 
and (ii) Milner's Context Lemma \cite{Milner77} for the operational 
semantics holds. 
To this end, define a logical relation $a\mathrel{R}A$
between values $a$ in $\EE$ and closed $\PCF^{(+)}$ terms $A$ 
by letting, for the type $\iota$, 
$a\mathrel{R_\iota}A\bYdef a\sqsubseteq_\iota{}$ the value to which $A$ 
operationally reduces, and show $\Dsem{A}\mathrel{R_\alpha}A$ for closed terms. 
(Thanks to Achim Jung who has drawn attention of the author 
to this proof of (i) and (ii).) 
The point is that only the $\Y$-property is used 
in the proof, and neither dcpo nor continuity properties 
of models considered are needed. We omit the details. 
Then the full abstraction property can be formulated in terms of operational 
semantics and thus leads to an 
operational characterisation of the relation $\Dsem{A}\sqsubseteq\Dsem{B}$ 
for $\PCF^{(+)}$ terms. 
}
models for \PCF\ and $\PCF^{+}$, respectively, satisfying the $\Y$-property and
in which all elements are definable 
from arbitrary type $\iota\arr\iota$ functions of the model 
where 
$\bbQ_{\iota\arr\iota}=\bbW_{\iota\arr\iota}=\bbD_{\iota\arr\iota}=
[\NN_\Undef\arr\NN_\Undef]$---all monotonic functions.\qed
\end{enumerate}

\end{thm}

\noindent
More general, in the latter uniqueness formulation 
we could consider for $\bbQ_{\iota\arr\iota}$ and $\bbW_{\iota\arr\iota}$ 
some other classes of type $\iota\arr\iota$ 
monotonic functions, say, all computable---as the minimal such a class. 
In the computable case only definability in pure $\PCF^{(+)}$ may be used, 
without reference to type $\iota\arr\iota$ functions in (b). 

\vspace{1em}

\section{Sequential Strategies}\label{sec-seq-stra}
\subsection{Definition, Informal Meaning and Examples}
\label{sec-seq-stra-def}

\subsubsection{Preliminary Definitions and Conventions}
\label{sec-seq-stra-def-prelim}

\noindent
Let $M$ be any set of abstract elements 
denoted as $m,m',m_1,m_2$, etc., 
each having a specified type (\eg, $m:\alpha$). 
That is, actually, $M$ is a disjoint union of sets $M_{\alpha}$ 
consisting of elements of the type $\alpha$. 
An additional structure on $M$ considered below 
will allow us to call these elements (computational) {\em strategies\/} 
(over~$M$).

For each type $\alpha$, let us also fix an infinite list of 
\emph{variables} of this type $x_1^{\alpha},x_2^{\alpha},\ldots$. 
We will use, $x$, $y$, $x'$, etc.\ as meta-variables. 
However, 
$x_i$ or $\bx=x_1,\ldots,x_n$ will usually refer 
to the numbering in the above lists, assuming some typing.  
That is, $x_i$ is $i$-th variable of a type which can 
be recovered from the context. 
Thus, given any types $\alpha_1,\ldots,\alpha_n$, 
we have the corresponding \emph{canonical} list 
of variables $x_1^{\alpha_1},x_2^{\alpha_2},\ldots,x_n^{\alpha_n}$ 
(first variable of the type $\alpha_1$, second 
variable of the type $\alpha_2$, etc.) 
or just $x_1,\ldots,x_n$ or $\bx$, for brevity.  
Well-typed \emph{applicative terms} over $M$ constitute the least 
set containing  
\emph{atomic terms} (i.e., variables $x:\alpha$ and constants $m:\alpha$), 
and closed under application: 
if~$A:\alpha\arr\beta$ and $B:\alpha$ then $AB:\beta$. 
Let $\mbox{Basic-Terms}(M)$ be the set of all well-typed applicative 
terms of the \mbox{Basic-type} (actually, $\iota$) built up from 
(typed) strategies of $M$ and (typed) variables. 
If $m:\alpha=(\alpha_1,\ldots,\alpha_n\arr\beta)$ then 
$mx_1\cdots x_n$ or $m\bx$ will denote 
the applicative term $(\cdots((mx_1)x_2)\cdots x_n)$ of the type $\beta$ 
with $x_i:\alpha_i$ (the $i$-th variable of the type $\alpha_i$). 
These notational agreements allow us to avoid 
type superscripts and related assumptions 
which, otherwise, would obscure the exposition. 
Strictly speaking, all variables, elements 
of $M$ and terms are typed.

Additionally, let us agree that, depending on the context, 
we can identify any variable $x:\alpha$ with some 
value in the corresponding set of values $E_{\alpha}$. 
This is in the same line as the tradition of using variables in 
ordinary mathematical texts. Again, this way we avoid 
extra complications in notation, relying on the context. 
Let us also assume that, by default, 
$v,v',v_1,v_2,\ldots$ range over $\NN$ 
whereas $u,w$ range over $\NN^{\ast}$.
Say, $v_1 v_2\cdots v_k\in\NN^{\ast}$ denotes the string 
of the length $k$, whereas $uw\in\NN^{\ast}$ is the concatenation 
of any two strings $u,w\in\NN^{\ast}$, and $uv\in\NN^{\ast}$ 
is the concatenation of any string $u$ with a one element string 
$v$, etc. We will use similar conventions for the case of 
$S^{\ast}$ for any other set $S$. 

\subsubsection{Main Definition}
\label{sec-seq-stra-def-main}
\begin{defi}\label{def:strategies}%
A {\em system of sequential computational strategies\/}%
\footnote{We will also consider, in Section~\ref{sec:wittingly-cons}, 
the more general concept of non-deterministic (non-sequential), 
wittingly consistent strategies. 
However, we will typically use the simple term ``strategy'' 
relying on the context. 
} 
is a pair $\tuple{M,\MM}$ consisting of 
the set $M$ of typed elements (strategies) 
and a {\em partial\/} 
function
\[
\MM:M \times \NN^\ast \mathrel{\dot{\arr}} \mbox{Basic-Terms}(M) \cup \NN,
\]
satisfying the following condition:
\begin{quote}
{\bf if} $m:\alpha = (\alpha_1,\ldots, \alpha_n \arr \Basictype)$ is a strategy, 
$x_1,\ldots, x_n$ is the canonical list of variables
of the types $\alpha_1,\ldots,\alpha_n$, respectively, so that
$mx_1\cdots x_n : \Basictype$, 
and $\MM(m,w)$ is defined {\bf then} either
\begin{enumerate}[(1)]
\item\label{item:strategies:variables}

$\MM(m,w) = A\{x_1,\ldots, x_n\}\in 
\mbox{Basic-Terms}(M)$ (written also as the query $\MM(m,w)$ $=$ ``$A\{\bx\}=\?$'') 
with all variables in $A$ contained in the list $x_1,\ldots, x_n$, or
\item
$\MM(m,w)\in \NN$ is a (defined) basic value. 
\end{enumerate}
\end{quote}
\end{defi}

\noindent
We also write $\MM(m,w)=\Undef$ if $\MM(m,w)$ is undefined.

\paragraph*{\bf Informal comments.}
Any applicative term of the form 
\[
m\bx = mx_1\cdots x_n=(\cdots(mx_1)\cdots x_n)
\] 
is considered as the \emph{query} or \emph{task} ``$m\bx=\?$'' of finding its 
(basic) value by means of the strategy $m$ with the help of an \emph{Oracle} as follows:
\begin{enumerate}[$\bullet$]
\item
by asking, in the case~1
above, 
{\em queries\/} of the form ``$A\{\bx\}=\?$'' 
(concerning~$\bx$) 
addressed to the Oracle,
assuming that a finite sequence of answers $w\in\NN^*$ 
to previous queries 
(called also a \emph{prompt} or \emph{computation history} 
for the strategy $m$) 
have been
received from the Oracle, and 
\item
by giving,
in the case~2, a {\em resulting value (solution)\/} for the
initial task ``$m\bx=\?$'', based on the previous computation history $w$. 
\end{enumerate}
In particular, it is possible that
$\MM(m,\Lambda) = v$ is a Basic-type value in $\NN$, 
or $\MM(m,\Lambda)$ is undefined, where
$\Lambda$ denotes the \emph{empty string} of the Oracle's replies 
to the previous queries
(i.e., when no queries to the Oracle have been asked yet---the 
empty history) 
and 
corresponds to the beginning state of the computation of strategy $m$.
In the case of $\MM(m,\Lambda)=v\in\NN$ we say that $m$  
defines (or is) a \emph{constant strategy\/} giving rise 
to a final result $v$ without asking the Oracle any questions. 
If $\MM(m,\Lambda)$ is undefined, then $m$ is called 
an {\em undefined constant strategy\/}. 
In each of these cases we write, respectively,
$m=v_{\alpha}$ or $m=\Omega_{\alpha}$ or even
$m=v$ or $m=\Omega$, especially when $\alpha$ is itself a basic type. 
Intuitively, a constant strategy $v_\alpha$ for 
$\alpha=(\alpha,\ldots,\alpha\arr\Basictype)$ defines (computes) 
the constant functional $\lambda x_1^{\alpha_1},\ldots,x_n^{\alpha_n}.v$ 
of the type $\alpha$. Analogously, $\Omega_{\alpha}$ denotes 
$\lambda x_1^{\alpha_1},\ldots,x_n^{\alpha_n}.\Undef$, the constant, 
undefined functional. 

However, typically, the strategy $m$ starts its computation 
by asking the Oracle sequentially some questions  
(concerning $\bx$) 
\begin{align*}
&\lqq A_1\{\bx\}=\?\rqq, 
\lqq A_2\{\bx\}=\?\rqq, 
\lqq A_3\{\bx\}=\?\rqq, 
\ldots; 
\\
&\MM(m,\Lambda)=A_1\{\bx\}, 
\MM(m,v_1)=A_2\{\bx\}, 
\MM(m,v_{1}v_{2})=A_3\{\bx\}, 
\ldots, 
\end{align*}
assuming that the Oracle replied 
\[
\lqq A_1\{\bx\}=v_1 \rqq, \lqq A_2\{\bx\}=v_2 \rqq, \ldots. 
\]
We assume that the strategy $m$ cannot continue computation until 
receiving the definite answer to the last asked query, if receiving 
any answer at all. 
This querying process can be either 
(i) finite with no result, if the Oracle does not answer a query, or 
(ii) infinite, or (iii) after some answers $v_{1},v_{2},\ldots, v_{k}$,  
$m$ could ``decide'' that it has already received 
all the ``required'' answers from the Oracle and stop asking queries  
by returning a resulting value 
\[
\MM(m,v_{1}v_{2}\cdots v_{k})=v\in\NN,
\]
instead of asking the next query $A_{k+1}\setof{\bx}$, 
if 
$\MM(m,v_{1}v_{2}\cdots v_{k})$ is defined at all.

We say that $m'$ is \emph{descendant} to $m$ if 
$\MM(m,w) = A\{x_1,\ldots, x_n\}$ for some $w$ 
and $m'$ occurs in $A\{x_1,\ldots, x_n\}$ 
(that is, $m$ \emph{asks} about $m'$, or $m'$ is a \emph{child strategy} of $m$)
or, recursively, $m'$ is descendant to a strategy occurring in 
$A\{x_1,\ldots, x_n\}$. Intuitively, only descendant strategies 
matter for the meaning of the given strategy $m$.

\subsubsection{Additional Requirements on Systems of Strategies}
\label{sec:additional-requirements}

Without restricting generality we can impose the following 
natural requirements on systems of strategies. 
\begin{enumerate}[$\bullet$]
\item
If 
$\MM(m,w)\in\NN_\Undef$ then $\MM(m,wu)$ is undefined  
for all non-empty $u\in\NN^*$. 
(Contraposition: If $\MM(m,wu)$ is defined then $\MM(m,w)$ defines a query.)
\item 
$\MM(m,w)$ is defined only for $m$-\emph{self-consistent}  
computational histories 
$w=v_1\cdots v_k$, i.e.\ for such $w$ which 
do not contain different answers 
to the same query by $m$: 
for all proper initial segments $w^i=v_1\cdots v_i$  
and $w^j=v_1\cdots v_j$, 
\[
\MM(m,w^i)=\MM(m,w^j)\in\mbox{Basic-Terms}(M)\Arrr v_{i+1}=v_{j+1}. 
\] 
\end{enumerate}
Note that only computational histories 
satisfying these properties are realizable in 
the interpreted computations considered below in 
Section~\ref{sec-seq-stra-den-sem}. The idea of consistency 
will be further generalized in 
Section~\ref{sec:wittingly-cons} 
when considering nondeterministic wittingly consistent strategies. 

Intuitively, each strategy $m:\alpha$ computes some functional 
$\Dsem{m}$ of the type $\alpha$. Let us first consider some simple examples.

\subsubsection{Examples of Strategies}
\label{sec-seq-stra-def-examples}

In these examples we assume that strategies compute functionals from the 
standard continuous model $\setof{\bbD_\alpha}$.
In the special case, when $\MM(m,\Lambda)=A\{\bx\}$ and
$\MM(m,v)=v$ for all basic values $v\in\NN$,  
we represent (the behaviour of) such a strategy $m$ by the formal equality
\[
m\bx=A\{\bx\}.
\]
This style of presentation allows us to avoid explicitly 
using $\MM$ when the behaviour of strategies is simple enough. 
It follows that the (typed) \PCF\ \emph{combinators} 
satisfying equalities%
\footnote{Strictly speaking, we should use the canonical list of variables 
$x_1,x_2,x_3,\ldots$ instead of $x,y,z$ and write, for example, 
$\Ss x_1 x_2 x_3 x_4 \cdots x_n = x_1 x_3 (x_2 x_3)x_4 \cdots x_n$ 
for the base type terms.
}
\[
\I x = x,\ \ \K x y = x,\ \ \Ss xyz = xz(yz), \mbox{  and  }\Y x = x(\Y x)
\]
\noindent may be also considered as strategies. 
In fact, we can consider \PCF\ \cite{Scott93,Plotkin77} as 
a system of strategies $\tuple{\PCF,\cPCF}$ where 
$\PCF=\{\I,\K,\Ss,\Y,\IF$, and some evident basic arithmetical operations$\}$ 
with typing 
omitted for brevity%
\footnote{this is actually an infinite system. 
}.
Note that the \emph{least fixed point operator} $\Y$ is an 
example of a \emph{recursive} strategy referring to itself. 
Another simple example
of a strategy is the {\em conditional\/} \PCF\ constant 
$\IF:\iota,\iota,\iota\arr\iota$ 
\[\IF\, xyz = \left\{\begin{array}{ll}
			y,      & \mbox{     if } x = \true\ (=1), \\
			z,      & \mbox{     if } x = \false\ (=0), \\
			\Undef, & \mbox{     otherwise.}
		   \end{array}
	    \right. 
\]
This strategy asks at most two questions: 
first ``$x =\?$'' and then, depending on the result 
$\true$ or $\false$,
it asks ``$y =\?$'' or ``$z =\?$'', respectively. 
The answer received from the Oracle to the second question 
on $y$ or $z$ will be 
returned by \IF\ as the final result of the computation. 
It is quite trivial to rewrite the above conditional equation for \IF\ 
in terms of $\MM_{\PCF}=\cPCF$---in the style of Definition~\ref{def:strategies}. 

Note that the following version of \IF, the
\emph{parallel conditional monotonic function} 
$\PIF_{\iota}:(o,\iota,\iota\arr\iota)$
(and analogously for $\PIF_o:(o,o,o\arr o)$)%
\footnote{Although we decided to avoid using the boolean type $o$ in 
the general theory of strategies, the examples considered here are a little simpler 
and more natural when this type is used. 
} 
defined as
\[
\PIF_{\iota}\ p\ \ELSE\ x\ \THEN\ y\ = 
\left\{
\begin{array}{cl}
x,      & \mbox{if } p=\true,\\
y,      & \mbox{if } p=\false,\\
x,      & \mbox{if } x=y,\\
\Undef, & \mbox{otherwise},
\end{array}
\right.
\] 
evidently has no computing it sequential strategy asking simple queries 
of the kind ``$p =\?$'', ``$x =\?$'', and ``$y =\?$'' (and, in fact 
no sequential strategy at all, asking arbitrary queries). 
Say, if the first query asked by such a strategy is ``$p =\?$'', 
it may happen that the answer is undefined, leading to an undefined 
result of the whole computation, whereas it can be $x=y\ne\Undef$ which should 
give a defined result. Analogously, such a strategy could not start with 
``$x =\?$'' or ``$y =\?$''.

Unlike \PIF, every \PCF\ constant 
can be considered as a sequential strategy. 
Say, the successor operation $x+1$ for $x:\iota$ 
is defined by the evident strategy which asks the question ``$x=\?$'' and, 
after getting a result $v\in\NN$ from the Oracle, returns the 
value $v+1$. 

As a less trivial example, consider the following strategy $m$  
computing the functional for 
the \emph{weak sequential existential quantifier} 
$\exists^{ws}:(\iota\arr o)\arr o$: 
\[
\exists^{ws} P = \left\{
			\begin{array}{cl}
			\true,  & \mbox{if } Px=\true\mbox{ for some }x, \\
			        & \quad\quad\mbox{with }P(y)=\false\mbox{ for all }y<x,\\
			\Undef, & \mbox{otherwise}
			\end {array}
		  \right.
\]
To compute $\exists^{ws} P$ (i.e., $mP$) 
this strategy starts by asking, sequentially, the queries  
\linebreak
``$P0=\?$'', ``$P1=\?$'', $\ldots$ to the Oracle.
The strategy keeps asking these queries in this order 
while all the currently received answers are \false. 
As soon as one of the answers obtained in this order 
is \true\ or $\Undef$, this value 
is the result of the computation. 
Alternatively, $m$ could be defined as follows. 
Again, $m$ starts with asking 
``$P0=\?$'' ($\MM(m,\Lambda)=P0$). If the answer is \true, 
$m$ returns the result \true\ ($\MM(m,\true)=\true$). 
Otherwise, $m$  
asks ``$m(\lambda x.P(x+1))=\?$''%
\footnote{This is a recursive query because $m$ asks about itself. 
}
and returns the answer of the Oracle to this query as the final result 
($\MM(m,\false)=m(\lambda x.P(x+1))$, $\MM(m,\false\;r)=r$). 
Here the lambda abstraction operator can be simulated, as usual, 
by combinatory strategies $\Ss$ and $\K$. 
Then, to compute $\exists^{ws}$, the system of strategies should also 
contain strategies $m,\Ss,\K$, and $+1$ (the successor).
The functional $\exists^{ws}$ can be also 
defined in \PCF\ by the recursive equation 
\[
\exists^{ws} P=\IF\ P0\ \THEN\ \true\ \ELSE\ \exists^{ws}(\lambda x.P(x+1)),
\]
or alternatively by using 
$\Y$: 
\[
\exists^{ws}=\Y\lambda P.\IF\ P0\ \THEN\ \true\ \ELSE\ \exists^{ws}(\lambda x.P(x+1)).
\]

Consider also the \emph{finite sequential existential quantifiers} 
$\exists_n^s:(\iota\arr o)\arr o$, $n=0,1,\ldots\,$ which can output 
both \true\ and \false:
\[
\exists_n^s P = \left\{
			\begin{array}{cl}
			\true, & \mbox{if } Px=\true\mbox{ for some }x\le n,\\
				 & \quad\quad\mbox{with }P(y)=\false
							\mbox{ for all }y<x,\\
			\false, & \mbox{if } P\Undef=\false, \\
			\Undef, & \mbox{otherwise}
			\end {array}
		  \right.
\]
The sequential strategy computing $\exists_n^s P$  
starts by asking $n$ queries 
``$P0=\?$'', ``$P1=\?$'', $\ldots\,$, ``$Pn=\?$''. 
As soon as one of the answers obtained in this order 
will be \true\ or $\Undef$ (undefined), this is the result of the computation. 
Otherwise, 
if all answers are \false, the strategy asks 
``$P\Undef=\?$'' and outputs the 
value of~$P\Undef$. 

The sequence of functionals $\exists_n^s$ is evidently increasing with a limit 
$\exists^s\bYdef\bigsqcup_n\exists_n^s$ which can be also defined as 
\[
\exists^s  P = \left\{
			\begin{array}{cl}
			\true, & \mbox{if } Px=\true\mbox{ for some }x,\\
				 & \quad\quad\mbox{with }P(y)=\false
						\mbox{ for all }y<x,\\
			\false, & \mbox{if } P\Undef=\false, \\
			\Undef, & \mbox{otherwise},
			\end {array}
		  \right.
\]
or in terms of \PCF:
\[
\exists^s P=P(\mu x.Px),\quad\mbox{as well as},\quad
\exists_n^s P=P(\mu x\le n.Px).  
\]
We omit the (well-known) definition in \PCF\ of the 
$\mu$-operator. 
The sequential strategy computing $\exists^s P$ reduces this task 
to the sub-task $P(\mu x.Px)$. The equation for $\exists_n^s$ 
gives an analogous strategy. 
The main point here is that strategies may be quite arbitrarily complicated. 
As we will 
see in Theorem~\ref{th:universal}, all (effectively computable) strategies, 
however general, can be simulated in \PCF, which 
characterises exactly its expressive power. 

\section{Interpreted Computations and the Denotational 
Semantics of Strategies}
\label{sec-seq-stra-den-sem}

\subsection{Preliminaries}
\label{sec-seq-stra-den-sem-prelim}

\noindent
Let us fix a given system of strategies 
$\tuple{M,\MM}$ 
and a monotonic, order extensional applicative structure 
$\EE=\{E_\alpha\}$ of finite type functionals, with $E_{\iota}=\NN_{\Undef}$ 
and $E_{\iota\arr\iota}=[\NN_\Undef\arr\NN_\Undef]$. 
Our current goal is to define a \emph{denotational semantics} 
of strategies 
\[
\Dsem{\defis}_\alpha=\Dsem{\defis}^{\MM}_\alpha:M_\alpha \arr E_\alpha, 
\textrm{ or briefly } \Dsem{\defis}:M \arr \EE, 
\]
as the least fixed point of some operator 
$\Dsem{\defis}\mapsto\Dsem{\defis}^+$, that is, 
the least solution of the equation $\Dsem{\defis}=\Dsem{\defis}^+$. 
This equation is also understood as the requirement of 
\emph{correctness} of the given semantics 
$\Dsem{\defis}$. 
In fact, $\Dsem{\defis}^+:M\arr \EE$ is defined via 
interpreted computations over $\EE$ 
performed by strategies of the system 
$\tuple{M,\MM}$ relative to~$\Dsem{\defis}$. 
The problem, however, concerns whether the operator 
$\Dsem{\defis}\mapsto\Dsem{\defis}^+$ is well-defined and 
whether the required least fixed point $\Dsem{\defis}$ exists. 
It does exist if $\setof{E_\alpha}$ is the standard continuous model 
$\setof{\bbD_\alpha}$.
It also exists for 
the monotonic model $\bbQ=\setof{\bbQ_\alpha}$ of hereditarily 
sequential functionals, which we will 
consider in Section~\ref{sec-seq-func}. 
In both the definition of a system of strategies $\tuple{M,\MM}$ and in 
earlier informal comments and examples 
it was \emph{implicitly}
assumed that both the Oracle and the strategy $m$ always give {\em correct\/}
(in a reasonable sense) answers/solutions 
to the queries/tasks they are
``resolving''. 
This can be further clarified as follows.
\nopagebreak
\subsection{Formal Definitions}
\label{sec:seq-stra-den-sem-formal-def}

Assume any semantic map $\Dsem{\defis}:M\arr \EE$ is given. 
We can extend $\Dsem{\defis}$ from $M$ to terms $\Dsem{A\{\bx\}}$ 
with variables from the list $\bx$ 
as usual, by induction, 
$\Dsem{CD}=\Dsem{C}\Dsem{D}$, assuming that each variable $x_i^{\alpha_i}$ 
has some associated value $\Dsem{x_i^{\alpha_i}}\in E_{\alpha_i}$. 
That is, $\Dsem{A\{\bx\}}$ depends on the values of~$\bx$.
Then, for any computational strategy 
$m:\alpha=(\alpha_1,\ldots,\alpha_n\arr\iota)$, 
we define that the initial task ``$m\bx=\?$'' (to be ``resolved'' 
by $m$) and 
all the queries 
``$A\{\bx\}=\?$'' asked by $m$ have
corresponding \emph{correct solutions} 
(with respect to~$\Dsem{\defis}$)---just the unique basic values 
$\Dsem{m\bx}$ and $\Dsem{A\{\bx\}}$ 
of these \mbox{Basic-Terms}, respectively.

Let us now give the formal definition of \emph{interpreted computation} of the basic 
value of 
$m\bx$ induced by a strategy $m$ 
in a system of strategies $\tuple{M,\MM}$
relative to some semantic map in 
$\EE$, $\Dsem{\defis}:M\arr \EE$, and some values of $\bx$ in $\EE$. 
This is a maximal finite or infinite 
sequence of pairs 
\begin{equation}\label{eq:interpreted-comp}
(A_1,v_1),(A_2,v_2),\ldots 
\end{equation}
of queries and Oracle's answers, 
i.e.\ of terms 
$A_i\setof{x_1^{\alpha_1},\ldots,x_n^{\alpha_n}}\in\mbox{Basic-Terms}(M)$ 
and basic values $v_i\in\NN$, which satisfy the following 
two conditions for 
each $(A_i,v_i)$:
\begin{description}
\item [$\Dsem{\defis}_1$] 
$\MM(m,v_1\cdots v_{i-1})=A_i$, 
\item [$\Dsem{\defis}_2$] 
$\Dsem{A_i\setof{x_1^{\alpha_1},\ldots,x_n^{\alpha_n}}}=v_i\ne\Undef$ 
(for the given values of $x_i^{\alpha_i}$ in $E_{\alpha_i}$). 
\end{description}
The latter means that \emph{Oracle's answers $v_i$ are correct} with respect to 
$\Dsem{\defis}$ and the values of~$\bx$.

A finite (maximal) interpreted computation $(A_1,v_1),\ldots,(A_t,v_t)\;,t\ge 0$, 
is called 
\emph{successful} with the \emph{result} $v\in\NN$ 
if, additionally, 
\begin{description}
\item 
[$\Dsem{\defis}_3$] $\MM(m,v_1\cdots v_{t})=v\in\NN$. 
\end{description}
As sequential strategies are ``deterministic'', 
the result $v\in\NN$ is determined uniquely, 
if it exists at all. 
If it does not exist, we also 
say that the result is \emph{undefined} ($\Undef$). 
This is possible in the following cases: 
\begin{enumerate}[(i)]
\item
 the computation is infinite, or 
\item
 it is finite and consisting of $t$ pairs, 
but unsuccessful, 
that is, $\MM(m,v_1,\ldots,v_{t})$ is either undefined, or $=$ some 
$%
A\setof{x_1^{\alpha_1},\ldots,x_n^{\alpha_n}}
$ 
with 
$\Dsem{A\setof{x_1^{\alpha_1},\ldots,x_n^{\alpha_n}}}=\Undef$
(for the given values of $x_j^{\alpha_j}$ in $E_{\alpha_j}$). 
\end{enumerate}

Now let $\Dsem{m}^+\bx$ denote 
the result $v$ in $\NN_\Undef$ of the interpreted 
computation (according to $\Dsem{\defis}_3$ above) 
of the value of $m\bx$ 
relative to $\EE$, $\Dsem{\defis}$ and any values of $\bx$ in $\EE$. 
Of course we would like to expect that $\Dsem{m}^+\bx=\Dsem{m}\bx$ 
(i.e.\ that \emph{the result of the computation is correct}) 
what, in general, 
is not true. For example, take $\Dsem{m}=\Undef$ for all $m$ of a non-trivial 
system of strategies. 
\begin{defi}%
$\Dsem{\defis}$ is called \emph{computationally correct} if 
the equation $\Dsem{m}^+\bx=\Dsem{m}\bx$ holds in $\EE$ wherever $m\bx:\iota$ 
or, briefly, $\Dsem{\defis}=\Dsem{\defis}^+$. 
\end{defi}
In general, $\Dsem{m}^+\bx$
is evidently monotonic on $\bx$ 
\label{page:monotonic}, as well as on $\Dsem{\defis}$, 
and defines 
a unique functional 
$\Dsem{m}^+:E_{\alpha_1}\times\cdots\times E_{\alpha_n}\stackrel{\rm mon}{\arrr} E_{\iota}$. 
But is this functional necessarily in 
$E_{\alpha}\subseteq E_{\alpha_1}\times\cdots\times E_{\alpha_n}\stackrel{\rm mon}{\arrr} E_{\iota}$?
If true for all $m$, this defines a new semantic map 
$\Dsem{\defis}^+:M\arr\EE$ and 
a monotonic operator \mbox{$\Dsem{\defis}\mapsto\Dsem{\defis}^+$} 
(probably defined not for all $\Dsem{\defis}$). 
In the case of the standard 
continuous model $\setof{\bbD_\alpha}$, this operator, 
being computable in the above sense, is evidently 
well-defined and also continuous 
and, therefore, has the least fixed point which we 
also denote as~$\Dsem{\defis}$. 
But in the general case of monotonic 
order extensional $\setof{E_\alpha}$ (and even of any continuous 
and directly complete 
$\setof{E_\alpha}$, but containing possibly not all continuous functionals)
the required value $\Dsem{m}^+$ might not exist in the model
and, even if 
it always exists, the 
monotonic operator $\Dsem{\defis}\mapsto\Dsem{\defis}^+$ 
might be not continuous (in the case of arbitrary monotonic 
$\setof{E_{\alpha}}$) and may have no least fixed point.%
\footnote{All of this seems quite plausible and desirable to confirm by example.}
But, when possible, we 
take $\Dsem{\defis}$ to be the least solution of the 
equation $\Dsem{\defis}=\Dsem{\defis}^+$. 
Thus, we are interested in the least computationally correct 
denotational semantics of strategies.

Moreover, for any model $\EE=\setof{E_\alpha}$ 
and arbitrary system of sequential strategies $\tuple{M,\MM}$,
let $\Dsem{m}^0\bYdef\Undef$ 
for all $m\in M$ 
and 
$\Dsem{\defis}^{n+1}\bYdef(\Dsem{\defis}^{n})^+$ 
assuming the latter is 
well-defined in $\EE$. 
Evidently, those $\Dsem{\defis}^{n}$ which exist are defined uniquely. 
It follows from the monotonicity of $^+$ and monotonicity and order extensionality 
of $\EE$ by induction on $n$ that 
$\Dsem{\defis}^{n}\sqsubseteq\Dsem{\defis}^{n+1}\sqsubseteq\Dsem{\defis}$ 
assuming $\Dsem{\defis}$ is an arbitrary computationally correct semantics. 
\begin{defi}\label{def:natural-defined}%
$\Dsem{\defis}$ is called \emph{naturally defined} in $\EE$ if 
all $\Dsem{\defis}^n$ exist and 
\[
\Dsem{\defis}=\pointwiselub_{n=0}^{\infty}\Dsem{\defis}^n,
\]
that is, 
$\Dsem{m}=\pointwiselub_{n=0}^{\infty}\Dsem{m}^n$ holds for each $m\in M$ 
where $\pointwiselub$ is the natural, 
or pointwise lub in $\EE$, as defined in Section~\ref{sec:natural}. 
\end{defi}
\begin{prop}\label{prop:natural-sem}\hfill
\begin{enumerate}[\em(a)]
\item
 If 
$\Dsem{\defis}:M\arr\EE$ is naturally defined in $\EE$ 
then it is $\sqsubseteq$ any computationally correct semantics in $\EE$. 
Thus, if $\Dsem{\defis}$ is also computationally correct then it is the least one. 

\item
 Moreover, 
if $\EE$ is naturally continuous and $\Dsem{\defis}$ is naturally defined then it 
is, indeed,  
the least computationally correct semantics of $\tuple{M,\MM}$.
\end{enumerate}

\end{prop}
\proof \hfill
\begin{enumerate}[(a)]
\item
The conclusion follows from the same statement on all $\Dsem{\defis}^n$. 

\item
Just the equality 
$\Dsem{\defis}=\pointwiselub_{n=0}^{\infty}\Dsem{\defis}^n$ 
implies that  
$\Dsem{m}\bx
\sqsubseteq\Dsem{m}^n\bx
\sqsubseteq\Dsem{m}^{n+}\bx
\sqsubseteq\Dsem{m}^+\bx$ 
holds whenever $m\bx:\iota$ for some $n$ depending on $\bx$.  
The converse inequalities 
$\Dsem{m}^+\bx\sqsubseteq\Dsem{m}^{n+}\bx\sqsubseteq\Dsem{m}\bx$ hold 
for appropriate $n$ depending on $\bx$ by using natural continuity of  $\EE$. 
It follows that $\Dsem{\defis}=\Dsem{\defis}^+$, as required.\qed
\end{enumerate}

\begin{defi}\label{def:sequentially-complete}%
If the naturally defined semantics $\Dsem{\defis}:M\arr\EE$ 
exists in $\EE$ and is (the least) computationally correct
for all sequential systems of strategies then $\EE$ 
is called \emph{sequentially complete}. 
\end{defi}

\noindent
Besides the evident example of the standard directly complete 
continuous model $\setof{\bbD_\alpha}$, 
the sequential completeness property holds also for 
the model $\setof{\bbQ_\alpha}$ of hereditarily sequential functionals
considered below in 
Sections~\ref{sec-seq-func}--%
\ref{sec:full-abs}. 
An analogous result takes place for another model $\setof{\bbW_\alpha}$ and a 
more general 
concept of nondeterministic (wittingly consistent)
strategies considered in Section~\ref{sec:full-abstr-PCF+}.

\begin{defi}\label{def:seq-func}%
Finite type functionals in $\EE$ of the form $\Dsem{m}$ 
for any strategy $m$ of any system $\tuple{M,\MM}$
(for the least computationally correct semantics $\Dsem{\defis}$, 
if it does exist)
are called 
\emph{sequential}.%
\footnote{We do not expect that this concept is really interesting 
for arbitrary~$\EE$. 
Although it is reasonable to restrict attention to naturally continuous and 
sequentially complete models, it may be unknown in advance that the given structure 
(such as $\bbQ$ or $\bbW$ considered below) satisfies these properties. 
Thus, for the sake of the argument, we need the general definition. 
}
If $m$ is a strategy from an (effectively) computable system 
of strategies $\tuple{M,\MM}$ (i.e. with computable 
$\MM$), 
then $\Dsem{m}$ is called
an \emph{effectively-sequential} functional.%
\footnote{A sequential functional can also be called 
\emph{sequentially computable}, although the corresponding strategy could 
be not (effectively) computable at all. 
That is, the concept of sequential computability is, in fact, a relative one 
(see also \cite{MTrakhte76TCS,Saz76d}).
}
\end{defi}

\noindent
This is the way that sequential and effectively-sequential finite type functionals 
(in appropriate $\setof{E_\alpha}$) can be defined in quite general terms of 
computational strategies~\cite{Saz76AL}. The same approach works for the 
type-free version of sequentiality~\cite{Saz76SMZH} in the Scott model 
$\bbD_\infty \cong [\bbD_\infty \arr \bbD_\infty]$%
\footnote{\label{foot:Dinfty}
Actually, a closely related and ``stronger'' isomorphism 
$\bbD_\infty \cong [\bbD_\infty\times \bbD_\infty\times\cdots\arr
\bbD_\iota]$ should be used. 
Note that this isomorphism evidently implies 
$\bbD_\infty \cong [\bbD_\infty\arr[\bbD_\infty\times\cdots\arr \bbD_\iota]]$
and hence 
$\bbD_\infty \cong [\bbD_\infty\arr\bbD_\infty]$.
This allows us to consider strategies asking (infinite) 
applicative queries over $\bbD_\infty$ 
of the basic type $\iota$, like in the typed approach. 
}. 
It could be also extended to more 
general type theories and models and also for more general kinds 
of basic values than the flat $\NN_\Undef$.

\section{Hereditarily Sequential Functionals}
\label{sec-seq-func}

\subsection{Canonical Strategies}
\label{sec:canonical}

\begin{defi}\label{def:canonical}
A system of strategies is said to be in the \emph{canonical form} if
all queries ``$A\{\bx\}=\?$'' asked by these strategies 
$m$ (with $m\bx=mx_1\cdots x_n$ of the basic type) have the form 
\begin{equation}\label{eq:canonical}
\mbox{``}x_i(m_1 x_1\cdots x_n)\cdots(m_{k_i}x_1\cdots x_n)=\?\mbox{''}
\end{equation}
\noindent
where each $m_k\bx=m_kx_1\cdots x_n$ 
has a type suitably depending on the type of the head variable $x_i$.
\end{defi}

\noindent
For example, a strategy $m$ of the type $(\iota\arr\iota)\arr\iota$ 
computing a functional $mf:\iota$ with $f:\iota\arr\iota$ 
can ask queries of canonical form 
``$f(m'f)=\?$'' or, in particular, ``$fn=\?$'' 
if $m'f$ is a constant functional having the integer value $n\in\NN$. 
Note that for sequential computability of such functionals 
it is insufficient to consider queries of the form ``$fn=\?$''. 
As we will see in Section~\ref{sec-quotient}, 
the canonical form of queries does not restrict the 
computational and denotational power of sequential strategies. 
Importantly, the descendant strategies $m_k$ in (\ref{eq:canonical}) 
have evidently \emph{the same, or lower, level} 
(of their types) than $m$.
This will serve below as the base for the inductive definition of 
hereditarily sequential functionals in terms of canonical systems 
of strategies. 

\subsection{The Main Inductive Definition}
\label{sec:ind-def}

\noindent
By using the above property of levels of strategies in canonical systems
we can give the
following inductive (level-by-level) definition of a 
\emph{monotonic order extensional structure
$\{\bbQ_{\alpha}\}$ of hereditarily sequential functionals} 
which will be shown later to be \emph{fully abstract model for\/ \PCF}. 
The initial part of this model for types up to level $l$ is denoted as $\bbQ^{\le l}$.
\begin{defi}\label{def:bbQ}%
For level 0, 
let $\bbQ_{\iota} \bYdef  \NN_{\Undef}$ 
be the flat basic domain. 
Assume, by induction, that 
the initial part of the model 
$\bbQ^{\le l}$ satisfying (\ref{eq:EE}) and (\ref{eq:App}) 
has been defined. 
For any $\alpha=(\alpha_1,\ldots,\alpha_n\arr\iota)$ of level $l+1$
take the minimal $k\le n$ such that 
$(\alpha_{k+1},\ldots,\alpha_n\arr\iota)$ is of the level ${}\le l$, and let, 
up to uncurrying%
\footnote{Note that the simpler definition 
$\bar{\bbQ}_\alpha
\bYdef
\{f:\bbQ_{\alpha_1}\times\cdots\times \bbQ_{\alpha_n}
\stackrel{\rm mon}{\arrr} \bbQ_{\Basictype}\}
$
does not work 
because we need to have below $\bar{\bbQ}^{\le l+1}$ to be an appropriate 
structure closed under application. 
}, 
\[
\bar{\bbQ}_\alpha\cong
(\bbQ_{\alpha_1}\times\cdots\times \bbQ_{\alpha_k}
\stackrel{\rm mon}{\arrr} 
\bbQ_{(\alpha_{k+1},\ldots,\alpha_n\arr\iota)}).
\]
\noindent
More precisely, let 
\begin{equation*}
\bar{\bbQ}_\alpha
\bYdef
\{f:\bbQ_{\alpha_1}\times\cdots\times \bbQ_{\alpha_n}
\stackrel{\rm mon}{\arrr} \bbQ_{\Basictype}
\mid
\forall\bx\in\bbQ_{\alpha_1}\times\cdots\times \bbQ_{\alpha_{k}} 
(
f\bx\in 
\bbQ_{(\alpha_{k+1},\ldots,\alpha_n\arr\iota)}
)
\}.
\end{equation*}
Then
\begin{equation}\label{eq:model}
\bar{\bbQ}{}^{\le l+1}\bYdef\bbQ^{\le l}\cup
\setof{\bar{\bbQ}_\alpha\mid\alpha\mbox{ is of level }l+1}
\end{equation}
can be considered as  
a monotonic, order extensional applicative structure up to level $l+1$ 
with the application operator defined by taking the residual map, 
as in~(\ref{eq:App}). 
Then, for any $\alpha$ of level $l+1$, define $\bbQ_\alpha\subseteq\bar{\bbQ}_\alpha$: 
\begin{align}
\bbQ_{\alpha}\bYdef\{\Dsem{m}\in\bar{\bbQ}_{\alpha} \mid {}
&m:\alpha\AND m\in M
 \AND\Dsem{\defis}:M\arr\bar{\bbQ}{}^{\le l+1}
\nonumber
\\
&\mbox{for some canonical system of strategies }M\}\label{eq:dom}
\end{align}
as the set of all monotonic 
mappings in $\bar{\bbQ}_\alpha$ which are computable/definable 
(as described in Section~\ref{sec:seq-stra-den-sem-formal-def}) 
by the strategies $m$ of the type
$\alpha$ of any system of strategies in canonical form%
\footnote{Without restricting generality, these systems 
may be evidently considered as 
containing only strategies of types $\tau$ up to level $l+1$.
}
for which 
the least correct semantics 
$\Dsem{\defis}$ 
in the structure $\bar{\bbQ}{}^{\le l+1}$ exists. 
In fact, we can equivalently%
\footnote{This will be clear later from isomorphic representation of 
$\bbQ$ as $\tQ$ and Theorem~\ref{th:least-correct-sem} (b). See also 
Proposition~\ref{prop:conditional-isomorphism}. 
}
require that $\Dsem{\defis}$ is \emph{naturally defined} 
(see Definition~\ref{def:sequentially-complete}). 

Alternatively, and equivalently (see the comments below), we can define 
for any $\alpha$ of level~\mbox{$l+1$}
\begin{align}
\bbQ_{\alpha}\bYdef\{\Dsem{m}\Dsem{m_1}\cdots&\Dsem{m_r}\in\bar{\bbQ}_{\alpha} 
\mid  m:(\gamma_1,\ldots,\gamma_r\arr\alpha) \AND m_i:\gamma_i
\nonumber
\\
&\AND \textrm{ level}(\gamma_i)\le l
 \AND m,m_i\in M
\nonumber
\\
&\AND\Dsem{\defis}:M\arr\bar{\bbQ}{}^{\le l+1}
\textrm{ correct and naturally defined}
\nonumber
\\
&\mbox{for some canonical system of strategies }M\}.\label{eq:dom'}
\end{align}
(See also Proposition~\ref{prop:natural-sem} (a).)
Sets of functionals $\bbQ_{\alpha}$ defined in this way 
for $\alpha$ of level $l+1$ are evidently nonempty and contain at least 
all the constant functionals. In particular, they contain 
the elements $\Undef_{\alpha}$ computable 
by the undefined strategies $\Omega_{\alpha}$. 
They are considered to be
partially ordered pointwise by $\sqle_{\alpha}$. 
This defines the extension $\bbQ^{\le l+1}$ of $\bbQ^{\le l}$ 
which satisfies (\ref{eq:EE}) and (\ref{eq:App}). 
(The latter property of the closure under application 
follows straightforwardly assuming (\ref{eq:dom'}).  
This is much more difficult to show, if the more intuitively plausible (\ref{eq:dom}) 
is assumed instead; see the comments below.)
 
This makes the induction step mathematically correct because we 
assumed, and used, the fact that $\bbQ^{\le l}$ satisfies only 
(\ref{eq:EE}) and~(\ref{eq:App}). 
Thus, (\ref{eq:dom'}) defines a monotonic order extensional structure $\bbQ$ by induction. 
\end{defi}

\paragraph*{\bf Comments} 
\begin{enumerate}[(1)]
\item
The induction step above defines
{\em simultaneously\/} all $\bbQ_{\alpha}$ of level $l+1$.
The canonical form of
strategies guarantees that {\em no\/} $\bbQ_{\tau}$ of a higher level
(not yet defined)
will be needed in the induction step.
By contrast, recall that, for example, Milner's definition of the fully abstract dcpo model, 
as well as later approaches to non-dcpo models, 
requires consideration of \emph{all} types and levels at once.

\item
Although (\ref{eq:dom}) and (\ref{eq:dom'}) are, in fact, equivalent definitions 
of $\bbQ_\alpha$ at the level \mbox{$l+1$}, unfortunately this is not so trivial and 
when taking the simpler equation~(\ref{eq:dom}) 
the proof of the \emph{correctness} of the whole definition would be rather involved.%
\footnote{Note that even for the standard definition of (hereditarily)
continuous functionals in $\setof{\bbD_\alpha}$ some correctness proof
is necessary.  Of course, the case of $\setof{\bbQ_\alpha}$ is more
complicated. Instead of contrasting the continuous case with the
sequential one we prefer to see some analogy here. Thus, both
approaches are essentially extensional with some intensional component in each
case, even if these intensional components have somewhat different
flavour and complexity.  } For the inductive step in
Definition~\ref{def:bbQ} to be legal in this case we must show that
the resulting $\bbQ^{{}\le l+1}$ satisfies both (\ref{eq:EE}) and
(\ref{eq:App}).  The condition (\ref{eq:EE}) holds by definition, and
(\ref{eq:App}) means that $\bbQ^{{}\le l+1}$ is closed under
application (also for results of the level $l+1$), that is under
taking residual maps, like for $\bar{\bbQ}{}^{\le l+1}$.  This is
quite straightforward in the case of~(\ref{eq:dom'}), unlike the case
of~(\ref{eq:dom}) although the latter looks more natural.  This is the
reason for our choice of~(\ref{eq:dom'}) in the above definition%
\footnote{Thanks to an anonymous referee for suggesting the formula~(\ref{eq:dom'}) 
which crucially simplified (made it just straightforward) 
correctness proof of the induction step 
of the definition of $\bbQ$. Based originally on~(\ref{eq:dom}) it required 
the full theory of sequential strategies of the next sections. 
But, anyway, this theory is still needed to prove the main 
properties of $\bbQ$. 
}.
The equivalence of (\ref{eq:dom}) and (\ref{eq:dom'}) 
will be shown later, 
as well as that 
an arbitrary system of sequential strategies, not necessary in the canonical form, 
has the least correct and even naturally defined denotational semantics $\Dsem{\defis}$ in $\bbQ$ 
(that is $\bbQ$ is sequentially complete), 
and that each element in 
$\bbQ_{\alpha}$ should have the form $\Dsem{m}$ for some (even canonical) strategy
$m : \alpha$. The latter means that $\bbQ$ consists of all, and only, sequentially 
computable functionals. 

\item
In general, we want to know that this structure is natural enough 
(although it is not a directly complete poset). 
That is it is a fully abstract model for \PCF, sequentially complete, 
naturally continuous, naturally algebraic and 
naturally bounded complete; we establish this later.  
But now we can prove a conditional 

\end{enumerate}

\begin{prop}\label{prop:conditional-isomorphism}
If some 
sequentially complete model $\bbQ'$ exists 
and each of its elements has the form 
$\Dsem{m'}$ for a strategy in some system of strategies 
in canonical form for $\Dsem{\defis}$ the (least) correct and naturally defined 
semantics in $\bbQ'$ then $\bbQ'\cong\bbQ$. 
It follows that in this case all the mentioned variations of the Definition~\ref{def:bbQ} 
give rise to the same $\bbQ$. 
\end{prop}
\proof 
Assuming 
that $\bbQ'$ (as well as $\bbQ$) satisfies 
(\ref{eq:EE}) and~(\ref{eq:App}) we can even show the identity $\bbQ'=\bbQ$. 
Thus, given by induction $\bbQ'^{\le l}=\bbQ^{\le l}$ 
(as is definitely true for $l=0$) 
and therefore 
$\bar{\bbQ}'_\alpha=\bar{\bbQ}_\alpha$, 
for $\alpha$ of level $l+1$, 
and $\bar{\bbQ}'^{\le l+1}=\bar{\bbQ}^{\le l+1}$, 
let us show that $\bbQ'_\alpha=\bbQ_\alpha$. 
But, according to (\ref{eq:dom'}) and our assumptions (in particular, the closure of 
$\bbQ'$ under applications as taking residuals), we have 
\begin{align*}
\bbQ'_\alpha & = 
\{\Dsem{m}\Dsem{m_1}\cdots\Dsem{m_r}\in\bbQ'_{\alpha}
\mid\ldots\bbQ'^{\le l+1}\ldots\} 
\\
 & = 
 \{\Dsem{m}\Dsem{m_1}\cdots\Dsem{m_r}\in\bar{\bbQ}'_{\alpha}
\mid\ldots\bar{\bbQ}'{}^{\le l+1}\ldots\} 
\\
 & = 
 \{\Dsem{m}\Dsem{m_1}\cdots\Dsem{m_r}\in\bar{\bbQ}_{\alpha}
\mid\ldots\bar{\bbQ}{}^{\le l+1}\ldots\} 
\\
 & \bYdef 
 \bbQ_{\alpha}
\end{align*} 
with the omitted parts ``$\ldots$'' as in (\ref{eq:dom'}).
In the second equality we use 
the routinely checked fact that the naturally defined and correct semantic map 
$\Dsem{\defis}$ in $\bbQ'^{\le l+1}$ is also naturally defined and correct in the extension 
$\bar{\bbQ}'^{\le l+1}\supseteq\bbQ'^{\le l+1}$ because $\bbQ'^{\le l+1}$ is closed 
under applications and all corresponding arguments and answers to all 
queries considered are evidently the same in both structures 
$\bbQ'^{\le l+1}$ and $\bar{\bbQ}'^{\le l+1}$. 
(Proposition~\ref{prop:natural-sem}~(a) shows that $\Dsem{\defis}$ is in fact the least 
correct semantics).\qed

\noindent
In particular, once the above shows $\bbQ=\bbQ'$, 
we have a simplified version of (\ref{eq:dom})
\[
\bbQ_\alpha  = 
 \{\Dsem{m}\!\in\!\bbQ_{\alpha}
\mid\Dsem{\defis}\textrm{ is the least correct semantics of a 
canonical system in }\bbQ^{\le l+1}\} 
\]
with the extensions 
$\bar{\bbQ}_{\alpha}$ and $\bar{\bbQ}^{\le l+1}$ no more necessary to mention.

\subsection{What Next?}
\label{sec:what-next}

\noindent
For showing the required properties of $\bbQ$ such as continuity 
and sequential completeness 
we will need more involved considerations and  
develop the corresponding general theory of sequential strategies 
\cite{Saz76SMZH,Saz76AL}
in Sections~\ref{sec-quotient}~and~\ref{sec:full-abs}. 

In particular, to represent the application operation in 
$\setof{\bbQ_\alpha}$
we will need to define corresponding operation
$\Osem{mm_1}$ for arbitrary strategies
$m:\alpha = (\alpha_1,\ldots,\alpha_n \arr \Basictype)$
and \mbox{$m_1 : \alpha_1$}, giving a ``residual'' strategy 
$\Osem{mm_1}$ of the type
$(\alpha_2,\ldots,\alpha_n \arr \Basictype)$, such that
$\Dsem{\Osem{mm_1}}=\Dsem{m}\Dsem{m_1}$; 
\cf\ Theorem~\ref{th:least-correct-sem} (a).
It is crucial here that $\Osem{\defis}$ serves as the operational semantics of strategies of arbitrary, not necessarily the basic types.

In fact, we will \emph{redefine} our model in a non-inductive, ``quotient'' form 
$\setof{\tQ_{\alpha}}\cong\setof{\bbQ_{\alpha}}$ where $Q=\bigcup\setof{Q_{\alpha}}$ 
is a unique \emph{universal} system of sequential strategies 
(containing in a sense all other systems---the unique up to isomorphism 
terminal object of the category 
of all systems of strategies) and will work mainly in terms of $Q$ and $\tQ$.

This general theory is based on the 
\emph{operational semantics} of strategies and will culminate in 
Sections~\ref{sec-quotient} in 
Theorem~\ref{th:least-correct-sem} and its Corollary~\ref{corr:isomorphism} 
(using the above Proposition~\ref{prop:conditional-isomorphism}) that $\tQ\cong\bbQ$. 
Moreover, 
we will also prove in Section~\ref{sec:full-abs} that $\{\tQ_{\alpha}\}$ is a fully
abstract 
model of~\PCF\ and has further good domain theoretic properties discussed in
Section~\ref{sec:prelim}.

\section{Sequential Functionals as Quotient Strategies}
\label{sec-quotient}

\noindent
According to \cite{Saz76SMZH,Saz76t}, 
there exists a {\em universal\/} system of
sequential strategies $\tuple{Q,\QQ}$ 
(with $Q$ of the cardinality of continuum)
such that for any other system
of strategies $\tuple{M,\MM}$ there exist a \emph{unique} homomorphism
$\varphi:\tuple{M,\MM}\arr\tuple{Q,\QQ}$. 
For the rest of this paper we will need only the existence of $\tuple{Q,\QQ}$, 
however its explicit construction is presented 
in Appendix~\ref{appendix:univ-sys-strategies}. 
In general, 
a \emph{homomorphism} $\varphi:\tuple{M,\MM}\arr\tuple{M',\MM'}$
is a map $\varphi:M\arr M'$ preserving types such that 
\[
\MM'(\varphi(m),w) = (\MM(m,w))^{\varphi}\quad\mbox{holds for all }
m\in M, w\in\NN,
\mbox{ where}
\]
\[(AB)^{\varphi} = (A^{\varphi})(B^{\varphi}),
\ %
m^{\varphi} = \varphi(m),
\ %
v^{\varphi}=v,
\ %
x^{\varphi}=x, 
\ %
\mbox{and}
\ %
\Undef^{\varphi}=\Undef\]
for any applicative terms $A,B$, strategy $m$, basic value $v$ and 
variable $x$. 
That is, a homomorphic image of a strategy has essentially ``the same'' behaviour.%
\footnote{\label{foot:approx-hom} 
In particular, $\Undef^\varphi=\Undef$ means that both 
$\MM'(\varphi(m),w)$ and $\MM(m,w)$ are defined, or not. 
A~more general concept of an \emph{approximating homomorphism} 
is obtained by allowing the requirement 
$\MM'(\varphi(m),w) = (\MM(m,w))^{\varphi}$ 
only in the case of $\MM(m,w)\ne\Undef$. 
That is, $\varphi(m)$ has ``the same or more definite'' behaviour than $m$.  
}
The fact that $\varphi$ can map different strategies in $M$ to the same strategy 
in $M'$ means that the latter is more ``abstract'' version of the former. 
Homomorphisms are evidently closed under compositions: 
$\MM''(\varphi\circ\psi(m),w) 
= (\MM'(\psi(m),w))^{\varphi} 
= ((\MM(m,w))^{\psi})^{\varphi}$.

Moreover, any strategy $m$ and its homomorphic image $\varphi(m)$ 
have the same denotational semantics in the following sense. 
\begin{prop}\label{prop:homomorphism-and-semantics}
Let $\varphi:\tuple{M,\MM}\arr\tuple{M',\MM'}$ be a homomorphism. 
\begin{enumerate}[\em(a)]
\item
 For any $\Dsem{\defis}':M'\arr\EE$ and its composition 
$\Dsem{\defis}\bYdef\Dsem{\varphi(\defis)}':M\arr\EE$
the corresponding results of 
the interpreted computations coincide: $\Dsem{m}^+\bx=\Dsem{\varphi(m)}'^+\bx$ 
wherever $m\bx:\iota$. 

\item
$\Dsem{\defis}^n=\Dsem{\varphi(\defis)}'^{\,n}$ holds
assuming 
$\Dsem{\defis}'^{\,n}$ exists.%
\footnote{For approximating homomorphisms defined in Footnote~\ref{foot:approx-hom} 
we rather have 
$\Dsem{m}^n\sqsubseteq\Dsem{\varphi(m)}'^{\,n}$ for all $n=0,1,\ldots$.
}

\item
 If $\Dsem{\defis}'$ is computationally correct 
(resp., naturally defined) then so is the composition 
$\Dsem{\defis}\bYdef\Dsem{\varphi(\defis)}'$. 

\item
 For $\EE$ 
sequentially complete, 
$\Dsem{\defis}=\Dsem{\varphi(\defis)}'$ holds 
for the (least) computationally correct and naturally defined semantics 
$\Dsem{\defis}$ and $\Dsem{\defis}'$ of these two systems, respectively. 
\end{enumerate}
\end{prop}
\proof \hfill
\begin{enumerate}[(a)]
\item follows from the similarity of the two interpreted computations 
via the homomorphism $\varphi$.

\setcounter{enumi}{0}
\item $\Arr$ (b)
(by induction):
\begin{align*}
\Dsem{\defis}^0=\Dsem{\varphi(\defis)}'^{\,0};
\quad
\Dsem{\defis}^n=\Dsem{\varphi(\defis)}'^{\,n}\Arrr 
\Dsem{m}^{n+}\bx=\Dsem{\varphi(m)}'^{\,n+}\bx, 
\end{align*}
\setcounter{enumi}{0}
\item[(a)] $\Arr$ 
the first part of (c):
\[
\Dsem{m}\bx
\bYdef\Dsem{\varphi(m)}'\bx
\stackrel{\rm corr.}{=}\Dsem{\varphi(m)}'^+\bx
\stackrel{\rm (a)}{=}\Dsem{m}^+\bx.
\]
\item[(b)] $\Arr$ 
the second part of (c):
\[ 
\Dsem{\defis}'=\pointwiselub_n\Dsem{\defis}'^{\,n}
\Arrr\Dsem{m}\bYdef\Dsem{\varphi(m)}'
=\pointwiselub_n\Dsem{\varphi(m)}'^{\,n}
\stackrel{\rm (b)}{=}\pointwiselub_n\Dsem{m}^{\,n}.
\]
\item[(c)] $\Arr$  
(d). (See also Proposition~\ref{prop:natural-sem}~(a).)
\qed
\end{enumerate}

\noindent
Therefore, it is natural to identify informally $m$ with ${\varphi(m)}$ 
and with their unique homomorphic image in $\tuple{Q,\QQ}$, 
and to consider the latter as a really universal system of strategies ``containing'' 
all possible strategies (up to homomorphism).

\medskip

Various strategies in $Q_{\alpha}\subseteq Q$ 
computing the same functional in $\bbQ_{\alpha}$, $\Dsem{q}=\Dsem{q'}$, 
may be identified via an equivalence relation $q\simeq_{\alpha}q'$ 
which will be also defined in Section~\ref{sec:def-of-tQ} by using 
operational semantics of strategies 
over $\tuple{Q,\QQ}$ 
so that we will actually have
$\bbQ_{\alpha}$ isomorphic to $\tQ_{\alpha}\bYdef Q_{\alpha}/_{\simeq_{\alpha}}$ 
and even could take the equality
$\bbQ_{\alpha} = \tQ_{\alpha}$ %
as (another) definition of $\bbQ_{\alpha}$.
Moreover, we will define a preorder relation $\lee_{\alpha}$ 
on the strategies in $Q_{\alpha}$ generating $\simeq_{\alpha}$ 
as the corresponding equivalence relation 
and inducing the approximation relation 
$\sqle_{\alpha}$ on $\tQ_{\alpha}$ 
(that is, ${\sqle_{\alpha}} = {\lee_{\alpha}/\simeq_{\alpha}}$) 
which, in fact, exactly corresponds to the pointwise approximation 
relation on $\bbQ_\alpha$ assumed in Section~\ref{sec:ind-def}.

\subsection{Operational Semantics for Strategies (Informally)}
\label{sec:operational-sem-informal}

\noindent
Following \cite{Saz76SMZH}, we will define an operation $\Osem{pq}$ 
of the application of strategies (having appropriate types) 
of the universal system $\tuple{Q,\QQ}$. 
More generally, given any combination 
$A$ of any type $\alpha$ 
consisting only of strategies, 
a new strategy can be defined $\Osem{A} \in Q$ of the same type $\alpha$ 
(also denoted in the op. cit. as $\hat{A}$). 
In particular, $A$ and $\Osem{A}$ should have the same denotational meaning 
in any ``reasonable'' model $\EE$, that is, 
\[
\Dsem{\Osem{A}}=\Dsem{A},\textrm{ or }
\Dsem{\Osem{pq}}=\Dsem{p}\Dsem{q}.
\]
This will be achieved in terms of 
a quite natural computation 
process induced by the strategies involved in $A$, without any reference 
to any model $\setof{E_\alpha}$. That is why this may be considered 
as an \emph{operational semantics} $\Osem{\defis}$ for the terms $A$, 
unlike the denotational semantics $\Dsem{\defis}$.

Therefore, let us consider the formal expression $\Osem{A}$ as a strategy 
(or we could take its unique homomorphic image in $Q$). 
We need to define the action of $\QQ(\Osem{A},u)$ 
for any string of the Oracle's answers $u\in\NN^*$. 
It is both simpler and instructive to first consider the case when $A$ and $\Osem{A}$ 
have the basic type~$\iota$. Such a strategy asks the Oracle no questions  
and ``computes'' some basic value $\QQ(\Osem{A},\Lambda)=v\in\NN$, 
if defined at all, for $u=\Lambda$ (the empty string of Oracle's answers). 
Thereby, the corresponding initial task ``$\Osem{A}={?}$'' or task 
``$A={?}$'' of finding this basic value $v$ will be resolved 
with the help of strategies participating in $A$ by reducing this task 
(by induction) to some sub-sub-$\,\cdots\,$-tasks ``$C={?}$''. 
Here all $C$ are terms of the basic type consisting only of strategies, and therefore 
having a numerical solution (if any) computed by induction in the same way 
until the original task ``$A={?}$'' is resolved.
In fact, each  sub-sub-$\,\cdots\,$-task $C$ has the form 
$C=mD_1 D_2\cdots D_k$, that is headed by a strategy $m$ 
which asks further queries (reduces $C$ to further immediate sub-tasks),
and continues the computation of the value of $C$ on the basis of the replies obtained. 
This generalizes the 
reduction process of lambda calculus or the natural (call-by-name) computation 
of the value of a closed \PCF\ term of the basic type.

In the general case, when the strategy $\Osem{A}$ or the term $A$ has 
an arbitrary, non-basic type 
$\alpha=(\alpha_1,\ldots,\alpha_k\arr\Basictype)$,
we need to consider the initial task ``$\Osem{A}\bar{y}={?}$'' 
or ``$A\bar{y}={?}$'' 
of the basic type $\iota$, with the variables $y_j:\alpha_j$. 
Then
it will be reduced to various sub-sub-$\,\cdots\,$-tasks 
$C:\iota$ 
which can now involve the variables~$\bar{y}$. If $C=mD_1 D_2\cdots D_k$ 
is headed by a strategy 
$m$ then the further computation (reduction to further immediate sub-tasks of $C$) 
proceeds as in the case above when all tasks considered had no variables. 
But it is also possible that 
$C=y_j D_1 D_2\cdots D_{n_j}$ is headed by a variable $y_j$ in $\bar{y}$. 
Here we assume that the computation continues with 
the help of an arbitrary (now non-empty) prompt $u$ by the Oracle 
because the head variable $y_j$ itself 
does not have the ``ability'' to continue the computation of $C$. 

For the initial task ``$A\bar{y}={?}$'' 
we actually want to know/compute: under which prompts $u$ from the Oracle, which 
sub-sub-$\cdots$-tasks $C$ headed by a variable, 
or which resulting values in $\NN$ can be generated? 
(The tasks $C$ headed by a strategy will continue the computation themselves.)
This is essentially the way (with many details omitted) 
how $\QQ(\Osem{A},u)$ can be defined (computed) 
by this process.

Formally, at each point 
we have a state of the computation 
like a ``stack'' (a finite string 
consisting of pending 
sub-sub-$\cdots$-tasks and basic values as the intermediate results) 
which may ``pulsate'' during time as we will see in the formal 
definition below.

\subsection{Operational Semantics for Strategies---Formal Definitions}
\label{sec:operational-sem-formal}

\noindent
Consider 
\begin{enumerate}[$\bullet$]
\item
a system of strategies $\tuple{M,\MM}$, 
\item
an applicative term 
$C=mD_1\cdots D_n\in\mbox{Basic-Terms}(M)$ 
(in the role of a currently considered task or sub-sub-$\cdots$-task 
of some initial task) 
with a head strategy $m\in M$ and possibly involving variables. 
\item
the canonical list of variables $\bx=x_1,\ldots,x_n$ for $m$ (such that 
$m\bx:\Basictype$), and 
\item
a prompt $w\in \NN^*$.
\end{enumerate}
Three cases are possible: 
\begin{enumerate}[($\MM$1)]
\item[($\MM1$)]
$\MM(m,w)=v\in \NN$, 
\item[($\MM2$)]
$\MM(m,w)$ is undefined, or 
\item[($\MM3$)]
$\MM(m,w)=B=B\setof{x_1,\ldots,x_n}\in\mbox{Basic-Terms}(M)$ 
\end{enumerate}
in which we will, respectively, say that the task 
$C=mD_1\ldots D_n$ (or ``$C=\?$'') 
is $w$-\emph{reducible} ($\MM1$) to the \emph{result} $v$, 
or 
($\MM2$) to the \emph{result} $\Undef$, 
or
($\MM3$) to 
the \emph{immediate sub-task} 
$C'=B\setof{D_1,\ldots,D_n}$ --- the result of substituting the terms 
$D_1,\ldots,D_n$ in $B$ $=$ 
$B\setof{x_1,\ldots,x_n}$ for 
its free variables $x_1,\ldots,x_n$. 

Now, given $\tuple{M,\MM}$, consider the set 
$\HH=\HH(M)\bYdef\mbox{Basic-Terms}(M)\cup\NN$.%
\footnote{Recall that the union is considered here to be disjoint, and 
$\mbox{Basic-Terms}(M)$ may also involve variables. 
}
As usual, $\HH^*$ denotes the set of finite strings over 
the set $\HH$ considered now as consisting of atomic data. These strings 
can serve as \emph{intermediate configurations} of a computation. 
Let the \emph{initial configurations} have the form $u(A\bar{y})$ 
where $u\in\NN^*\subseteq\HH^*$ is a numerical string (the potential Oracle's answers) 
and $\bar{y}$ shown are the only occurrences of 
variables in $A\bar{y}:\iota$. 
We use parentheses around $A\bar{y}$ 
to emphasize that this is a single element of $\HH$. 

Define a computational procedure 
consisting of a transformation of finite strings in $\HH^*$ 
by the following rules defining inductively a transformation relation 
${\vdash}\subseteq\HH^*\times\HH^*$. For any $C,C'\in\mbox{Basic-Terms}(M)$, 
$h\in\HH^*$, $w\in\NN^*$, and $v\in \NN$ the 
following transformations (derivations) are allowed: 
\begin{enumerate}[($\HH$1)]
\item
$hCw\vdash hv$, if $C$ is $w$-reducible to $v$; 
\item
$hCw\vdash hCwC'$, if $C$ is $w$-reducible to the immediate sub-task $C'$; 
\item
$vhC\vdash hv$, if $C$ has a head variable, i.e., has the 
form $y_{j}D_1D_2\cdots D_{n_j}$%
\footnote{Here $v$ is considered as the Oracle's prompt for the 
variable-headed task $C=y_{j}D_1D_2\cdots D_{n_j}$. 
Thus, query $C$ is replaced by the the prompt $v$ which, actually, originates 
from an element in $u$ of the initial configuration $u(A\bar{y})$. 
If $u=u'vu''$ with $u',u''\in\NN^*$ and $v\in\NN$ then, before applying this 
rule to the occurrence of $v$, 
the initial segment $u'$ should have been used analogously as the Oracle's 
answers on the previous steps of the computation. 
}; 
\item
 Transitivity: if $h\vdash h'$ and $h'\vdash h''$ then 
$h\vdash h''$. 
\end{enumerate}
Note, that no two of the rules ($\HH1$--$\HH3$) are applicable 
simultaneously to a string in~$\HH^*$. It follows that $\vdash$ 
determines a \emph{deterministic} (sequential) computation process. 
The term $C$ in the rules $(\HH1)$, $(\HH2)$ should be necessarily 
headed by a strategy, i.e., should have a form 
$mD_1 D_2\cdots D_{n_m}$ with $m\in M$.
A derivation terminating in a string of the form $hCw$, 
with $C$ $w$-reducible to $\bot$, is called 
\emph{dead-ended}. 

For any initial configuration 
$u(A\bar{y})\in\HH^*$, 
exactly one 
of three cases is possible:
\begin{enumerate}[(\rlap{$\hat{\phantom{0}}$}1)]
\item
$u(A\bar{y})\vdash v$ (with $u$ completely ``exhausted'' by using $(\HH3)$), 
where $v\in \NN$; 
\item
$u(A\bar{y})\vdash (A\bar{y})hC$ \ \ 
(with $u$ completely ``exhausted'' by using $(\HH3)$), where $h\in \HH^*$ and 
``sub-sub-$\cdots$-task'' $C\in\mbox{Basic-Terms}(M)$ is headed by a variable; 
\item
either there exists an infinite or dead-ended derivation starting with 
$u(A\bar{y})$, or 
$u'(A\bar{y})\vdash v$ 
holds for some initial segment $u'\ne u$ of the string $u$ 
(i.e.\ not all prompts from $u$ are used). 
\end{enumerate}
Given any applicative term $A$ of a type $\alpha$ without variables 
consisting of strategies in $M$, 
consider a formal expression of the form $\Osem{A}$ as a new strategy 
of the same type. 
Define a new system of strategies $\tuple{\hat{M},\,\hat{\!\MM}}$ where 
$\hat{M}$ is the set of all such formal expressions  
$\Osem{A}$ and $\,\hat{\!\MM}$ is a function 
making $\hat{M}$ a system of strategies which is defined below 
with the help of a ``splicing'' function 
$\delta:\mbox{Basic-Terms}(M)\arr\mbox{Basic-Terms}(\hat{M})$. 
We set $\delta(C)$ to be the result of grouping in the term 
$C$, with the aid of $\Osem{\defis}$, all the maximal sub-terms 
not containing variables. For example, 
\[
\delta(m_1 m_2 y_1(m_3(m_4 y_2))y_3 y_4)=
\Osem{m_1 m_2} y_1(\Osem{m_3}(\Osem{m_4} y_2))y_3 y_4.
\]
Finally, we define $\,\hat{\!\MM}$ by setting, for any $\Osem{A}\in\hat{M}$ 
and $u\in\NN^*$, 
\[
\,\hat{\!\MM}(\Osem{A},u)\bYdef\left\{
\begin{array}{ll}
v\in\NN, & \mbox{if } (\hat{1}), \\
\delta(C)\in\mbox{Basic-Terms}(\hat{M}), & \mbox{if } (\hat{2}), \\
\bot, & \mbox{if } (\hat{3}). 
\end{array} \right.
\]
Thus, 
the system of strategies $\tuple{\hat{M},\,\hat{\!\MM}}$ 
is based on the computation process ($\,\vdash$) 
induced by the strategies of $\tuple{M,\MM}$. 
By (implicit) use of the unique homomorphism from $\tuple{\hat{M},\,\hat{\!\MM}}$ 
into the universal system of strategies $\tuple{Q,\QQ}$, this 
gives $\Osem{A}\in Q$ for any applicative term $A$ over $Q$ without variables. 
In particular, $\Osem{pq}\in Q$ for any two strategies $p,q\in Q$. 
For $A$ of the basic type $\iota$, the strategy 
$\Osem{A}$ computes a constant value $v\in\NN_\Undef$ 
of this type (defined or not). This is also written as $\Osem{A}=v$. 

\begin{note}\label{note:canonical}\em
For the case of arbitrary type, 
the resulting strategy $\Osem{A}$ only asks queries headed by a variable 
(see ($\hat{2}$) above) and 
may be slightly redefined in such a way that 
all these queries will be in the \emph{canonical}
form (\ref{eq:canonical}) 
(by the evident use of combinators \Ss\ and \K\ and the splicing function 
$\delta$), 
even if the strategies participating in $A$ were not canonical. 
Alternatively, we could trivially extend $\Osem{\defis}$ to the case of 
$\lambda$-terms as $\Osem{\lambda\bx.A}$ for $A$ involving no $\lambda$ 
and use these $\lambda$-terms to get the canonical form. 
\end{note}

\subsection{Relating Denotational and Operational Semantics 
of Strategies for the Standard Continuous Model 
\texorpdfstring{$\{\bbD_\alpha\}$}{D}}
\label{sec:denot-oper}

\noindent
The main result of \cite{Saz76SMZH} relates
the {\em denotational\/} and {\em operational\/} 
semantics, $\Dsem{\mbox{-}}$
and $\Osem{\mbox{-}}$, of strategies in the
standard dcpo model $\{\bbD_\alpha\}$%
\footnote{more precisely,---in an untyped model
	$\bbD_{\infty} \cong [\bbD_{\infty}\arr \bbD_{\infty}]$;
	the case of typed model $\{\bbD_\alpha\}$ is quite similar
	and a corresponding result like (\ref{eq:oper-denot-sem}) 
	is formulated without proof in 
        \cite{Saz76AL} (see also Footnote~\ref{foot:Dinfty})
	 }
of all continuous finite type functionals over the given basic flat domain
$\bbD_{\iota}=\NN_{\Undef}$. 
It consists in the
following equality which holds for any 
typed applicative combination $A$ of
strategies containing no variables:
\begin{equation}\label{eq:oper-denot-sem}
\Dsem{\Osem{A}} = \Dsem{A}\mbox{  or, in particular, }
\Dsem{\Osem{pq}} = \Dsem{p}\Dsem{q}. 
\end{equation}
Here the right-hand side of the equality 
is the ordinary denotational semantics of an
applicative term defined by the application operator 
in the model $\{\bbD_\alpha\}$
and by $\Dsem{\mbox{-}}$ eventually applied to the strategies comprising $A$.
We will show in Theorem~\ref{th:least-correct-sem}~(a) 
that the same equality holds in the model $\setof{\tQ_\alpha}$ 
(and therefore in its isomorphic version~$\setof{\bbQ_\alpha}$).

The equality (\ref{eq:oper-denot-sem}) is essentially based on the 
\emph{associativity law} for  
$\Osem{\mbox{-}}$:
\begin{equation}\label{eq:assoc}
\Osem{{\mathcal A}} = \Osem{A}\mbox{  or, in particular, }
\Osem{\Osem{B}\Osem{C}} = \Osem{BC}
\end{equation}
where $A,B,C$ are  any combinations of strategies in $Q$, and ${\mathcal A}$ is
obtained from $A$ by grouping some sub-terms of $A$ with the help of
the operation $\Osem{\mbox{-}}$. 
The associativity law allows us to eliminate any nesting of $\Osem{\mbox{-}}$ 
and can be proved by a
thorough  analysis of $\vdash$-computations defined by strategies
$\Osem{\mathcal A}$ and $\Osem{A}$; \cf~\cite{Saz76SMZH} for a detailed proof 
(for the untyped case and for more general non-deterministic strategies).

\subsection{Definition of 
\texorpdfstring{$\lee\,,$ $\simeq\,$ and $\tQ$}{tilde-Q}}
\label{sec:def-of-tQ}

\noindent
Having the operational semantics $\Osem{\mbox{-}}$, we can define a
relation $\lee_\alpha$ on strategies of the same type $\alpha$ as follows.
\begin{equation}\label{eq:approx-strat}
q\lee_\alpha q\pr \bYdef
\forall \bar{q}.(\Osem{q\bar{q}}\lee_\iota \Osem{q\pr\bar{q}})
\end{equation}
where $p\lee_\iota p'$ relates the (constant) strategies of 
basic type $\iota$ and 
means that the strategy $p$ outputs 
the same basic value as the strategy $p'$, if the first value is
defined at all. 
	To simplify notation we will often omit the external $\Osem{\mbox{-}}$
	in inequalities $\Osem{A}\lee\Osem{B}$ for applicative terms
	$A$ and $B$ writing simply $A\lee B$.
Evidently, $\lee_\alpha$ is a preorder on the set of strategies 
$Q_\alpha$ of the type $\alpha$. The corresponding
equivalence relation is denoted as $\simeq_\alpha$, and the ``undefined'' strategy 
$\Omega_\alpha$ is the
$\lee$-least element in each type. 
Due to the above associativity law, we
have $\Osem{\Osem{q}\bar{q}}\simeq\Osem{q\bar{q}}$ and, hence, 
$\Osem{q}\simeq q$. Therefore, 
\begin{prop}
Any strategy $q$ is $\simeq$-equivalent to a
strategy in canonical form
\em (see Note~\ref{note:canonical}%
).\qed
\end{prop}
\begin{lem}\label{lemma:monotonic-op-sem}
Operational semantics is monotonic in the sense that 
for any applicative term $A\setof{q}$ without variables 
which involves a strategy $q$, 
\[
q\lee q'\Arr
\Osem{A\setof{q}}\lee\Osem{A\setof{q'}}. 
\]
\end{lem}
\proof  We can evidently consider that $A$ has the basic type. 
Then the proof proceeds by induction on the length $t$ of the computation 
$\Osem{A\setof{q}}=v\ne\Undef$. 
Let us write $A$ for $A\setof{q}$ and $A'$ for $A\setof{q'}$, etc. 
Two cases are possible. 
\begin{enumerate}[(1)]
\item 
$A=sA_1\cdots A_n$ and 
$A\pr=sA\pr_1\cdots A\pr_n$ for the same head strategy $s$. 
The case if $s$ is a constant strategy (with the value $v$) is trivial. 
Otherwise, $s$ reduces
the computation of the value $v$ of $A$ to some length ${}< t$ sub-computations of the
(basic) values $v_i$ of some sub-tasks $B_i$. By the induction hypothesis, 
corresponding $B\pr_i$ evaluate to
the same results $v_i$. It follows that $A\pr$ also evaluates 
to $v$ by the strategy $s$, as required.
\item 
$A=qA_1\cdots A_n$ and $A\pr=q\pr A\pr_1\cdots A\pr_n$ 
for the above $q$ and $q'$. 
Then, as
it was just proved,
$q A_1\cdots A_n$ and $q A\pr_1\cdots A\pr_n$ evaluate both to $v$, 
and it suffices to note that $q\lee q\pr$ and to use the 
definition of $\lee$ with $\bar{q}=\Osem{\bar{A}'}$ and associativity 
of~$\Osem{\mbox{-}}$.%
\qed
\end{enumerate}

\noindent
The following Lemma 
(Theorem 6.4 in \cite{Saz76SMZH}) 
corresponds to the \emph{context lemmas} in~\cite{Milner77}. 

\begin{lem}\label{lemma:context}
 Given any types $\alpha$ and $\beta$,
\[q\lee_{\alpha} q\pr
\iff
\forall p:\alpha\arr\beta.(\Osem{pq}\lee_{\beta} \Osem{pq\pr}).
\]
In particular, 
\[q\lee_{\alpha} q\pr
\iff 
\forall p:\alpha\arr\iota.(\Osem{pq}\lee_{\iota} \Osem{pq\pr}).
\]
\end{lem}
\proof\hfill
\begin{itemize}  
\item[($\Arr$)] follows from Lemma~\ref{lemma:monotonic-op-sem}. 

\item[($\Larr$)]
Let us assume (for contraposition) that 
$q\bar{q}\simeq v\not\lee_{\alpha} q\pr\bar{q}$. 
For any basic value $c$, define a strategy $p$ by 
\[
px\bar{y}=\IF\;x\bar{q}=v\;\THEN\;c_{\beta}\bar{y}\;\ELSE\;\Omega_{\beta}\bar{y}.
\]
Then $pq\simeq c_{\beta} \not\lee_{\beta} \Omega_{\beta}\simeq pq\pr$, 
as required.%
\qed
\end{itemize}

\noindent
Now, our goal is to show that 
$\bbQ_\alpha$ (\cf\ Definition~\ref{def:bbQ}) 
is isomorphic to the quotient 
$\tQ_\alpha\bYdef Q_\alpha /_{\simeq_\alpha}$ where each $q\in Q_\alpha$ 
generates the equivalence class $[q]\in \tQ_\alpha$ and $\sqle_\alpha$ 
is the partial order on $\tQ_\alpha$ induced by $\lee_\alpha$. 
The natural (typed) application operation in
$\tQ$ is defined by 
\begin{equation}\label{eq:appl-in-tQ}
[p][q]\bYdef[\Osem{pq}]
\end{equation}
which does not depend on
representatives $p$ and $q$ of the equivalence classes. 
So defined structure $\tQ$ 
is \emph{monotonic and order extensional} by 
Lemma~\ref{lemma:monotonic-op-sem} and definition 
(\ref{eq:approx-strat}) of~$\lee$.
\nopagebreak
\subsection{Denotational Semantics of Strategies in 
\texorpdfstring{$\tQ$}{tilde-Q} 
and the isomorphism \texorpdfstring{$\tQ\cong\bbQ$}{of tilde-Q and inductively defined Q}}
\label{sec:den-sem-isomorphism}

Let us consider $[\defis]$ as the denotational semantics of $Q$ in $\tQ$.

\begin{thm}\hfill  %
\label{th:least-correct-sem}
\begin{enumerate}[\em(a)]
\item
 Denotational semantics $s\mapsto [s]$
of the universal system of strategies $Q$ in $\tQ$ is coherent 
with the operational one%
\footnote{Compare this with the equation (\ref{eq:oper-denot-sem}) for 
the case of $\setof{\bbD_\alpha}$. 
}:
$
[\Osem{A}]=[A] 
$. 

\item
$\tQ$ is sequentially complete 
(in particular, satisfying the $\Y$-property (\ref{eq:Y-property})) 
with~$[\defis]$ the least correct denotational semantics 
which is also naturally defined. 
\end{enumerate}
\end{thm}
\proof\hfill  %
\begin{enumerate}[(a)]
\item
Apply associativity of $\Osem{\defis}$ and the definition 
(\ref{eq:appl-in-tQ}) of application in $\tQ$. 
For example, $[\Osem{p(qr)}]=[\Osem{p\Osem{qr}}]=[p][\Osem{qr}]=
[p]([q][r])]\bYdef [p(qr)]$. 

\item
 First, show correctness of $[\defis]$. 
Consider the interpreted computation by a strategy $q\in Q$ 
associated with the task ``$qx_1\cdots x_n=\?$'' 
of the basic type
with some fixed values $[q_i]$ in $\tQ$ 
for the arguments $x_i$ (and
$q_i\in Q$). We should assume that $q$ receives correct replies to its queries
``$A\{x_1,\ldots,x_n\}=\?$'' where $A$ is a combination of strategies
$s\in Q$ and the variables $x_i$. According to the assignment $s\mapsto [s]$ 
and (a),
the correct replies are obtained
just by replacing all strategies $s$ in $A$ by $[s]$ or, equivalently, by replacing
$A\{x_1,\ldots,x_n\}$ by 
$[A\{q_1,\ldots, q_n\}]=[\Osem{A\{q_1,\ldots, q_n\}}]$.
Then we must show that the resulting basic value $v$ 
(possibly ${}=\Undef$) 
of the interpreted 
computation coincides with the value of the combination
$[q][q_1]\cdots [q_n]= [q q_1\cdots q_n]=[\Osem{q q_1\cdots q_n}]$.
However, the latter value is
obtained by $\vdash$-computation, i.e.\ by 
essentially the same interpreted computation as above 
plus $\vdash$-sub-computations of the values
$[A\{q_1,\ldots, q_n\}]=[\Osem{A\{q_1,\ldots, q_n\}}]$ for all queries.
The required correctness follows.

Let us show 
that $\tQ$ is sequentially complete.
First, we present a general consideration on the ``approximating'' semantics $\Dsem{\defis}^k$ 
in any monotonic and order extensional structure $\EE$. 
Given any system of strategies $\tuple{M,\MM}$, define its 
``approximating'' 
version $\tuple{M^{A},\MM^{A}}$ by letting 
\begin{align*}
&M^{A}=\setof{m^k\mid m\in M\AND k\in\NN}, 
\\
&\MM^{A}(m^0,w)\bYdef\Undef,
\\
&\MM^{A}(m^k,w)\bYdef(\MM(m,w))^{k-1},
\end{align*}
where $m^k$ is considered as a formal expression (a pair of $m$ and $k$), 
$v^k = v$ for $v\in\NN_\Undef$, $(AB)^k=A^k B^k$ for applicative terms, 
and $x^k=x$ for variables. 
For any structure~$\EE$, if a computationally correct 
$\Dsem{\defis}^{A}:M^{A}\arr\EE$ exists 
then all $\Dsem{\defis}^k:M\arr\EE$, $k\in\NN$, exist too and 
$\Dsem{m^k}^{A}=\Dsem{m}^k$ holds for all $m\in M$, and vice versa. 
In particular, $\Dsem{\defis}^{A}$ is uniquely defined, if exists at all 
(iff all $\Dsem{\defis}^k:M\arr\EE$, $k\in\NN$, exist).

Now, let $\EE=\tQ$, and 
$\varphi:M\arr Q$ and $\varphi^{A}:M^{A}\arr Q$ be the unique homomorphisms. 
Then both $\Dsem{\defis}\bYdef[\varphi(\defis)]$ and 
$\Dsem{\defis}^{A}\bYdef[\varphi^{A}(\defis)]$ 
are computationally correct semantics of $M$ and $M^{A}$ in $\tQ$ 
by the correctness of $[\defis]$ 
and Proposition~\ref{prop:homomorphism-and-semantics}~(c). 
It follows from the latter that all $\Dsem{\defis}^k:M\arr\tQ$ exist, and, for 
sequential completeness of  $\tQ$, it remains to show that 
$\Dsem{m}=\pointwiselub_k\Dsem{m}^k=\pointwiselub_k\Dsem{m^k}^{A}$, that is 
$[\varphi(m)]=\pointwiselub_k[\varphi^{A}(m^k)]$, or equivalently, that 
for all strategies $\bq$ of appropriate types
$\varphi(m)\bq\simeq_\iota\varphi^{A}(m^k)\bq$ holds for some $k$. 
But the latter holds because, in each $\vdash$-computation 
giving a defined result in $\NN$, 
$m^k$ behaves as $m$ for sufficiently large $k$ and gives the same result.

It follows that $\Dsem{\defis}$ and therefore its special case 
$[\defis]$ are naturally defined and computationally correct and hence (by 
Proposition~\ref{prop:natural-sem}~(a))
both are the least correct semantics of $M$ and $Q$, respectively, in $\tQ$.\qed
\end{enumerate}

\begin{cor}\label{corr:isomorphism}
$\setof{\bbQ_\alpha}\cong\setof{\tQ_\alpha}$.
\end{cor}
\proof 
Use Proposition~\ref{prop:conditional-isomorphism}.\qed

\section{Main Results on Full Abstraction and Domain Theoretic Properties 
of \texorpdfstring{$\bbQ$}{Q}}
\label{sec:full-abs}

\subsection{Full Abstraction, Universality and 
\texorpdfstring{\PCF}{PCF}-Definability}
\label{sec:main-res}

\begin{thm}\label{th:full-abstr}
$\bbQ\cong\tQ$ is fully abstract model of\/ \PCF. 
The same holds for $\PCF^-$ (\PCF\ with $\Y$ omitted). 
\end{thm}
\proof 
Assume $q,q'\in Q_\alpha$ and $Cq\lee_\iota Cq'$ holds 
for all \PCF\ combinations $C:\alpha\arr\iota$. Then, in particular, 
$q\bc\lee_\iota q'\bc$ for all $\PCF^-$ definable terms $\bc$ 
of appropriate types. Let us infer $q\lee q'$, or equivalently 
that $q\bq\lee_\iota q'\bq$ holds for all strategies $\bq$ of appropriate types. 
Indeed, according to Section~\ref{sec:observation} below, 
if $q\bq\vdash v$ for some $v\in\NN$ then $q\bc\vdash v$ holds also 
for some \emph{finite} (and even \emph{finitary ranked}) and 
therefore definable in $\PCF^-$ strategies $\bc\lee\bq$ 
(see Lemma~\ref{lemma:computations-finitary}~(a) 
and Theorem~\ref{th:continuity} (b) below). 
It follows $q'\bc\vdash v$ and $q'\bq\vdash v$, as required.%
\qed

\noindent
As in \cite{Saz76AL} (the case of $\setof{\bbD_\alpha}$), 
\cite{Abramsky-Jagadeesan2000,Hyland-Ong2000} 
and also \cite{Longley-Plotkin} (the effective case), we have
\begin{thm}\label{th:universal}
For any type $\alpha$ there exists a\/ \PCF-definable functional
\[
U_{\alpha}\in \bbQ_{(\iota\arr \iota)\arr \alpha}
\]
which is universal in the
sense that its range is the whole set $\bbQ_{\alpha}$ of sequential 
functionals. Moreover, there
exists\/ \PCF-definable
$U^{\mbox{\scriptsize\rm eff}}_{\alpha}\in \bbQ_{\iota\arr \alpha}$
which enumerates all elements of $\bbQ$ definable by
computable strategies (i.e. those in systems $\tuple{M,\MM}$
with computable~$\MM$).

In particular,\/ \PCF\ exactly grasps sequential computability over $\bbQ$, 
that is, \PCF\ definable = sequentially computable. 
\end{thm}
\proof  As in \cite{Saz76AL} for the case of 
$\setof{\bbD_\alpha}$. It is omitted here, but see the proof 
in Section~\ref{sec:full-abstr-PCF+} 
of analogous result for $\setof{\bbW_\alpha}$ and $\PCF^+$.%
\qed

\begin{thm}[Normann \cite{Normann2004}]
\label{th:normann}
The (unique up to isomorphism) directly complete and continuous 
fully abstract model $\setof{\dQ_\alpha}$
for \PCF\ defined by Milner {\em \cite{Milner77}}%
\footnote{which is, more precisely, isomorphic to the limit (ideal) completion $\dQ_\alpha$ 
of $\bbQ_\alpha$; 
\cf\ Section~\ref{sec:completion}. 
}
cannot be exhausted by sequentially computable functionals 
in $\bbQ_\alpha\embed\dQ_\alpha$, 
i.e.\ by those definable in \PCF\ $+$ all monotonic $f:\NN_\bot\arr \NN_\bot$.\qed
\end{thm}

\noindent
More precisely, the proof in \cite{Normann2004} shows that 
$\bbQ_{\alpha}$ is \emph{not} an $\omega$-complete domain for some $\alpha$ 
of level~3. 
We know from Theorem~\ref{th:least-correct-sem}~(b) that $\bbQ$ 
is only sequentially complete.

\subsection{Deriving Domain Theoretic Properties 
of \texorpdfstring{$\bbQ$}{Q}}

We need to use Lemma~\ref{lemma:algebraicity}, and this requires 
to work out appropriate versions 
of ``finite'' approximations of strategies. 

\subsubsection{Finite, Finitely Restricted and Finitary Sequential Strategies}
\label{sec:ranked-finitary}

\begin{defi}\label{def:restricted}%
We say that a system of strategies $\tuple{M\pr,\MM\pr}$ 
is a \emph{restriction}, or \emph{subsystem } 
(or \emph{approximation}%
\footnote{but in a different sense than considered above 
system $\tuple{M^{A},\MM^{A}}$
})
of
another system $\tuple{M,\MM}$ if $M\pr\subseteq M$, 
and, as partial functions, $\MM\pr\subseteq\MM$.%
\footnote{Note, that the embedding $M'\embed M$ is  
an approximating homomorphism; \cf\ Footnote~\ref{foot:approx-hom}. 
}
A restriction $\tuple{M\pr,\MM\pr}$ is called \emph{finite} if both 
the set $M\pr$ and the function $\MM\pr$ are finite. 
Strategies (if any) from finite restrictions of $\tuple{M,\MM}$ are called 
\emph{finite}. 
\end{defi}

\noindent
If $\tuple{M\pr,\MM\pr}$ is a restriction of $\tuple{M,\MM}$ 
then the (unique)
homomorphic image $q\pr$ in $Q$ of any $m$ in
$\tuple{M\pr,\MM\pr}$ is called a 
\emph{restriction}, or \emph{sub-strategy} (or \emph{approximation}) 
of the homomorphic
image $q$ of the same $m$ considered as a strategy of $\tuple{M,\MM}$. 
By abusing notation, we write $q\pr\subseteq q$.%
\footnote{By considering the explicit construction of $Q$ 
(\cf\ \cite{Saz76t} or Appendix~\ref{appendix:univ-sys-strategies}), the relation $\subseteq$ on $Q$ may be treated, indeed, 
as set inclusion between strategies considered as graphs of partial functions 
of a special kind and is therefore 
a partial order. However, we will not need this fact. 
}
Then, evidently, $[q\pr]\sqle [q]$ (i.e.\ $q\pr\lee q$), holds in~$\tQ$.

Let us introduce a more general concept than a finite strategy. 

\begin{defi}\label{def:finitely-restricted}%
Given any system of strategies $\tuple{M,\MM}$, 
let $\MM^{[k]}(m,w)$ be defined and equal to $\MM(m,w)$ 
if, and only if, (i) the string $w$ consists only of numbers ${}\le k$ 
and (ii)~$\MM(m,w)\le k$ in the case of $\MM(m,w)\in\NN$. 
The system $\tuple{M,\MM^{[k]}}$ 
is called \mbox{$k$-\emph{restriction}} of $\tuple{M,\MM}$. 
If, in fact, $\MM^{[k]}=\MM$ then 
the original system is called \mbox{$k$-\emph{restricted}}. 
Then \emph{finitely restricted} means $k$-restricted for some $k$. 
A strategy $q\in Q$ is called $k$-\emph{restricted} if it is contained in the homomorphic 
image of some $k$-restricted system of strategies. 
\end{defi}
\noindent
Evidently, $\MM=\bigcup_k\MM^{[k]}$, and also 
any finite $\tuple{M,\MM}$ is finitely restricted (but not vice versa). 
$k$-restricted strategies ``understand'' only basic values ${}\le k$, 
as if it was our basic domain $\NN_\Undef$ so restricted to 
$\setof{0,1,\ldots,k}_\Undef$. 
Strategies from the original and restricted versions 
of a system of strategies, although formally having the same names, 
behave differently. Therefore, to emphasize that a restricted version 
is assumed, we will write $m^{[k]}$ instead of $m$ and 
$\tuple{M^{[k]},\MM^{[k]}}$ instead of $\tuple{M,\MM^{[k]}}$, 
whereas $m$ will typically be considered as a strategy of the 
non-restricted system $\tuple{M,\MM}$.%
\footnote{Note that, although there is a kind of analogy between the strategies $m^{[k]}$ 
considered here, and $m^k$ considered in the proof of 
Theorem~\ref{th:least-correct-sem}, the behaviour of these strategies is different. 
}
In the following Lemma 
we identify strategies with their homomorphic images in $Q$ 
and relate $k$-restriction with the projection maps $\Psi^{[k]}$ 
defined in Section~\ref{sec:natural}. 

\begin{lem}\label{lemma:finitely-restricted}
Functionals $[m^{[k]}]$ in $\tQ$ defined by finitely ($k$-)restricted
strategies are also finitely ($k$-)restricted 
\em (as defined in Section~\ref{sec:natural}). 
\end{lem}
\proof 
Consider projection functionals $\Psi^{[k]}$, $k=0,1,\ldots$  
and computing them sequential strategies 
$\psi^{[k]}$. 
Their behaviour can be described by the equality (in the basic type $\iota$, 
assuming $f:\alpha$ and $\psi^{[k]}:\alpha\arr\alpha$) 
\begin{equation*}\label{eq:proj-strategy}
\psi^{[k]} f\bx \;\; = \;\; f\bpsi^{[k]} (\bx),\textrm{ if the result is bounded by }k, 
\textrm{ and }=\Undef\textrm{ otherwise}.
\nonumber
\end{equation*}
Here $\bpsi^{[k]} (\bx)$ means the application of $\psi^{[k]}$ (of appropriate type) 
to each $x_i$ in~$\bx$. 
Let us show that $m^{[k]}\simeq\Osem{\psi^{[k]} m}$. 
The task ``$\psi^{[k]} m\bx=\?$''
is reducible to ``$m\bpsi^{[k]}( \bx)=\?$''.
By assuming that $m$ asks queries in canonical form 
``$x_i(m_1\bx)\cdots(m_n\bx)=\?$'', 
the task ``$m\bpsi^{[k]}( \bx)=\?$'' is further reducible 
by $m$ to the sub-task  
\begin{equation*}
\textrm{``}(\psi^{[k]} x_i)(m_1\bpsi^{[k]}( \bx))\cdots(m_n\bpsi^{[k]}( \bx))=\?\textrm{''}, 
\end{equation*}
and then by $\psi^{[k]}$ to 
\begin{equation*}
\textrm{``}
x_i\psi^{[k]}(m_1\bpsi^{[k]}( \bx))\cdots\psi^{[k]}(m_n\bpsi^{[k]}( \bx))=\?^{[k]}
\textrm{''} 
\end{equation*}
with the head variable $x_i$, 
where $\?^{[k]}$ assumes that only answers ${}\le k$ will be taken into account. 
As $m_1\bpsi^{[k]}( \bx)$ is $\vdash$-computationally equivalent 
to $(\psi^{[k]} m_1)\bx\;$%
\footnote{if to replace the variables $\bx$ by arbitrary strategies $\bq$ 
of the same types
},
the latter query is equivalent to 
\begin{equation*}
\textrm{``}x_i((\psi^{[k]} m_1)\bx)\cdots((\psi^{[k]} m_n)\bx)=\?^{[k]}\textrm{''}.
\end{equation*}
All of this means that $\psi^{[k]} m$ behaves computationally as $m^{[k]}$ 
which asks similar queries 
``$x_i(m^{[k]}_1\bx)\cdots(m^{[k]}_n\bx)=\?^{[k]}$'' and reacts to the answers 
in the same way as $m$ and $\psi^{[k]} m$, 
except considering the integer values bigger than~$k$ as if they were undefined. 
It follows that $m^{[k]}\simeq\Osem{\psi^{[k]} m}$, as required. 
Moreover, $[m^{[k]}]=[\psi^{[k]}] [m]=\Psi^{[k]} [m]$. 
If the original system is $k$-restricted then $m\simeq m^{[k]}$, 
and therefore the functional $[m]=\Psi^{[k]} [m]$ is $k$-restricted in~$\tQ$.%
\qed

\begin{defi}\label{def:ranked}%
A system of sequential strategies $\tuple{M,\MM}$ is called \emph{ranked} if 
(ignoring types)
$M$ is a disjoint union $\bigcup_{i\in \NN} M_i$
such that any strategy
in $M_i$ can ask queries only concerning the strategies in $M_{i+1}$.  
\end{defi}
\noindent
We have actually considered a similarly ranked systems in the proof of 
Theorem~\ref{th:least-correct-sem}~(b) but with the \emph{inverse} ranking order. 
Our choice of the \emph{ranking order} as in Definition~\ref{def:ranked} is based  
on the following Lemma. Independently of the choice of this order, 
ranked systems of strategies evidently remain ranked under restriction.

\begin{lem}\label{lemma:ranked}
Any system of strategies $\tuple{M,\MM}$ is 
homomorphic image of a ranked system. 
\end{lem}
\proof 
Indeed, $\tuple{M,\MM}$ is homomorphic image of a ranked system 
$\tuple{M\times \NN, \MM\pr}$ with $\MM\pr$ defined for all $m,w,n$ as 
\begin{equation*}
\MM\pr(\tuple{m,n},w)
\bYdef\textrm{sub}_{n+1}(\MM(m,w))
\end{equation*}
where $\textrm{sub}_{k}A$ is obtained from $A$, for $A$ any term, by replacing each 
occurrence of a strategy $m'$ in $A$ by
$\tuple{m',k}$, and 
$\textrm{sub}_{k}v=v$ for any resulting basic output value $v$.  
The required homomorphism is $\pi:\tuple{m,n}\mapsto m$.\qed

\noindent
Moreover, if $\varphi:\tuple{M_1,\MM_1}\arr\tuple{M_2,\MM_2}$ is a 
homomorphism 
then $\varphi^{\rm R}(\tuple{m_1,n})\bYdef\tuple{\varphi(m_1),n}$ is 
also a homomorphism of corresponding ranked systems 
\[
\varphi^{\rm R}:\tuple{M_1\times \NN,\MM_1\pr}\arr
\tuple{M_2\times \NN, \MM_2\pr},  
\]
and the resulting square diagram commutes: 
$\varphi \circ\pi=\pi \circ\varphi^{\rm R}$.

\begin{defi}\label{def:finitary}%
Strategies from
ranked systems of strategies $\tuple{M,\MM}$ with  
both $M$ and $\MM$ finite are called \emph{finitary}. 
(That is, essentially, finitary = finite $\&$ ranked, also = finite well-founded).) 
Equivalently, only $\MM$ may be required to be finite. 
\end{defi}

\begin{lem}\label{lemma:finitary-finitely-restricted}
Finitary strategies are special case of finite strategies 
which, in turn, are special cases of finitely restricted ones and therefore 
define (in fact all; \em see Theorem~\ref{th:continuity}\em) finitely restricted 
functionals in $\tQ\cong\bbQ$.\qed
\end{lem}
\noindent
If $M$ is finite and $M=\bigcup_i M_i$ is the ranking then all $M_i$ 
are empty for $i$ large enough.
In a reasonable sense 
finitary strategies are considered as \emph{non-recursive}. 
Homomorphic images in $Q$ of finitely restricted (resp., finitary) strategies 
can also be unofficially called \emph{finitely restricted (resp., finitary)} ones. 
Any (finitary) strategy $m_k\in M_k$ from a finite ranked system 
$\tuple{M,\MM}$ with the ranking $M=\bigcup_{i\in \NN} M_i$ 
has a finite \emph{rank} 
which is the length $r$ 
of a maximal chain $m_k,\ldots,m_{k+r}$ of strategies 
(in $M_k,\ldots,M_{k+r}$, respectively) starting with given $m_k$ 
such that each 
$m_{k+i}$, $0\le i<r$, asks a query on $m_{k+i+1}$ 
(i.e.\ $m_{i+1}$ is a child of~$m_i$). 
Now, K\"onig's Lemma entails more general
\begin{prop}\label{prop:finitary}
All strategies in $\tuple{M,\MM}$ are finitary%
\footnote{each in an appropriate finite ranked subsystem of $\tuple{M,\MM}$
} 
iff for each $m\in M$ 
there is only a finite number of computational histories $w\in\NN^*$ such that 
$\MM(m,w)$ is defined and there are no infinite chains 
$m=m_0,m_1,m_2,\ldots$ where $m_i$ asks a query on $m_{i+1}$ 
(i.e.\ $\tuple{M,\MM}$ is well-founded). 
\end{prop}
\proof
``Only if'' case is trivial. For ``if'' case assume its condition, and let 
\[
M_{{}\ge r}\bYdef\setof{m\in M\mid \exists m_0=m,m_1,\ldots m_r\in M
\;\forall i<r\;(m_{i+1}\textrm{ is a child of }m_i)}. 
\]
Then  
$M_r\bYdef M_{{}\ge r}\setminus M_{{}\ge r+1}$ is an \emph{inverse ranking} 
of $\tuple{M,\MM}$ (in the evident sense dual to Definition~\ref{def:finitary}). 
By K\"onig's Lemma, each $m$ has only a finite set $M^{[m]}\subseteq M$ 
of improper descendants (including $m$ itself) 
which, if intersected with each $M_r$, gives a finite (inverse) ranked 
subsystem of $\tuple{M,\MM}$, as is essentially required.\qed

\noindent
The finitary strategies of rank 0 are either constant strategies of any type 
(asking no queries to the Oracle) or strategies which can ask 
in each of finitely many possible ways of computation 
only (finitely many) queries which are applicative terms 
consisting of variables only. 
The finitary strategies of rank 1 are defined analogously, except that they can ask 
queries involving, besides variables, only strategies of rank 0. 
Etc., for finitary strategies of any rank. 

\medskip

But we need to be careful with such verbal descriptions. 
For example, the functional
$F(f)=\IF\;f(0)=0\;\THEN\;0\;\ELSE\;1$ 
(and $F(f)=\Undef$ if $f(0)=\Undef$)
computable by the evident rank 1 strategy 
is 
\emph{not finitary} because, 
in its computation, the query $f(0)$ can have \emph{any} answer ${}\ne0$ 
leading to the definite result $1$. 
In fact, $\MM$ describing the evident strategy computing functional $F$ has an infinite 
domain.

\subsubsection{Observation on Computations and Finitary Strategies}
\label{sec:observation}
It follows from Lemma~\ref{lemma:ranked} that 
in computations only countable ranked systems of strategies 
$\tuple{M,\MM}$ matter.

\begin{lem}\label{lemma:computations-finitary}\hfill
\begin{enumerate}[\em(a)]
\item
For any combination of strategies $A:\iota$ over 
$\tuple{M,\MM}$, 
if $A\vdash_\MM v$ then also $A\vdash_{\MM'} v$ over a 
finite restriction $\tuple{M',\MM'}$ of $\tuple{M,\MM}$. 

\item
For any countable 
system of strategies, $\tuple{M,\MM}=\bigcup_k\tuple{M^{(k)},\MM^{(k)}}$ holds 
for some monotonic by set inclusion sequence of finite restrictions of $\tuple{M,\MM}$.

\item
 For any system represented as a monotonic union 
$\tuple{M,\MM}$ $=$ 
$\bigcup_k\tuple{M^{(k)},\MM^{(k)}}$ of some 
restrictions, 
any resulting computation $A\vdash v$ over $\tuple{M,\MM}$ is, in fact, 
a computation over some $\tuple{M^{(k)},\MM^{(k)}}$, or equivalently 
over some $\tuple{M,\MM^{(k)}}$. 

\item
Let the strategy $m^{(k)}$ be just $m\in M$ considered as a strategy 
of $\tuple{M,\MM^{(k)}}$ with $\MM^{(k)}$ as in (c).%
\footnote{$m^{(k)}$ may be \emph{finite}, or even 
\emph{finitary} in the case of (b) and ranked $\tuple{M,\MM}$, 
or \emph{finitely restricted} in the case 
$\MM^{(k)}=\MM^{[k]}$ from 
Definition~\ref{def:finitely-restricted} 
with $m^{(k)}$ denoted there as~$m^{[k]}$.
} 
By identifying these strategies with their homomorphic images 
in $Q$, this gives rise to the $\sqsubseteq$-increasing sequence $[m^{(k)}]$ with the 
natural lub $[m]=\pointwiselub_k[m^{(k)}]$. 

\item
In particular, any functional in $\tQ$ is the natural lub of an increasing sequence of 
finitary presented functionals (and the same for any of the version of ``finite'' considered in Section~\ref{sec:ranked-finitary}). 
\end{enumerate}
\end{lem}
\proof\hfill
\begin{enumerate}[(a)]
\item
Let $M'$ consist only of those finitely many strategies in $M$ which 
participate in the original derivation $A\vdash_\MM v$ and 
(the finite) $\MM'(m,w)$ be defined if, and only if, $m$ and 
the computational history $w$ for $m$ 
was really used in the derivation $A\vdash_\MM v$. 

\item
 Let $M=\bigcup_k M_k$ with $M_k$ any increasing sequence of finite subsets exhausting~$M$. 
Let $\NN_k\bYdef\setof{0,1,\ldots,k}$ and 
$\MM^{(k)}\bYdef\MM\restr (M_k\times\NN_k^{\le k})$, and define $M^{(k)}$ 
to consist of all strategies 
participating in the domain and range of $\MM^{(k)}$. 

\item
 Like in (a), construct finite $\tuple{M',\MM'}$ and embed it in appropriate 
$\tuple{M^{(k)},\MM^{(k)}}$. 

\item
 Use (c) with the equation (\ref{eq:pointwise-lub}) 
defining the natural lub as the ordinary 
pointwise defined lub $\bigsqcup$ in the basic type 
by using an appropriate list of arguments. 

\item
 Use Lemma~\ref{lemma:ranked} and (d) with $\MM^{(k)}$ as in (b).\qed
\end{enumerate}

\begin{thm}\label{th:continuity}\hfill
\begin{enumerate}[\em(a)]
\item
The model of sequential functionals $\bbQ\cong\tQ$ 
is naturally continuous, 
naturally $\omega$-algebraic and naturally finitely bounded complete. 
Naturally finite elements of each $\bbQ_\alpha$ are exactly 
finitely restricted ones\/ {\em(in the sense of Definition~\ref{def:proj-finitely-restricted})} 
or, equivalently, definable 
by finitary strategies or, equivalently, 
by finite strategies or, equivalently, 
by finitely restricted strategies.%
\footnote{In \cite{Saz76SMZH}, special non-deterministic (non-sequential) 
strategies $\xi_a$ played the role analogous to that of 
sequential finitely restricted/finitary strategies considered here 
to define finite elements in $\bbD_\infty$ (or in $\setof{\bbD_\alpha}$ in the 
typed case), and $\bbD_\infty$ was also represented as a quotient of a universal 
system of (consistent) non-deterministic strategies. 
}
\item
Naturally finite elements of $\bbQ\cong\tQ$ are definable in 
\PCF\ (even without using $\Y$). 
\end{enumerate}
\end{thm}

\proof\hfill
\begin{enumerate}[(a)]
\item
follows from Lemma~\ref{lemma:algebraicity} whose condition (*) 
is satisfied because of the above observations and  
Lemmas~\ref{lemma:finitely-restricted}, \ref{lemma:ranked}, 
\ref{lemma:finitary-finitely-restricted}, and, most important, 
\ref{lemma:computations-finitary} (a).
Also recall that the naturally finite natural lub of an increasing sequence 
in $\bbQ$ must stabilize. 

\item
Use straightforward induction on the rank of finitary sequential strategies.  
Alternatively, apply the general Theorem~\ref{th:universal} concerning definability 
in \PCF\ (having much more involved proof).\qed
\end{enumerate}

\begin{note}\label{note:finite-simeq-finitary}\em
It follows from the definition of naturally finite elements in $\bbQ\cong\tQ$ 
that any finitely restricted or finite 
(possibly recursive) strategy is $\simeq$ 
to some finitary (ranked, non-recursive) strategy, by representing the former as  
the natural (in fact, stabilizing up to $\simeq$) lub of finitary strategies. 
But this proof is non-constructive, and by appropriate adaptation 
of the technique of Loader \cite{LoaderTCS2001} it should be possible to show that 
the \emph{there is no corresponding ``canonization'' algorithm} 
finite $\mapsto$ finitary as there is no way to determine the moment 
of stabilization in the above lub. Also the related \emph{problem ``$p\simeq q$?'' 
even for finitary (ranked) strategies should be undecidable}. 
\end{note}

\noindent
Note also that Theorem~\ref{th:continuity}~(b)  
and Lemma~\ref{lemma:computations-finitary}~(a) were 
actually used in the proof of Theorem~\ref{th:full-abstr} 
that the model $\bbQ\cong\tQ$ is fully abstract for \PCF\ 
which was incomplete till this moment.

\medskip

We conclude this section by proving that \emph{the class of finitary strategies 
is effectively closed under taking applications}. This was actually used in  
Section~\ref{sec:efficiency} in representation of naturally finite functionals 
in $\bbQ$ by finitary strategies (and, similarly, for $\bbW$). 
\begin{note}\label{note:closure-under-appl-of-finite-strategies}\em
On the other hand, the closure of finitely ($k$-) restricted strategies under 
application is trivial. But, unlike the finitary strategies, they are not necessary finite 
(and can be recursive). Also, \emph{arbitrary finite strategies are 
probably not closed under application} 
(note that ranking is essentially used in the proof of the following theorem), 
however evidently giving rise to finitely restricted strategies. 
\end{note}

\begin{thm}\label{th:applications-of-finitary}
For any applicative term $A$ consisting of finitary strategies, the strategy 
$\Osem{A}$ is finitary, too, and\/ {\em(as a finite object understood in the evident sense)} 
can be effectively computed from $A$ and 
comprising its strategies. 
\end{thm}
\proof 
Let us slightly generalize the concept of the initial configuration $uA\bY$ 
from Section~\ref{sec:operational-sem-formal} 
(where $u\in\NN^*$ and $\bY$ is a list of variables making the term $A\bY$ 
be of the basic type $\iota$) 
by allowing the term $A$ to contain any variables.  
The statement which we will actually prove is a kind of 
\emph{normalization (termination) property}: 
for each applicative term $A$ involving only finitary strategies 
and any variables 
\begin{quotation}
(*) \emph{ for any list of variables $\bY$ making $A\bY$ a term of the basic type 
there exists only a finite number of finite non-dead-ended computations%
\footnote{This requirement also means that for each numerical answer 
(either computed or taken from $u$) 
to a strategy question during such a computation the strategy should be able 
to react in a definite way 
giving either a result in $\NN$, as in the case of 
($\HH1$), or a new query, as in ($\HH2$). 
If dead-ended computations would be allowed then we might have an infinite number of them 
for $u\in\NN^*$ with large values in $\NN$. 
Indeed, only finitely many strategies---all being finitary descendants 
of those occurring in $A$---can participate in such computations, 
and they ``do not understand'' large numerical values. 
}
(sequences of derivation steps) starting from 
\mbox{$uA\bY\vdash\cdots$} for various 
$u\in\NN^*$ obtained by the rules ($\HH1$--$\HH3$) with $u$ completely ``exhausted''%
\footnote{Exhaustion 
is necessary, otherwise infinitely many $u$ 
of unbounded length would be admitted.
}. 
}
\end{quotation}
Then appropriate application of K\"onig's Lemma will imply that $\Osem{A}$ is indeed 
finitary and computable from $A$. 

Following Tait~\cite{Tait67} and the presentation by Barendregt~\cite{Barendregt85} of 
the normalizability proof for typed calculi, 
(*) can be shown 
for any $A$ as follows.%
\footnote{We give the detailed proof to show the specifics of the concept of 
strategies.
}
Define classes of typed terms 
consisting of finitary strategies and variables:
\begin{align*}
\CC_\iota&=\setof{A:\iota\mid A \textrm{ satisfies (*)}}, \\
\CC_{\alpha\arr\beta}&=\setof{A:\alpha\arr\beta\mid\forall B\in 
\CC_\alpha(AB\in \CC_\beta)}, \\
\CC&=\bigcup_\sigma \CC_\sigma.
\end{align*}
Evidently, 
\[
A\in \CC\iff\forall \bar{C}\in \CC (A\bar{C}:\iota\Arr A\bar{C}\textrm{ satisfies (*)}),
\]
and $\CC$ is closed under taking applications of terms. 
Any variable satisfies~(*) and belongs to~$\CC$. 
Also any finitary strategy trivially satisfies (*).  
It belongs to $\CC$ if its rank is 0, i.e.\ it is either a constant 
(defined or undefined) strategy or a strategy whose all possible 
(basic type) queries involve only variables. This is because $\CC$ is closed under 
applications, and therefore $\CC$-substitution cases of such queries 
satisfy (*). 
(That, in fact, all finitary strategies belong to~$\CC$ 
can be concluded from the following considerations.) 

Then we show by induction on the type of $A$ that 
\begin{align}\label{eq:finitary-1}
A\in \CC
\Arr A\textrm{ satisfies (*)}. 
\end{align}
Indeed, the base case $A:\iota$ holds by definition. 
For $A\in \CC_{\alpha\arr\beta}$ and any variable $y_1:\alpha$ 
we have $y_1\in \CC_\alpha$, $Ay_1\in \CC_\beta$, 
and hence $Ay_1$ satisfies (*) by induction hypothesis. 
Then it follows straightforwardly that $A$ itself satisfies (*). 

Finally, we show by induction on $k$ that for any term $A$ whose
participating strategies have rank ${}\le k$  
\begin{equation}\label{eq:finitary-2}
\textrm{any $\CC$-substitution case of $A$ belongs to }\CC.
\end{equation}
The case $k=0$: That (\ref{eq:finitary-2}) holds for atomic terms (variables 
and rank 0 strategies) was, in fact, shown above. The rest follows from 
the closure of $\CC$ and therefore of the class of $A$  satisfying 
(\ref{eq:finitary-2}) under applications. For $k>0$ it again suffices to show 
(\ref{eq:finitary-2}) for atomic terms. The main case is finitary strategies $m$ 
of rank~$k$ for which we should show that $m\in\CC$. We need to show that 
$m\bar{C}:\iota$ satisfies (*) for any $\bar{C}\in\CC$ of appropriate types. 
But this follows from the fact that $m\bY$ asks a bounded number of queries $B_i\setof{\bY}:\iota$, $i<N$, 
involving only variables $\bY$ and 
strategies of the rank ${}<k$ and which therefore satisfy 
(\ref{eq:finitary-2}) by induction hypothesis, 
and hence $B_i\setof{\bC}\in\CC$ so that all such 
$B_i\setof{\bC}$ satisfy~(*). 
Finally, this implies that $m\bC$ satisfies~(*). 
Indeed, from our requirements on the computations $um\bC\vdash\cdots$ 
each value in $u$ should be used either by $m$ or by (its child strategies from) the  
subcomputations generated by $B_i\setof{\bC}$. Thus, $u$ should have bounded both 
the length and participating numerical values. This concludes the proof.\qed

\section{Fully Abstract Model for \texorpdfstring{$\PCF^+$}{PCF+}}
\label{sec:full-abstr-PCF+}

\noindent
For the case of $\PCF^+$, let us consider the more general 
concept of a \emph{nondeterministic system of 
strategies} \cite{Saz76SMZH} extending the Definition~\ref{def:strategies}
of sequential (deterministic) strategies by letting 
\[
\MM:M \times \NN^\ast \arr \mbox{Basic-Terms}(M)\, \cup\NN\, 
\cup\setof{\#},
\]
and adding the clause (third possibility for $\MM$)
\begin{enumerate}[(1)]
\stepcounter{enumi}
\stepcounter{enumi}
\item %
$\MM(m,w)=\#$ (\emph{the nondeterministic state of computation}).
\end{enumerate}
The nondeterministic state can be also considered as representing a specific query 
``$\#=\?$''. The ``correct'' answer from the Oracle to this query 
is \emph{any} numerical value $r\in \NN$. 
However, such an extended concept of nondeterministic strategies 
is too general to grasp $\PCF^+$ 
(unlike $\PCF^{++}$ --- the case which we will not consider 
in full detail). 
Thus, we need to appropriately restrict nondeterministic strategies 
to fit them with $\PCF^+$.

\subsection{Wittingly Consistent Strategies}
\label{sec:wittingly-cons}

\noindent
First, without restricting generality we can assume that the 
requirements from Section~\ref{sec:additional-requirements} 
hold also for non-deterministic systems of strategies. 
Further, a pair of prompts (computational histories) 
$w=r_1\cdots r_k$ and $u=s_1\cdots s_n\in\NN^*$ 
for a strategy $m$ is called $m$-\emph{consistent}  
if they do not contain different answers 
to the same query by $m$, i.e.\ if 
for all proper initial segments 
$w^i=r_1\cdots r_i$ and $u^j=s_1\cdots s_j$, 
\[
\MM(m,w^i)=\MM(m,u^j)\in\mbox{Basic-Terms}(M)\Arrr r_{i+1}=s_{j+1}. 
\] 
In this paper, we will additionally require for systems of 
nondeterministic strategies $\tuple{M,\MM}$ that they 
should be 
\emph{wittingly consistent} (\cite{Saz76t}, Chapter II, \S4). 
This means 
that, for any $m\in M$ and any $m$-consistent pair 
of prompts $w$ and $u$, the strategy $m$ cannot output two contradictory 
final results:
\[
\MM(m,w)\in\NN\AND\MM(m,u)\in\NN\Arrr\MM(m,w)=\MM(m,u).
\]
Sequential (deterministic) systems of strategies are evidently 
wittingly consistent (assuming the first requirement 
of Section~\ref{sec:additional-requirements}).

Consider one example of such a wittingly consistent 
strategy $\mPIF$ computing 
a parallel conditional monotonic function 
$\PIF_{\iota} = \Dsem{\mPIF}:(o,\iota,\iota\arr\iota)$ 
defined in Section~\ref{sec-seq-stra-def-examples}: 

$\MM(\mPIF,\Lambda)=\#$, 

$\MM(\mPIF,0)=\mbox{``$p=\?$''}$,  

$\MM(\mPIF,0\,\true)=\mbox{``$x=\?$''}$, 
$\MM(\mPIF,0\,\false)=\mbox{``$y=\?$''}$, 

$\MM(\mPIF,0\,\true\,v)=v$, 
$\MM(\mPIF,0\,\false\,v)=v$, 

$\MM(\mPIF,1)=\mbox{``$x=\?$''}$, 
$\MM(\mPIF,1\,v)=\mbox{``$y=\?$''}$, $\MM(\mPIF,1\,v\,v)=v$. 

\noindent
In all other cases $\MM(\mPIF,w)$ is undefined. 

Consider also 
\emph{parallel disjunction} $\vee:(o,o\arr o)$ 
(used in infix notation) 
\[
p\vee q \bYdef \PIF\ p\ \THEN\ \true\ \ELSE\ q.
\]
It is parallel (as well as \PIF) because it is \true\ if 
\emph{any one} of the arguments is \true\ 
while the other may even be undefined~($\Undef$). Thus, there is no 
sequential way of evaluating the arguments, but an appropriate 
wittingly consistent strategy exists.

For wittingly consistent strategies, 
the \emph{interpreted (nondeterministic) computation} is defined as before 
in Section~\ref{sec:seq-stra-den-sem-formal-def}.  
All the successful computations under 
\emph{any} given interpretation of strategies $\Dsem{\defis}$ should 
evidently lead to a unique value $v\in\NN$ independently of 
the non-deterministic steps. This gives rise, as before, to 
the concept of the (least correct and naturally defined) denotational semantics 
$\Dsem{\defis}$ for any system of wittingly consistent strategies. 
As to operational semantics, $\Osem{\defis}$, we can easily show 
that the (appropriately defined as in Section~\ref{sec:operational-sem-formal}) 
system of strategies 
$\tuple{\hat{M},\hat{\MM}}$ is wittingly consistent if 
$\tuple{M,\MM}$ is.

In the most general case of nondeterministic strategies 
(the least) denotational semantics may give rise to $\Dsem{m}=\top$, 
the ``over-defined'' or ``contradictory'' value,  
for some ``contradictory'' $m$ because for some 
values $\bx$ the interpreted computation of the value $\Dsem{m}^+\bx$
gives different final results in $\NN$ for various paths of the computation. 
A weaker concept of consistency \cite{Saz76SMZH,Saz76t} of a system of nondeterministic 
strategies (in a structure) means the mere possibility of giving 
(the least) denotational semantics with $\Dsem{m}\bx=\Dsem{m}^+\bx\ne\top$ 
for all strategies in $M$ independently of the ways of computation. 
Witting consistency is a kind of guarantee, or sufficient condition, of the 
existence (say, in $\setof{\bbD_\alpha}$) of ``non-contradictory'' 
semantics. Otherwise this existence 
would be either somewhat accidental and unpredictable, 
or just fail, because of nondeterminism.

The theory for sequential strategies vs.\ \PCF\ considered so far can be 
naturally and, in many cases, straightforwardly 
extended for the case of wittingly consistent nondeterministic strategies 
vs.\ $\PCF^+$ (${}= \PCF + \PIF$), 
giving a fully abstract and naturally continuous order extensional 
model $\bbW=\setof{\bbW_\alpha}$ 
consisting exactly of all functionals definable in 
$\PCF^+$ $+$ all monotonic functions 
\mbox{$f:\NN_{\bot}\arr \NN_{\bot}$}. 
(Corresponding results for $\setof{\bbD_\alpha}$, instead of the case 
$\setof{\bbW_\alpha}$ considered here for the first time, 
were announced without proof 
in \cite{Saz76t}.) 
This model can be defined, like $\bbQ$, both inductively, level-by-level 
of types, 
and as a quotient $\tW$ of the universal system $\tuple{W,\WW}$ 
of wittingly consistent strategies. The universal functionals 
$U^+_{\alpha}\in\bbW_{(\iota\arr\iota)\arr\alpha}$ for each type 
can be constructed as for $\bbQ\cong\tQ$ (and $\bbD$) for sequential functionals. 
This gives a reasonable answer to a 
question of Longley and Plotkin in \cite{Longley-Plotkin} 
concerning the mere possibility of a general approach 
to a fully abstract model for $\PCF^+$ with 
definability properties like the above. 
(Cf. Introduction for a quotation.) 

Everything for wittingly consistent strategies goes almost as smoothly 
as for sequential strategies, except we should make some additional 
technical 
considerations needed for the definability of universal functionals 
$U^+_{\alpha}$
with the range being the whole~$\bbW_{\alpha}$. 
(We mean additional considerations 
in comparison with the case of sequential functionals and \PCF\ 
\cite{Saz76AL} --- what is unchanged 
is presented below without proof.) 
Note, that universal functionals for 
a (countable) fully abstract term model for $\PCF^+$ 
(of types $\iota\arr\alpha$, rather than $(\iota\arr\iota)\arr\alpha$) 
have also been defined in \cite{Longley-Plotkin}. But we use our old 
technique for \PCF\ and $\setof{\bbD_{\alpha}}$
(here --- for the model $\setof{\bbW_{\alpha}}$) 
with appropriate additions. 

Constructing $U^+_\alpha$ is the primary goal of 
Section~\ref{sec:full-abstr-PCF+}. 
However, for better understanding both of the nature of wittingly consistent 
strategies, and that witting consistency is an essential restriction, 
it makes sense to consider first some example demonstrating 
that $\bbW$ is not $\omega$-complete and thus does not coincide with 
the standard continuous model $\bbD$. 
Otherwise, the reader can well skip the following subsection.

\subsection{\texorpdfstring{$\tbbW_{(\iota\arr o)\arr o}$}{W} is not 
\texorpdfstring{$\omega$}{omega}-Complete}
\label{seq:non-complete}

\noindent
Although the undefinability result of this section is essentially 
well-known (in slightly different form) for the case of $\setof{\bbD_\alpha}$ 
(\cf\ \cite{Plotkin77,Saz76d,Saz76AL}), 
it makes sense to present its proof in terms of wittingly consistent 
strategies which was not published yet, except in~\cite{Saz76t}. 
Applied to the case of $\setof{\tbbW_\alpha}$, this implies that 
$\tbbW_{(\iota\arr o)\arr o}$ is \emph{not $\omega$-complete} and, therefore, 
it is a \emph{proper subset} of $\bbD_{(\iota\arr o)\arr o}$. 

Let us define functionals $\exists$ and $\exists_n\in \bbD_{(\iota\arr o)\arr o}$, 
$n\ge0$, with $P\in \bbD_{\iota\arr o}=\tbbW_{\iota\arr o}$ any argument for them, 
by the following equation: 
\[
\exists_{(n)} P = 
\left\{
\begin{array}{cl}
\true & \mbox{if }Px=\true\mbox{ for some }x\ (\le n),\\
\false & \mbox{if }P\Undef=\false, \\
\Undef & \mbox{otherwise}.
\end{array}
\right.
\]
Recall that $\PCF^{++}=\PCF^+ + {\;\exists}$ defines exactly all computable 
functionals (computable --- in terms of recursive enumerability of finite approximations) 
in the standard continuous model $\setof{\bbD_{\alpha}}$, and, 
by using arbitrary (actually, only strict) 
functions 
$f\in \bbD_{\iota\arr\iota}$, 
this language defines all continuous functionals of this model 
\cite{Plotkin77,Saz76d,Saz76AL}. 
On the other hand, each $\exists_n$ is definable in $\PCF^+$ by using 
the (wittingly consistent) parallel disjunction $\vee$: 
\[
\exists_n P = \IF\ P0\vee P1\vee\cdots\vee Pn\ \THEN\ \true\ 
\ELSE\ P\Undef, 
\]
and therefore $\exists_n\in\tbbW_{(\iota\arr o)\arr o}$. 
Moreover, $\exists=\pointwiselub_{n\ge0}\exists_n$ (pointwise), but 
$\exists\not\in\tbbW_{(\iota\arr o)\arr o}\subseteq\bbD_{(\iota\arr o)\arr o}$ because of the following
\begin{prop}
$\exists$ 
is not a wittingly consistent functional 
and hence not definable in $\PCF^+$. 
In particular, $\tbbW$ is not $\omega$-complete at the level 2. 
\end{prop}
\proof  
Let us assume on the contrary that $\Dsem{m}=\exists$ holds 
for some strategy $m$ of the type 
\mbox{$(\iota\arr o)\arr o$} from a wittingly consistent system of strategies 
$\tuple{M,\MM}$. We may consider that 
all queries asked by the strategy $m$ computing $mP$ have the canonical form 
``$P(m' P)=\?$'' for some $m':(\iota\arr o)\arr \iota$ in $M$. 

For each $i\in \NN$, define 
$P_i$ by $P_i x\bYdef\IF\ x=i\ \THEN\ \true\ \ELSE\ \Undef$. 
Let us show that \emph{for various $i$, the sets of sub-tasks in any successful 
interpreted computations for $mP_i$ do not intersect}.
To this end, consider two successful interpreted computations of $mP$ 
for $P=P_i$ and $P=P_j$, $i\ne j$, both giving a result (actually ${}=\true$ 
by $\Dsem{m}=\exists$), and 
assume on the contrary that the initial task ``$mP=\?$'', 
for $P=P_i$ and $P=P_j$, is reduced to the same task ``$P(m'P)=\?$'' 
(i.e., with the same $m'$) 
in the course of these two computations. 
As both the computations should continue further to the result, 
we would have 
$P_i(\Dsem{m'}P_i)\ne\Undef$, $P_j(\Dsem{m'}P_j)\ne\Undef$, 
both actually ${}=\true$ by the definition of $P_i$ and $P_j$,  
and hence 
$i=\Dsem{m'}P_i=\Dsem{m'}(P_i\sqcup P_j)=\Dsem{m'}P_j=j$, 
contrary to $i\ne j$. 

Now, let us consider an arbitrary $m$-prompt $w$ giving a defined boolean result 
\linebreak
$\MM(m,w)=r$, and show that the only possibility is $r=\true$.  
Indeed, the corresponding $m$-computation along $w$
involves only finite number of queries 
($\MM(m,w')\in\mbox{Basic-Terms}(M)$ for $w'$ initial segments of $w$)
which, 
by the above consideration, 
may also participate in successful interpreted computations of $\Dsem{m}P_i$ 
only for a finite number of $i$.  Therefore, for $i$ outside this 
finite set, $m$-prompt $w$ 
is $m$-consistent with the $m$-prompt $w^{(i)}$ arising 
in some interpreted computation of the value $\Dsem{m}P_i$ 
giving a defined result, 
which should be $\true$ by the assumption $\Dsem{m}=\exists$. 
From the definition of witting consistency, 
it follows that $r=\true$, as required. 

Thus, the values of 
$\Dsem{m}P$ for any predicate $P:\iota\arr o$ 
may only be $\Undef$ or $\true$, and $\Dsem{m}$ cannot be $\exists$ 
(for which $\exists P=\false$ is possible), contrary to the main assumption.\qed

\bigskip

\noindent
As to sequential functionals, the 
increasing sequence $\exists_n^s\in\bbQ_{(\iota\arr o)\arr o}$, 
$n=0,1,\ldots\,$ analogous to $\exists_n\in\tbbW_{(\iota\arr o)\arr o}$ 
cannot demonstrate that $\bbQ$ is not $\omega$-complete because 
this sequence 
has the limit $\exists^s$ existing also in $\bbQ_{(\iota\arr o)\arr o}$, as we 
have shown in Section~\ref{sec-seq-stra-def}. 
Thus, demonstrating the incompleteness of $\bbQ$ 
requires the more subtle considerations of \cite{Normann2004} at the level 3.

It is useful to note that \emph{strictly sequential functionals} of
the type $(\iota\arr o)\arr o$, i.e.\ those computable by the
sequential strategies asking only simple queries of the form
``$Pi=\?$'' with $i\in \NN$, \emph{are closed under $\omega$-limits}.
(Hint: first note, that if $F$ is strictly sequential then so is any
$F'\sqsubseteq F$, and consider limits of finite, in the sense of
$\setof{\bbD_\alpha}$, strictly sequential functionals.)  Further, for
a functional of the type $(\iota\arr\iota)\arr\iota$ or $(\iota\arr
o)\arr o$, \emph{to be strict {\em(see below)} and sequential is
equivalent to be strictly sequential}.  Moreover, looking for limits
of sequences of more complicated, non necessarily strict sequential
functionals of this type (based on the general queries of the form
``\mbox{$P(m'P)=\?$}'') will also fail.  In fact, the minimal level of
$\bbQ_\alpha$ where non-$\omega$-completeness holds is 3
\cite{Normann2004}.

\subsection{Definability in \texorpdfstring{$\PCF^+$}{PCF+} 
of Strict Continuous Functionals 
\texorpdfstring{\mbox{$F:(\iota\arr\iota)\arr\iota$}}{F:(i->i)->i}}
\label{sec:strict-functionals}

\noindent
Here we will consider strict level~2 functionals. 
We will also rely on some definability concepts 
and ideas due to Plotkin \cite{Plotkin77}.  
A similar definability technique was assumed also in 
the corresponding results announced in 
\cite{Saz76d,Saz76AL}, but without 
presenting details and proofs. 

A function $f\in \bbD_{\iota\arr\iota}$ is called \emph{strict} if 
$f\Undef=\Undef$.
Given any 
$a_i,b_i\in \NN$,  
$i<n$, $n\ge 0$, with all $a_i$ different, 
let 
$	%
		 \left[\,^{b_0,\ldots,b_{n-1}}
		 _{a_0,\ldots,a_{n-1}}\right]
$
denote a \emph{strict (naturally) finite function} in $\bbD_{\iota\arr\iota}$ 
such that 
\[
		 \left[\,^{b_0,\ldots,b_{n-1}}
		 _{a_0,\ldots,a_{n-1}}\right]
		x
		=\left\{\begin{array}{cl}
		b_i,  & \mbox{if } x=a_i \mbox{ for some } i<n, \\
 		\Undef, & \mbox{otherwise}
			   \end{array}
		\right.
\]
or, equivalently, 
\begin{equation}\label{eq:fin}
		 \left[\,^{b_0,\ldots,b_{n-1}}
		 _{a_0,\ldots,a_{n-1}}\right]
		x
		=\bigsqcup_{a_i \sqsubseteq x} b_i.  
\end{equation}
Recall that more general \emph{finite} (not necessarily strict) functions 
in $\bbD_{\iota\arr\iota}$ are defined by such tables with 
$a_i,b_i$ arbitrary elements of $\bbD_{\iota}$, possibly ${}=\Undef$, 
satisfying 
a natural consistency requirement, and defined by equation~(\ref{eq:fin}), 
and analogously (by induction) for finite elements of 
arbitrary $\bbD_{\alpha\arr\beta}$ with 
$a_i$ and $b_i$ being finite elements, respectively, 
of $\bbD_{\alpha}$ and $\bbD_{\beta}$. 
Note, that any (constant) function in $\bbD_{\iota\arr\iota}$ 
such that $f\Undef\ne\Undef$ is also 
finite ($f=\left[\,_{\Undef}^{c}\right]$ for some $c$), but not strict. 
Let $\varphi_a$, $a\in \NN$, be an \emph{effective numbering} of all 
strict finite functions in $\bbD_{\iota\arr\iota}$ 
such that, given $a$, the numbers $n,a_i,b_i$ (all ${}\ne\Undef$) can be recovered. 

We can also consider \emph{strict finite functionals} of the form 
$\left[\,_{\varphi}^{b}\right]\in \bbD_{(\iota\arr\iota)\arr\iota}$ 
with $\varphi$ strict finite and $b\ne\Undef$:
\[
\left[\,_{\varphi}^{b}\right]f=\left\{
				\begin{array}{cl}
				b, & \mbox{if }\varphi\sqsubseteq f, \\
				\Undef, & \mbox{otherwise}.
				\end{array}
				\right.
\] 
In general, any continuous functional $F\in \bbD_{(\iota\arr\iota)\arr\iota}$ is 
called \emph{strict} if, 
for all $f,f':\mbox{$\iota\arr\iota$}$, the coincidence of $f$ and $f'$ 
on all type $\iota$ 
arguments ${}\ne\Undef$ implies $Ff=Ff'$. 
Equivalently, $F$ is strict if for each $f$ 
there exists a strict (and therefore exists a strict finite) 
$\varphi\sqsubseteq f$ such that $Ff=F\varphi$. 

\begin{lem}\label{lemma:strict-def}\hfill
\begin{enumerate}[\em(a)]
\item
All strict functionals $F\in \bbD_{(\iota\arr\iota)\arr\iota}$ are 
(uniformly) definable in $\PCF^+$ from strict 
functions of type 
$\iota\arr\iota$ and are, in fact, wittingly consistent. 

\item
The same holds for the functionals 
$G:\iota,(\iota\arr\iota)\arr\iota$ 
which are strict in the first type $\iota$ argument 
and either constant or strict in the second 
type \mbox{$\iota\arr\iota$} argument
{\em(and can be identified with arbitrary sequences 
$G_m:(\iota\arr\iota)\arr\iota$, 
$m=0,1,\ldots\,$, of constant or strict functionals)}. 
\end{enumerate}
\end{lem}
\proof\hfill  %
\begin{enumerate}[(a)]
\item
First, note that parallel disjunction can be generalized 
to \emph{bounded quantification}. This can be defined in $\PCF^+$ 
recursively (for $P:\iota\arr o$): 
\[
(\exists i<n.Pi) = 
\IF\ n=0\ \THEN\ \false\ \ELSE\ (\exists i<n-1.Pi)\vee P(n-1).
\]
In particular, $(\exists i<\Undef.Pi)=\Undef$. 
This allows us to define in $\PCF^+$ a functional 
\begin{align*}
&\#:(\iota,(\iota\arr\iota)\arr o),  
\\
&\#c f
	=
	\exists i<n(f(a_i)\ne b_i), 
\end{align*}
assuming that 
$\varphi_c=
		 \left[\,^{b_0,\ldots,b_{n-1}}
		 _{a_0,\ldots,a_{n-1}}\right]
$ and 
$\ne$ is understood as a strict predicate. 
Here we rely on the simple fact that the number $n$ and functions 
$a_i$ and $b_i$ of $i<n$ are computable and \PCF-definable from $c\in\NN$. 
The value of $\#c f$ is \true\ if the strict finite function 
$	
	\varphi_c =
		 \left[\,^{b_0,\ldots,b_{n-1}}
		 _{a_0,\ldots,a_{n-1}}\right]
$
is inconsistent with $f$; $\#c f=\false$ if  
$	%
		 \left[\,^{b_0,\ldots,b_{n-1}}
		 _{a_0,\ldots,a_{n-1}}\right]
\sqsubseteq f
$; 
otherwise, $\#c f=\Undef$. Also, $\#\Undef f=\Undef$.

Now, any strict $F$ can be evidently represented as   
$F=\bigsqcup_{\alpha(k)\ne\Undef}
\left[\,_{\varphi_{\alpha(k)}}^{\beta(k)}\right]$, or as 
\[
F=F_0\mbox{ where }
F_r=\bigsqcup_{k\ge r,\alpha(k)\ne\Undef}
\left[\,_{\varphi_{\alpha(k)}}^{\beta(k)}\right]
=
\left[\,_{\varphi_{\alpha(r)}}^{\beta(r)}\right]\sqcup F_{r+1}
,\; 
r\ge0,
\]
with appropriate strict one place numeric functions $\alpha,\beta:\iota\arr\iota$ 
such that $\beta(k)=F\varphi_{\alpha(k)}$. Although we can take $\alpha(k)=k$, 
we will need the general case. Note that for arbitrary $\alpha$ and $\beta$ 
this lub may not 
exist if $\varphi_{\alpha(k)}$ and $\varphi_{\alpha(k')}$ 
are consistent, but $\beta(k)\ne\beta(k')$ 
for some $k$. 
We can evidently assume that $\alpha$ and $\beta$ are 
defined (${}\ne\Undef$) on the \emph{same} 
initial segment of $\NN$, finite or the whole $\NN$. 
(In fact, only two cases suffice here: the whole $\NN$, or 
the empty segment, if $F=\Undef$. 
But the case of an arbitrary segment will be needed later.)
Then for arbitrary $\alpha$ and $\beta$ 
for which the lub $F_r$ exists
we have $F_r\varphi_{\alpha(k)}=\beta(k)$ if $\alpha(k)\ne\Undef$, 
and
also 
$F_r=\Undef$ if $\alpha(r)=\Undef$. 
Then $F_r$ is also definable 
in $\PCF^+$ recursively on $r$ and thus by using the least fixed point 
operator $\Y$ as well as 
the parallel conditional function $\PIF$: 
\[
F'_r f=
	\PIF\ \#\alpha(r)f\ \THEN\ F'_{r+1}f\ \ELSE\ \beta(r).
\]
Let us show that the two definitions are equivalent ($F_r = F'_r$). 
First note that 
$F_r$ satisfying the first definition should also satisfy this formula 
with $=$ replaced by $\sqsupseteq$, thus giving $F'_r \sqsubseteq F_r$. 
Indeed, the value of the right-hand side, when defined, is equal either to 
$\beta(r)$, if 
$\varphi_{\alpha(r)}\sqsubseteq f$, 
or to 
$F_{r+1}f$. 
In both cases the left-hand side, $F_r f$, has evidently the same value. 
For the converse, $F_r \sqsubseteq F'_r$, it suffices to show that 
for the second definition 
we have 
$F'_r\varphi_{\alpha(k)}=\beta(k)$, 
for all $k\ge r$ with defined $\alpha(k)$, 
assuming that the above union does exist, and $\alpha$ and $\beta$ are 
defined on the same initial segment of $\NN$. 
This can be shown by induction on $k-r$: 
if $\varphi_{\alpha(r)}$ contradicts $\varphi_{\alpha(k)}$ then 
$F'_r\varphi_{\alpha(k)}=F'_{r+1}\varphi_{\alpha(k)}=\beta(k)$; 
otherwise, $F'_{r+1}\varphi_{\alpha(k)}=\beta(k)=\beta(r)$, and hence again  
$F'_r\varphi_{\alpha(k)}=\beta(k)$.

We can define, in \PCF, 
the \emph{correction operator} $\alpha,\beta\mapsto\alpha',\beta'$ 
with $\alpha'\sqsubseteq\alpha$ and $\beta'\sqsubseteq\beta$ 
by restricting $\alpha'=\alpha\restr\setof{k\in\NN\mid k\le n}$, and the same 
for $\beta$, for the maximal $n$ (possibly ${}=\infty$) such that 
the union 
$\bigsqcup^{k=n}_{k\ge0,\alpha'(k)\ne\Undef}
\left[\,_{\varphi_{\alpha'(k)}}^{\beta'(k)}\right]$
exists. 
Evidently, if the unrestricted union exists for the original 
$\alpha$ and $\beta$ then $\alpha'=\alpha$ and $\beta'=\beta$. 
This, together with the definition of $F_r$, 
constructs, in $\PCF^+$, a \emph{universal functional} 
$
\tilde{U}\alpha\beta:
((\iota\arr\iota)\arr\iota)
$
for all strict continuous functionals of the type $((\iota\arr\iota)\arr\iota)$.

Finally, for $F=\tilde{U}\alpha\beta$, the functional 
$Ff$ can be computed by the strategy $s$ whose behaviour 
is definable from the functions $\alpha'(k)$ and $\beta'(k)$ as follows: 
\[
\begin{array}{lcl}
\MM(s,\Lambda) & \bYdef & \#
\\
\MM(s,k) & \bYdef & \mbox{``}f(a_0)=\?\mbox{''},
\\
\MM(s,kb_0) & \bYdef & \mbox{``}f(a_1)=\?\mbox{''},
\\
\MM(s,kb_0 b_1) & \bYdef & \mbox{``}f(a_2)=\?\mbox{''},
\\
& \ldots &
\\
\MM(s,kb_0 b_1\cdots b_{n-2}) & \bYdef & \mbox{``}f(a_{n-1})=\?\mbox{''},
\\
\MM(s,kb_0 b_1\cdots b_{n-2} b_{n-1}) & \bYdef & {\beta'(k)},
\end{array}
\]
where 
$
	\varphi_{\alpha'(k)} =
		 \left[\,^{b_0,\ldots,b_{n-1}}
		 _{a_0,\ldots,a_{n-1}}\right]
$.
It is easy to see that $s$ is wittingly consistent.

\item[(b)] 
Define, essentially, 
\begin{align*}
G_m f  \bYdef &\; \IF\ G_m\Undef\ \textrm{is constant }c_m\ \THEN\ c_m\ 
\\
 & 
\;\ELSE\ \textrm{as in the proof of (a),}
\textrm{ by using some }\alpha_m,\beta_m:\iota\arr\iota. 
\end{align*}
This leads to an universal functional for the required class of 
type $(\iota,(\iota\arr\iota)\arr\iota)$ functionals.\qed
\end{enumerate}

\noindent
These definability considerations were devoted   
mainly to strict type 
\mbox{$(\iota\arr\iota)\arr\iota$} 
functionals of the standard continuous model $\setof{\bbD_{\alpha}}$. 
For the monotonic non-dcpo model $\setof{\tbbW_{\alpha}}$ we have isomorphisms   
$\tbbW_{\iota}\cong \bbD_{\iota}$, $\tbbW_{\iota\arr\iota}\cong \bbD_{\iota\arr\iota}$
(and also for all level 1 $\tbbW_\alpha$),
but 
$\tbbW_{(\iota\arr\iota)\arr\iota}\not\cong \bbD_{(\iota\arr\iota)\arr\iota}$, 
(by Section~\ref{seq:non-complete}). 
(The same holds for $\tbbQ_{\iota}$ and $\tbbQ_{\iota\arr\iota}$, 
whereas 
$\tbbQ_{(\iota\arr\iota)\arr\iota}$ is strictly embeddable in 
$\tbbW_{(\iota\arr\iota)\arr\iota}$ which is also strictly embeddable in 
$\bbD_{(\iota\arr\iota)\arr\iota}$ and consisting, thereby, of continuous 
functionals only.) 
Moreover,  $\tbbW_{(\iota\arr\iota)\arr\iota}$ contains all (but not only) 
strict continuous functionals. 
The latter holds because the above Lemma~\ref{lemma:strict-def} 
on the (relative) definability of strict continuous functionals holds 
in the $\PCF^+$-model $\tbbW$, as well as in $\bbD$.

\subsection{On Denotational Semantics of Wittingly Consistent Strategies}
\label{sec:den-sem-wit-strat}

\noindent
Let us look again at denotational semantics of 
any wittingly consistent 
system of strategies $\tuple{M,\MM}$.

For any 
strategy $m\in M$ of the type 
$\alpha=(\alpha_1,\ldots,\alpha_n\arr\iota)$  
define a 1-1 computable enumeration 
of the basic terms  
$A_{ma}\setof{\bx}$, 
$a\in\NN$, over $M$ with variables from the canonical list 
$\bx$ = $x_1:\alpha_1,\ldots,x_n:\alpha_n$ only 
which contains all queries to the Oracle 
potentially ``asked'' by the strategy $m$.  

For any such system $\tuple{M,\MM}$, let us construct  
a system of  
continuous functionals 
$\G^{\MM}_m:(\iota\arr\iota)\arr\iota$, $m\in M$,  
such that the denotational 
semantics $\Dsem{\defis}$ of the system $\tuple{M,\MM}$ 
in the model $\{\bbD_{\alpha}\}$ (respectively, in $\{\tbbW_{\alpha}\}$) 
may be equivalently defined 
(instead of explicitly using the interpreted computations) 
as the least solution of the system of equations%
\footnote{This means that the fixed point equation 
$\Dsem{\defis}=\Dsem{\defis}^+$ considered formerly 
can be represented in this form for appropriate $\G^{\MM}_m$. 
}
\begin{equation}\label{eq:den-sem-w}
\Dsem{m}\bx=\G^{\MM}_{m}(\lambda a.\Dsem{A_{ma}\{\bx\}}),\; m\in M.
\end{equation}
Here $x_i$ are ranging over $\bbD_{\alpha_i}$ 
(or, alternatively, over $\tbbW_{\alpha_i}$) 
and, for all $m\in M$, $\lambda a.\Dsem{A_{ma}\{\bx\}}$ are considered 
as strict functions in $\bbD_{\iota\arr\iota}=\tbbW_{\iota\arr\iota}$. 

The required functionals $\G^{\MM}_m$ can be defined as 
$\G^{\MM}_m(f)=v\in\NN$ if, and only if, for some 
$w=r_1\cdots r_k\in\NN^*$ the following two conditions hold:
\begin{enumerate}[(1)]
\item 
$\MM(m,w)=v$ (with $\MM(m,w')\not\in\NN$ for all initial segments 
$w'$ of $w$), and 
\item
for all $i<k$, 
if $\MM(m,r_1\cdots r_i)=A_{ma}$ (${}\ne\#$) then $r_{i+1}=f(a)$.
\end{enumerate}
This definition is correct ($v$ does not depend on the choice of 
$w$) because the system of strategies $\tuple{M,\MM}$ is 
wittingly consistent. 
Indeed, let $u=s_1\cdots s_n$ satisfy the analogous condition 
as $w$ with $\MM(m,u)=v'\ne v$. It follows that 
the pair $u,w$ is not $m$-consistent and for some proper 
initial segments $w'=r_1\cdots r_i$ and $u'=s_1\cdots s_j$,  
$\MM(m,w')=\MM(m,u')=A_{ma}\in\mbox{Basic-Terms}(M)$ and 
$f(a)=r_{i+1}\ne s_{j+1}=f(a)$ --- the contradiction.

The functional $\G^{\MM}_m(f)$ is also computable by a strategy 
$\sm:(\iota\arr\iota)\arr\iota$ 
\linebreak
induced by $m$: $\Dsem{\sm}=\G^{\MM}_m$. 
It behaves in the same way as $m$, except that 
instead of the queries 
``$A_{ma}={?}$'' it asks ``$f(a)={?}$'' for $a\in\NN$. 
The resulting system of 
strategies is denoted as $\tuple{\sM,\sMM}$. 
Evidently, $\tuple{\sM,\sMM}$ is sequential/wittingly 
consistent if $\tuple{M,\MM}$~is. 

The equation (\ref{eq:den-sem-w}) and its versions 
(\ref{eq:den-sem-w-special}) and (\ref{eq:den-sem-w-special-u}) below 
considerably simplify the corresponding equation in \cite{Saz76AL}%
\footnote{for a functional denoted there as $H$
} 
for the sequential case.
They will be needed for the construction in $\PCF^+$ of a universal 
functional $U^+_\alpha$ in Section~\ref{sec:univ-wittingly-consis}.  

\subsection{Definability of \texorpdfstring{$\G^{\MM}_m(f)$}{G-M-m(f)}}
\label{sec:def-of-F}

\noindent
Without restricting generality we can consider that 
the given wittingly consistent 
system of strategies $\tuple{M,\MM}$ 
is countable. 
Elements of $M$ may be numbered, or even identified with 
the natural numbers: $M=\NN$. Our current goal is to define 
the functional 
$\G^{\MM}_m(f)$ in \PCF$^+$ from some type 
$\iota\arr\iota$ numerical functions 
which can be computed from $\sMM$ (so that if $\MM$ 
is effectively computable, such are these numerical 
functions, too).

According to the strategy $m$ in $\tuple{M,\MM}$ or $\sm$ in 
$\tuple{\sM,\sMM}$, the functional $\G^{\MM}_m(f)$ is 
evidently either constant $c_m$ or strict. 
Therefore $\lambda mf.\G^{\MM}_m(f)$ is definable 
in $\PCF^+$ from some strict type $\iota\arr\iota$ functions 
by Lemma~\ref{lemma:strict-def} (b). 
Note, that the constants $c_m$, the (partial) predicate  
``$\G^{\MM}_m$ is a constant functional ${}\ne\bottom$'' 
and the corresponding numerical functions  
$\alpha_m,\beta_m:\iota\arr\iota$
for the strict $\G^{\MM}_m(f)$ 
used in the Lemma are effectively computable from $\sMM$ and $\sm$.

\subsection{A Universal Functional for Special Wittingly Consistent 
Systems of Strategies}
\label{sec:univ-special}

\noindent
Let us fix an arbitrary Basic-term $A\setof{j,\bar{y},\bx}:\iota$ 
constructed from 
\begin{enumerate}[$\bullet$]
\item 
symbols of the language \PCF,  
\item
a variable $j:\iota$ and a fixed list of variables
$\bar{y}=y_1,\ldots y_s$ of the same type 
$\gamma=(\gamma_1,\ldots,\gamma_n\arr \iota)$, and 
\item
a fixed list of variables 
$\bx=x_1:\gamma_1,\ldots,x_n:\gamma_n$. 
\end{enumerate}
Let us also fix a set $M=\PCF\cup\setof{\mu_0,\mu_1,\mu_2,\ldots}$ 
of strategies (the constant symbols) with all $\mu_p$ of the same type $\gamma$. 
Consider the class $\KA$ of all wittingly consistent systems of strategies 
$\tuple{M,\MM}$, with $M$ fixed as above  
and $\MM$ varying, but with the ordinary reductions for the 
constants of \PCF\ and such that the terms $\mu_p x_1,\cdots x_n$ 
can only be $\MM$-reduced to terms of the form 
\[
A\setof{\jj,\mu_{p_1},\mu_{p_2},\ldots,\mu_{p_s},x_1,\ldots,x_n}, 
\mbox{ or shortly }
A\setof{\jj,\bar{\mu}_{\bar{p}},\bx}
\]
with the same fixed $A$, 
where $\jj$ is a numeral ($0+1+\cdots+1$) and $p_1,\ldots,p_s$ 
are arbitrary  natural numbers. 
The class of effective systems in $\KA$ is called 
$\KAeff$.
\begin{lem}
\label{lemma:KA-universal}
Both for $\setof{\bbD_{\alpha}}$ and $\setof{\tbbW_{\alpha}}$, 
a universal functional 
\mbox{$U^A_{\gamma}:(\iota\arr\iota)\arr\gamma$} 
for some superset of $\KA$-computable functionals is definable in $\PCF^+$. 
Specifically, $U^A_\gamma f$ ranges over some superset of $\KA$-computable 
($\KAeff$-computable) type $\gamma$ functionals, 
if $f$ ranges over all (respectively, all effective) strict monotonic 
functions of the type 
$\iota\arr\iota$. 
In particular, each $\KA$-computable 
($\KAeff$-computable) type $\gamma$ functional 
is definable in $\PCF^+$ from some (effective, in the case of $\KAeff$) 
$f:\iota\arr\iota$. 
\end{lem}
\proof 
The above recursive equation (\ref{eq:den-sem-w}) becomes now 
\begin{equation}\label{eq:den-sem-w-special}
\Dsem{\mu_p}\bx=
G^{\MM}(p,\lambda j\bar{p}.
\Dsem{A\{j,\bar{\mu}_{\bar{p}},\bx\}})
\end{equation}
with $x_i$ ranging over $\bbD_{\alpha_i}$ (respectively, over $\tbbW_{\alpha_i}$).
It is inessential that $G^{\MM}$ here has a slightly different 
type than in~(\ref{eq:den-sem-w}). 
So, it is still definable in $\PCF^+$ from some type $\iota\arr\iota$ strict 
functions computable from $\MM$. 

Now, consider a variable $u:\iota\arr\gamma$ and the following 
version of the above recursive equation 
\begin{equation}\label{eq:den-sem-w-special-u}
up\bx=
G^{\MM}(p,\lambda j\bar{p}.
{A\{j,up_1,\ldots,up_s,\bx \}})
\end{equation}
By using combinators $\Ss$ and $\K$ to simulate lambda abstraction, and 
the least fixed point combinator $\Y$ of an appropriate type, 
this gives rise to a $\PCF^+$-term 
$\hat{U}^A_{\gamma}\setof{\bar{f}}:\iota\arr\gamma$ 
(corresponding to the above variable $u:\iota\arr\gamma$)
depending on some, actually strict, functions $\bar{f}:\iota\arr\iota$ which 
were involved in the $\PCF^+$-definition of $\G^{\MM}$. 
By some trivial encoding this gives rise to the required 
$\PCF^+$-term $U^A_{\gamma}:(\iota\arr\iota)\arr\gamma$ 
involving no variables $\bar{f}$ at all.\qed

\begin{note}\em
Lemma~\ref{lemma:KA-universal} may be easily generalized to the case 
of any finite number of terms $\bar{A}\setof{j,\bar{y},\bx }:\iota$ 
with the same variables, giving rise to the universal functional 
$U^{\bar{A}}_{\gamma}$ for $\KbarA$-computable functionals 
of the type $\gamma$ . 
\end{note}

\subsection{A Universal Functional for all Wittingly Consistent 
Functionals of a Given Type}
\label{sec:univ-wittingly-consis}

\noindent
The general universal $\PCF^+$-definable functional 
$U^+_{\alpha}\in\tbbW_{(\iota\arr\iota)\arr\alpha}$, or its version 
$\in \bbD_{(\iota\arr\iota)\arr\alpha}$, for all wittingly consistent 
functionals of any given type $\alpha$
can be obtained from $U^{\bar{A}}_{\gamma}$ for suitable $\gamma$ 
and $\bar{A}$ by using only \PCF. 
Here we also employ the fact that, without restricting generality, 
we can consider only systems of strategies $m$ asking queries 
in the canonical form (\ref{eq:canonical}). 
Given any such $m$, this allows us to ``concentrate'', by some encoding 
most of the strategies descendant to $m$ 
(having levels $\le$ the level of $m$) 
in a finite number of types, and, even in only one type $\gamma$, 
(and, analogously, to further restrict 
the form of queries). That is, the general wittingly 
consistent systems of strategies can be 
reduced to the special systems of some class $\KbarA$ considered above. 
We omit the details which are presented in \cite{Saz76AL}.

\section{Conclusion}
\label{sec:conclusion}

\noindent
A generalized non-dcpo domain theoretic framework for finite type functionals 
which are not necessarily closed under directed limits was presented in this paper 
in terms of pointwise (natural) least upper bounds, and corresponding 
natural continuity, natural algebraicity and natural bounded completeness 
properties. 

An inductive definition of a monotonic fully abstract 
model $\bbQ$ for \PCF\ satisfying the above properties and 
based on a quite general concept of 
sequential strategies was also given. 
This model consists hereditarily of all finite type functionals 
computable by the sequential 
strategies which also prove to be uniformly definable in \PCF\ from 
(strict)
functions of the type $\iota\arr\iota$. 
This is the universality property also characterising 
precisely the expressive power of \PCF. 
Thereby we have demonstrated 
that the old concept of sequential strategies \cite{Saz76SMZH,Saz76AL} 
can be used quite naturally for defining the fully abstract 
model along with 
the more recent game approach 
\cite{Abramsky-Jagadeesan2000,Hyland-Ong2000,Nickau96}. 
The uniqueness of $\bbQ$ was also shown. 
The essential feature of our definition is its straightforward, 
inductive and computational character. 
For each level we just hereditarily restrict the class of monotonic 
functionals to those that are sequentially computable. 
However, either the correctness proof of the induction step of 
this definition, if based on (\ref{eq:dom}), 
or (in the case of alternative definition based on (\ref{eq:dom'}) with a simpler 
correctness) proving the main properties 
of $\bbQ$ is more complicated 
and requires developing a general and quite involved theory of all 
computational strategies 
with their generalized operational semantics 
coherent with the denotational one. 
In this way the above ``natural'' 
non-dcpo domain theoretic continuity and other properties of 
$\bbQ\cong\tQ$ are also shown. 

Quite analogous inductive definition of a fully abstract model 
$\bbW\cong\tW$ for $\PCF^+ = \PCF+{}$``\mbox{parallel OR}'' satisfying 
the above non-dcpo domain theoretic properties 
+ the universality property relative to $\PCF^+$ 
was also briefly outlined in terms 
of wittingly consistent nondeterministic strategies.  
The model $\bbW$ proves to be not $\omega$-complete, as well as  
the model of sequential functionals $\bbQ$ 
for which this was shown in \cite{Normann2004}.  

As the future perspective, it would be interesting 
to develop a game semantics version of wittingly consistent strategies. 
Recall also several domain theoretic hypotheses from Section~\ref{sec:natural} 
on the model $\bbQ$ (equally applicable to $\bbW$) 
related with the fact that it is not $\omega$-complete, as well as the 
hypotheses concerning effectiveness of representation of naturally finite functionals 
in Section~\ref{sec:efficiency} and the related Notes~\ref{note:finite-simeq-finitary} 
and~\ref{note:closure-under-appl-of-finite-strategies} on finite and finitary strategies.

\section*{Acknowledgement} 

\noindent
The author is grateful to Gordon Plotkin 
for fruitful discussions on the subject, to Achim Jung for his comments on 
the domain theoretic part,  
and to Michael Fisher for his kind help in polishing the English. 
Thanks to the referees for numerous useful comments helping to considerably improve 
the exposition and in particular for the amending Definition~\ref{def:bbQ} 
which made its correctness proof just straightforward. 

\bibliographystyle{plain}
\bibliography{lmcs-PCF}

\appendix 
\section{Universal System of Sequential Strategies 
\texorpdfstring{$\tuple{Q,\QQ}$}{}}
\label{appendix:univ-sys-strategies}

\newcommand{\BT}{\mbox{BT}_\Box}
\newcommand{\qq}{q}
\newcommand{\pp}{p}
\newcommand{\type}{\textrm{\bf type}}
\newcommand{\Types}{\textrm{Types}}
\newcommand{\Erase}{\Box}

\noindent
Here we give a construction of the typed version of the universal system of strategies 
$\tuple{Q,\QQ}$ \cite{Saz76SMZH} 
(with the details which are a bit more complicated than in the untyped case 
presented formerly only in \cite{Saz76t}). 

Let $\bNN$ consist of duplicates of natural numbers $\bar{0},\bar{1},\bar{2},\ldots$ 
so that $\bNN$ is disjoint with $\NN$ (and with any other set considered below), 
and $\Box_\alpha$ be the ``empty'' constant (placeholder for a strategy 
from~$Q_\alpha$) of a type $\alpha$ for each~$\alpha$. 
Then, according to Section~\ref{sec-seq-stra-def-prelim}, 
$\BT\bYdef\mbox{Basic-Terms}(\setof{\Box_\alpha\mid\alpha\in\Types})$ 
is the set of basic terms (possibly with variables) over 
the set of constants $\Box_\alpha$. 

Define $Q$ (recursively) as the set of all functions 
\[
\qq:(\NN\cup\bNN)^*\cup\setof{\type}\arr \NN_\Undef\cup\BT\cup\Types
\]
(considered as partial due to $\Undef\in \NN_\Undef$) 
satisfying the following conditions for all $u,w\in(\NN\cup\bNN)^*$ and 
$\bj\in\bNN$:
\begin{enumerate}[(1)]
\item 
$\qq((\NN\cup\bNN)^*)\subseteq\NN_\Undef\cup\BT$.
\item
$q(\type)\in\Types$. 

\noindent
We write $q:\alpha$ if $q(\type)=\alpha$ and take 
$Q_\alpha\bYdef\setof{q\in Q\mid q:\alpha}$.  
If we have a map $\qq':(\NN\cup\bNN)^*\arr \NN_\Undef\cup\BT$ 
(i.e.\ $\qq'$ is undefined on $\type$) then writing 
$\qq':\alpha$ also means ``assignment'' of the type $\alpha$ to $\qq'$, i.e.\ 
adding $\setof{\type\mapsto\alpha}$ to the graph of $\qq'$ so that 
$\qq=(\qq':\alpha)$ is a map with $\qq(\type)=\alpha$. 
\item
$\qq(u)\in \NN_\Undef\Arr \qq(uw)=\Undef$ for non-empty $w$.

\item
{\bf If} $\qq(u)=B\in\BT $ and $u\in\NN^*$ {\bf then} all variables in $B$ are 
from the canonical list $x_1,\ldots,x_n$ for the type of $\qq$ 
(the condition similar to that in 
Definition~\ref{def:strategies}(\ref{item:strategies:variables})).

\item
{\bf If} $\qq(u)=B\in\BT$ (with $u\in(\NN\cup\bNN)^*$) 
{\bf and} $B$ contains ${}<j$ occurrences 
of the symbol $\Box$ 
{\bf then} 
$\qq(u\bj w)=\Undef$, 
{\bf otherwise}, if the $j$-th occurrence of $\Box$ in $B$ has the type 
$\beta_j$ then 
$
q_j=((\lambda w.\qq(u\bj w)):\beta_j)\in Q
\ (\textrm{in fact, }{}\in Q_{\beta_j}). 
$
\end{enumerate}
More precisely, we take the set $Q$ to be the \emph{largest} 
one whose elements $\qq$ satisfy the above conditions, i.e.\ the largest set  
satisfying 
\[
Q\subseteq\setof{\qq:(\NN\cup\bNN)^*\cup\setof{\type}\arr \NN_\Undef\cup\BT\cup\Types
\mid\Phi(q,Q)}
\] 
where $\Phi(q,Q)$ is the (universally quantified by $u,w$ and $\bj$) 
conjunction of the above conditions (1)--(5), which is 
monotonic on $Q$. 
(Note that the least such set is just empty. 
Thus, the definition of $Q$ is, in fact, \emph{co-recursive}.) 

Define a function $\QQ:Q\times \NN^*\arr \NN_\Undef\cup\mbox{Basic-Terms}(Q)$, 
making the pair $\tuple{Q,\QQ}$ a system of strategies, 
by taking for all 
$\qq\in Q$ and $u\in\NN^*$
\begin{equation}\label{eq:univ-system-map}%
\QQ(\qq,u)\bYdef\left\{
\begin{array}{lll}
r,                              & \textrm{if}&\qq(u)=r\in \NN_\Undef, \\
A[\qq_1,\qq_2,\ldots]\in\mbox{Basic-Terms}(Q),& \textrm{if}&\qq(u)=A\in\BT 
                                                                    \textrm{ and}\\
                                                    &            &
q_j=(\lambda w.\qq(u\bj w)):\beta_j,\ \\                     
                                                    &            &                      
                                                                      j\ge1.
\end{array}\right.
\end{equation}
Here $A[\qq_1,\qq_2,\ldots]$ 
is the term obtained as the result of the substitution 
in $A$ of the strategies $\qq_1,\qq_2,\ldots$, 
respectively, in place of the first, second, etc.\ occurrences of $\Box$ in $A$, and 
$\beta_1,\beta_2,\ldots$ are the types of these occurrences.

Our goal is to show the universality of the defined system of strategies 
$\tuple{Q,\QQ}$. First, define $\mu_j(B)$, for any $B\in\textrm{Basic-Terms}(M)$, 
as the $j$-th occurrence of an element from $M$ in term~$B$. If $B$ has ${}<j$ 
occurrences of elements from $M$ then $\mu_j(B)$ is undefined. 
(If~$B\in\BT$ then $\mu_j(B)$ is the $j$-th occurrence of a $\Box$-symbol in $B$.)
Denote by 
$\Erase_M(B)$ the result of ``erasing'' in $B$ of all occurrences of elements 
from $M$, i.e.\ the result of replacement of all such occurrences by the symbol 
$\Box$ (of appropriate type). 
Evidently, 
\[
B=\Erase_M(B)[\mu_1(B),\mu_1(B),\ldots] 
\]
(and dually for $B\in\BT$). 
It will also be convenient to define $\Erase_M(r)=r$ for $r\in\NN_\Undef$ 
and $\Erase_M(\alpha)=\alpha$ for $\alpha\in\Types$. 

Given any system of strategies $\tuple{M,\MM}$ and $m\in M$, define
two functions 
\begin{align*}
&\bMM_m:(\NN\cup\bNN)^*\cup\setof{\type}\arr 
                              \NN_\Undef\cup\textrm{Basic-Terms}(M)\cup\Types, \\
&\MM_m: (\NN\cup\bNN)^*\cup\setof{\type}\arr \NN_\Undef\cup\BT\cup\Types,
\end{align*} 
by letting 
$
\bMM_m(\type)=\MM_m(\type)=\textrm{the type of }m, 
$
and, iteratively, for any 
$u\in\NN^*$, $w\in(\NN\cup\bNN)^*$ and $\bj\in\bNN$,  
\begin{equation}\label{eq:bMM}
\begin{array}{lll}
\bMM_m(u)     &=&\MM(m,u), \\[0.5em]
\bMM_m(u\bj w)&=&\left\{
                      \begin{array}{lll}
                      \bMM_{m_j}(w), & \textrm{if} & \MM(m,u)=B\in\textrm{Basic-Terms}(M)\\
                                     &             &\textrm{and } m_j=\mu_j(B),\\
                      \Undef,  & \textrm{if} & \MM(m,u)\not\in\textrm{Basic-Terms}(M) \\
                      & & \textrm{or }\mu_j(B)\textrm{ is undefined},
                      \end{array}
                 \right. \\[2.5em]
\MM_m(w)      &=&\Erase_M(\bMM_m(w)).
\end{array}
\end{equation}
For any system of strategies $\tuple{M,\MM}$ the elements $\qq$ of the set $Q_\MM\bYdef\setof{\MM_m\mid m\in M}$ 
satisfy the conditions (1)--(5) above. Therefore $Q_\MM\subseteq Q$ 
(preserving types). 
\begin{lem}\label{lemma:hom-universal}
If $\varphi:\tuple{M,\MM}\arr\tuple{M',\MM'}$ is a homomorphism then 
$\MM'_{\varphi(m)}=\MM_m$ for all $m\in M$. 
\end{lem}
\proof
First note that for all $m\in M$ and $w\in(\NN\cup\bNN)^*$ the equality 
$
\bMM'_{\varphi(m)}(w)=(\bMM_m(w))^\varphi
$
evidently holds. 
It follows that 
$\MM'_{\varphi(m)}(w)$
$=$
$\Erase_{M'}(\bMM'_{\varphi(m)}(w))$
$=$
$\Erase_{M'}(\bMM_m(w))^\varphi$
$=$
$\Erase_{M}(\bMM_m(w))$
$=$
$\MM_m(w).$
\qed
\begin{lem}\label{lemma:universal}
$\QQ_q=q$ for all $q\in Q$. Therefore 
$Q=\bigcup_{\tuple{M,\MM}} Q_\MM$. 
\end{lem}
\proof
As $\QQ_q=\Erase\circ\bQQ_q$ 
and $\QQ_q(\type)=\bQQ_q(\type)=q(\type)$, 
it evidently suffices to show that, for $u\in(\NN\cup\NN)^*$ 
and $q\in Q$, 
\begin{equation}\label{eq:lemma:universal}
\bQQ_q(u)=\left\{
\begin{array}{lll}
r,                              & \textrm{if}&\qq(u)=r\in \NN_\Undef, \\
A[\qq_1,\qq_2,\ldots]\in\mbox{Basic-Terms}(Q),&\textrm{if}&\qq(u)=A\in\BT 
                                                                    \textrm{ and}\\
                                                    &            &
q_j=(\lambda w.\qq(u\bj w)):\beta_j,\ \\                     
                                                    &            &                      
\beta_j=\textrm{type of }\mu_j(A),\; j\ge1. %
                                               \end{array}\right.                 
\end{equation}
This equality is proved by induction on the number of occurrences of symbols from $\bNN$ 
in the string $u$. If $u\in\NN^*$ 
then the equality evidently follows from (\ref{eq:univ-system-map}) and (\ref{eq:bMM}). 
Let $u=y\bi z$ where $y\in\NN^*$, $\bi\in\bNN$ and $z\in(\NN\cup\bNN)^*$. 
Two cases are possible. 
\begin{enumerate}[(1)]
\item
$q(y)=B\in\BT$. Then, using (\ref{eq:univ-system-map}), 
$\QQ(q,y)=B[\pp_1,\pp_2,\ldots]$ 
for $\pp_i=\lambda t.q(y\bi t):\gamma_i$ and an appropriate type $\gamma_i$,
and hence $\bQQ_q(y\bi z)=\bQQ_{\pp_i}(z)$ 
according to (\ref{eq:bMM}). 
The number of occurrences from $\bNN$ in string $z$ is less than in $u$. 
Therefore the formula (\ref{eq:lemma:universal}) is applicable by induction:
\begin{equation}%
\nonumber
\bQQ_{\pp_i}(z)=\left\{
                 \begin{array}{lll}
                 r,              & \textrm{if} & \pp_i(z)=q(y\bi z)=q(u)=r\in\NN_\Undef,\\
                 A[q_1,q_2,\ldots],%
                                 & \textrm{if} & 
                                                  \pp_i(z)=q(u)=A\in\BT,\textrm{ and} \\
                                 &              & 
q_j=\lambda w.\pp_i(z\bj w):\beta_j \\
                                 &              & 
\hspace{1.15em}   =\lambda w.q(y\bi z\bj w):\beta_j \\
                                 &              & 
\hspace{1.15em}   =\lambda w.q(u\bj w):\beta_j, \\
                                 &              & 
\beta_j=\textrm{type of }\mu_j(A),\; j\ge1.
                 \end{array}
           \right.
\end{equation}
Since $\bQQ_{q}(u)=\bQQ_q(y\bi z)=\bQQ_{\pp_i}(z)$, this gives exactly the equality 
(\ref{eq:lemma:universal}). 

\smallskip

\item
$q(y)\in\NN_\Undef$. Then $\QQ(q,y)\not\in\textrm{Basic-Terms}(Q)$ and therefore 
$\bQQ_q(u)=\bQQ_q(y\bi z)=\Undef$ by (\ref{eq:bMM}). Then $q(y)\in\NN_\Undef$ 
entails $q(u)=q(y\bi z)=\Undef$ by the definition of $Q$. Therefore 
(\ref{eq:lemma:universal}) holds in this case as well.\qed
\end{enumerate}
\begin{thm}\label{th:universal-system}
$\tuple{Q,\QQ}$ is the unique up to isomorphism universal system of strategies. 
For any system $\tuple{M,\MM}$, the map $m\mapsto\MM_m$ is the unique 
homomorphism $\tuple{M,\MM}\arr\tuple{Q,\QQ}$.
\end{thm}
\proof
The uniqueness of the homomorphism follows from Lemmas~\ref{lemma:hom-universal} 
and~\ref{lemma:universal}. 
For, if $\varphi:\tuple{M,\MM}\arr\tuple{Q,\QQ}$ is a homomorphism then 
$
\varphi(m)=\QQ_{\varphi(m)}=\MM_m
$
for any $m\in M$. 

To establish that $m\mapsto\QQ_m$ is a homomorphism we need to show that 
for all $m\in M$ and $u\in \NN^*$ that 
\begin{enumerate}[(i)]
\item
$\MM(m,u)=r\in\NN_\Undef\Arrr\QQ(\MM_m,u)=r$, and
\item
$\MM(m,u)=A[m_1,m_2,\ldots]\in\textrm{Basic-Terms}(M)\Arrr
                 \QQ(\MM_m,u)=A[\MM_{m_1},\MM_{m_2},\ldots]$. 
\end{enumerate}
Here we assume that $A\in\BT$ and $m_j:\beta_j$, $j\ge1$. 
The first implication is easy. 
In the second, we need to show by the definition of $\QQ$ that 
$\MM_m(u)=A$, and $\MM_{m_j}=\lambda w.\MM_m(u\bj w):\beta_j$, 
$j\ge1$. Both equalities follow from (\ref{eq:bMM}). The first is easy. For the second, 
we get $\bMM_m(u\bj w)=\bMM_{m_j}(w)$ 
for all $w\in(\NN\cup\bNN)^*$ and $j\ge1$, 
and apply the erasing operator~$\Box_M$. \qed

\end{document}